\documentclass[aps,preprint,nofootinbib,preprintnumbers,eqsecnum]{revtex4-1}

\usepackage[usenames, dvipsnames]{color}

\newcommand{\nb}[1]{\color{blue}}

\newcommand{\HL}[1]{{\textcolor{magenta}{#1}}}

\newcommand{\hl}[1]{\color{magenta}}

\usepackage[
	  pagebackref=false,
	  colorlinks=true,
      linkcolor=blue,
      urlcolor=blue,
      filecolor=black,
      citecolor=red,
      pdfstartview=FitV,
      pdftitle={},
        pdfauthor={},
        pdfsubject={},
        pdfkeywords={},
        pdfpagemode=None,
        bookmarksopen=true
      ]{hyperref}

\usepackage[normalem]{ulem}
\usepackage{amsmath}
\usepackage{enumerate}
\usepackage{amsfonts}
\usepackage{epsfig}
\usepackage{mathbbol}

\def\Tr{\mathop{\rm Tr}}

\def\Im{\mathop{\rm Im} }

\newcommand\half{{\ensuremath{\frac{1}{2}}}}
\newcommand\p{\ensuremath{\partial}}

\newcommand\field[1]{{\ensuremath{\mathbb{{#1}}}}}

\newcommand\vev[1]{{\ensuremath{\left\langle{#1}\right\rangle}}}

\newcommand\ket[1]{\ensuremath{\lvert{#1}\rangle}}
\newcommand\bra[1]{\ensuremath{\langle{#1}\rvert}}

\newcommand{\RR}{\field{R}}

\newcommand{\be}{\begin{equation}}
\newcommand{\ee}{\end{equation}}
\newcommand{\bea}{\begin{eqnarray}}
\newcommand{\eea}{\end{eqnarray}}
\newcommand{\bega}{\begin{gather}}
\newcommand{\eega}{\end{gather}}

\newcommand{\bi}{\begin{itemize}}
\newcommand{\ei}{\end{itemize}}
\newcommand{\ben}{\begin{enumerate}}
\newcommand{\een}{\end{enumerate}}
\newcommand{\bca}{\begin{cases}}
\newcommand{\eca}{\end{cases}}
\newcommand{\bln}{\begin{align}}
\newcommand{\eln}{\end{align}}
\newcommand{\bst}{\begin{split}}
\newcommand{\est}{\end{split}}
\def\ie{\begin{equation}\begin{aligned}}
\def\fe{\end{aligned}\end{equation}}
\newcommand{\bma}{\le(\begin{matrix}}
\newcommand{\ema}{\end{matrix}\ri)}

\newcommand\al{{\alpha}}
\def\b{{\beta}}
\newcommand\ep{\epsilon}
\newcommand\sig{\sigma}
\newcommand\Sig{\Sigma}
\newcommand\lam{\lambda}
\newcommand\Lam{\Lambda}
\newcommand\om{\omega}
\newcommand\Om{\Omega}

\newcommand\ga{{\ensuremath{{\gamma}}}}
\newcommand\Ga{{\ensuremath{{\Gamma}}}}
\newcommand\de{{\ensuremath{{\delta}}}}
\newcommand\De{{\ensuremath{{\Delta}}}}

\newcommand\ze{\zeta}

\newcommand\da{{\dagger}}

\def\th{{\theta}}

\newcommand\ov{\over}
\newcommand\ha{{\half}}

\def\le{\left}
\def\ri{\right}

\newcommand\sA{{\ensuremath{{\mathcal A}}}}
\newcommand\sB{{\ensuremath{{\mathcal B}}}}
\newcommand\sC{{\ensuremath{{\mathcal C}}}}

\newcommand\sF{{\ensuremath{{\mathcal F}}}}

\newcommand\sH{{\ensuremath{{\mathcal H}}}}
\newcommand\sK{{\ensuremath{{\mathcal K}}}}
\newcommand\sL{{\ensuremath{{\mathcal L}}}}
\newcommand\sM{{\ensuremath{{\mathcal M}}}}
\newcommand\sN{{\ensuremath{{\mathcal N}}}}
\newcommand\sO{{\ensuremath{{\mathcal O}}}}
\newcommand\sP{{\ensuremath{{\mathcal P}}}}

\newcommand\sJ{{\mathcal J}}
\newcommand\sR{{\mathcal R}}
\newcommand\sS{{\mathcal S}}
\newcommand\sT{{\mathcal T}}

\newcommand\sX{{\mathcal X}}
\newcommand\sY{{\mathcal Y}}

\newcommand\vx{{\vec x}}

\newcommand{\fb}{{\mathfrak{b}}}

\newcommand{\rb}{{\rm{b}}}

\newcommand{\rk}{{\mathrm{k}}}

\newcommand{\rt}{{\rm t}}

\newcommand{\ktfd}{{\ket{\Psi_\b}}}
\newcommand{\tfd}{{\rm TFD}}

\newcommand{\bid}{\mathbf{1}}

\begin{document}

\title{Emergent times in holographic duality}

\preprint{MIT-CTP/5382}

\author{Sam Leutheusser and Hong Liu}
\affiliation{Center for Theoretical Physics, 
Massachusetts
Institute of Technology, \\
77 Massachusetts Ave.,  Cambridge, MA 02139 }

\begin{abstract}

 \noindent 
 
 In holographic duality an eternal AdS black hole is described by two copies of the boundary CFT in the thermal field double state. 
In this paper we provide explicit constructions in the boundary theory of infalling time evolutions which can take bulk observers behind the horizon. The constructions also help to illuminate the boundary emergence of the black hole horizons, the interiors, and the associated causal structure. A key element is the emergence, in the large $N$ limit of the boundary theory, of a type III$_1$ von Neumann algebraic structure from the type I boundary operator algebra and the half-sided modular translation structure associated with it.

\end{abstract}

\today

\maketitle

\tableofcontents

\section{Introduction}


\begin{figure}[h]
\begin{centering}
	\includegraphics[width=4.5in]{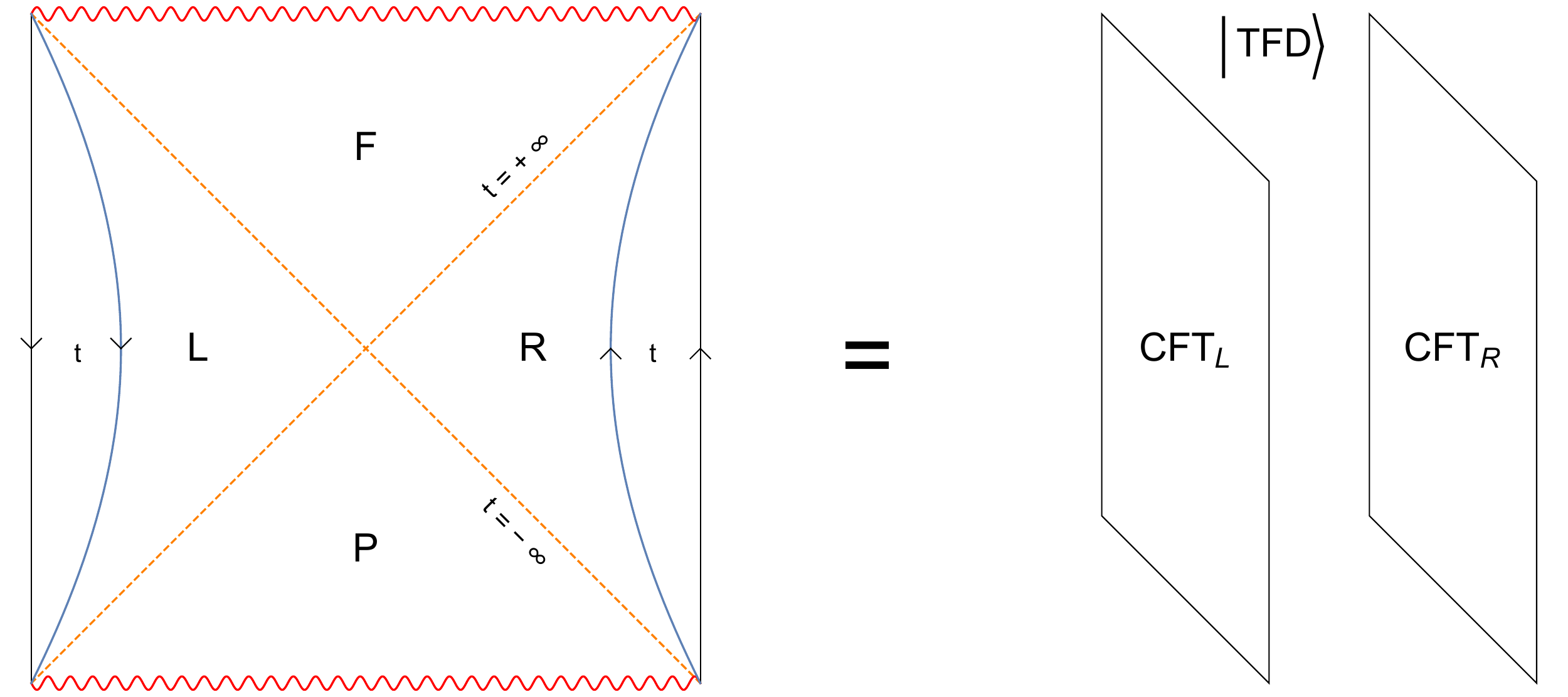}
\par\end{centering}
\caption{
The Penrose diagram of an eternal black hole. The dashed lines are event horizons and the wavy lines are the singularities. 
}
\label{fig:casu1}
\end{figure}

Time is a baffling concept in quantum gravity. While it plays an absolute role in the formulation of quantum mechanics, in gravity it can be arbitrarily reparameterized by gauge diffeomorphisms and hence lacks a definite meaning.  In an asymptotic anti-de Sitter (AdS) spacetime, a sensible notion of boundary time can be established in the asymptotic region as  gauge transformations generating time reparameterizations are required to vanish at spatial infinities. For static spacetimes with a global timelike Killing vector, the asymptotic time can be extended to the interior with the help of the symmetry. 
But for spacetimes without such a symmetry, whether it is possible to describe time flows in the interior in a diffeomorphism invariant way is a subtle question whose understanding is important in many contexts.  

\begin{figure}[h]
\begin{centering}
	\includegraphics[width=4.5in]{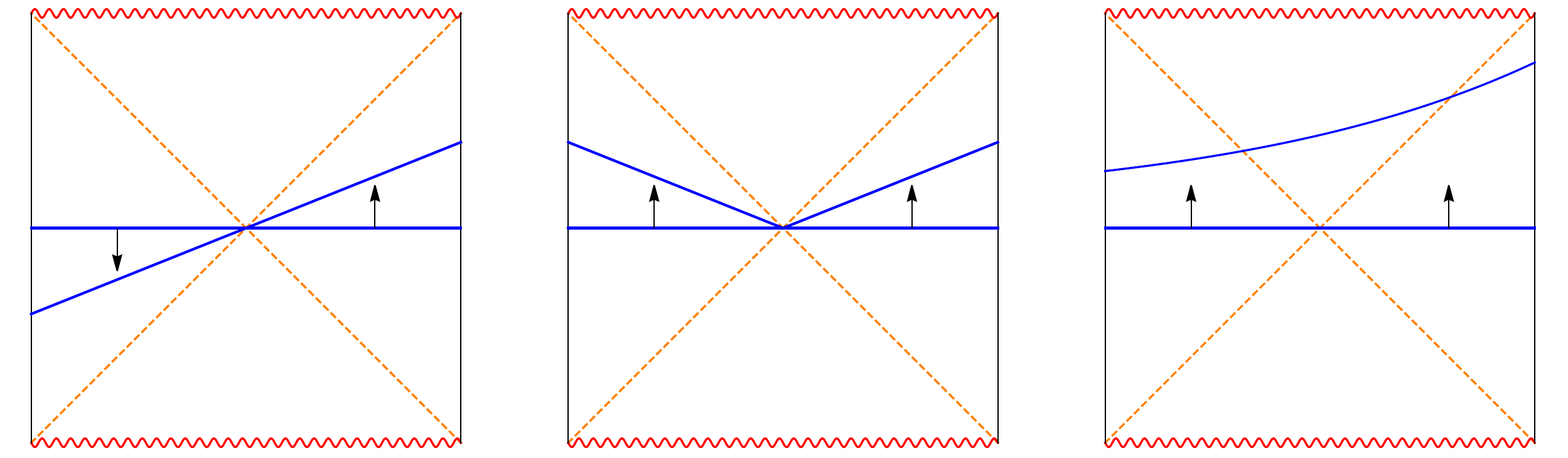} \qquad 
\par\end{centering}
\caption{Left: Evolution of the $\rt =0$ bulk slice under $H_R - H_L$, where $H_{R, L}$ denote the Hamiltonians of the boundary theories. Center: Evolution of the $\rt =0$ slice under $H_R + H_L$, the action of which is singular at the horizon. 
In fact any linear combination of $H_R$ and $H_L$ other than $H_R - H_L$ is expected to have a singular action at the horizon. Right: A smooth Kruskal-like evolution. If such an evolution can be described in a diffeomorphism invariant way, it must be emergent in the boundary theory. 
}
\label{fig:evolutions}
\end{figure}

For this purpose an eternal black hole in AdS, which is dual to two copies of the boundary CFT in the thermal field double state~\cite{Maldacena:2001kr} (see Fig.~\ref{fig:casu1}), offers perhaps a simplest nontrivial example. 
The black hole spacetime possesses a time-like Killing vector in the exterior $R$ and $L$ regions. The associated time $\rt$, which can be considered as the extension of the boundary time, however, ends at the event horizon, with no timelike Killing vector inside the horizon. A natural question is whether the boundary theory can describe an ``infalling'' time evolution, which we define as any evolution which can take the Cauchy slice at $\rt=0$ to Cauchy slices which go inside the horizon. 
Such a time, if it exists, must be emergent, as the evolutions of the usual boundary times do not probe the interior, see Fig.~\ref{fig:evolutions}.

 There have been many different ways that boundary observables can probe regions behind the horizon see e.g.~\cite{Kraus:2002iv,Fidkowski:2003nf,Festuccia:2005pi,Hartman:2013qma,Liu:2013iza,Liu:2013qca,Susskind:2014rva,Grinberg:2020fdj,Zhao:2020gxq,Haehl:2021emt,Haehl:2021prg,Haehl:2021tft}, but in these discussions neither an infalling time evolution nor the casual structure of the horizon was visible from the boundary, except in systems with symmetries~\cite{Maldacena:2018lmt,Lin:2019qwu}.
 Similarly, ER$=$EPR type arguments~\cite{VanRaamsdonk:2010pw,Maldacena:2013xja} are largely concerned with a single time slice. 
 While it is possible to express bulk operators in the black hole interior regions in terms of boundary operators~\cite{Hamilton:2005ju,Hamilton:2006fh,Papadodimas:2012aq,Papadodimas:2013jku,Papadodimas:2015xma}, such ``bulk reconstructions'' require either evolving bulk equations of motion or analytic continuation around the horizon, and thus are not intrinsically boundary constructions. See 
 also~\cite{Jafferis:2020ora,Gao:2021tzr} for an interesting recent discussion of keeping track of the proper time of an in-falling observer using modular flows and~\cite{Nomura:2018kia,Nomura:2019qps,Nomura:2019dlz,Langhoff:2020jqa,Nomura:2020ska}
for a description of the black hole interior from the perspective of coarse-graining.

In this paper we provide an explicit construction of infalling time evolutions from the boundary theory.\footnote{A summary of the main idea and results has appeared earlier in~\cite{Leutheusser:2021qhd}.}  
It should be emphasized that our goal is not to describe  in-falling geodesic motion of some localized bulk observers, which in general cannot be formulated in a diffeomorphism invariant way. The goal is to construct ``global'' evolutions of a Cauchy slice 
as in Fig~\ref{fig:evolutions}(c). 
Understanding such emergent evolutions also helps to illuminate the emergence in the boundary theory of the bulk horizon and the associated  causal structure.

The key to our discussion is the emergence, in the large $N$ limit of the boundary theory, of a type III$_1$ von Neumann algebraic structure\footnote{For reviews on the classification of von Neumann algebras see chapter III.2 of~\cite{Haag:1992hx} or section 6 of~\cite{Witten:2018zxz}.} from the type I boundary operator algebra and the half-sided modular translation structure associated with it. 
A distinctive property of the ``evolution operators'' $U(s)= e^{- i G s}, \, s \in \RR,$  resulting from this construction is that  
the Hermitian {generator} $G$ has a spectrum that is {\it bounded from below},
\be \label{soen}
G \geq 0 \ .
\ee
The spectrum property is natural from the following perspectives: 
(i) It distinguishes $G$, as a generator of ``time'' flow,  from an operator generating other unitary transformations, e.g. spacelike  displacements or internal symmetries, whose spectrum is not bounded from below. 
(ii) If we interpret the eigenvalues of $G$ as energies
associated with the ``global'' infalling time $s$, they {should}  be bounded from below to ensure stability. The existence of the singularity means that such evolution may only have a finite ``lifetime,'' but there should nevertheless exist  a well-defined quantum mechanical description before hitting the singularity. Also by construction $G$ involves degrees of freedom from both CFT$_R$ and CFT$_L$.\footnote{The necessity of left/right couplings has previously been discussed. For example, see~\cite{Mathur:2014dia}.} 

Our discussion will be restricted to leading order in the $1/N$ expansion, but we expect the structure uncovered should be present to any finite order in the expansion. New structure from incorporating $1/N$ corrections to all orders is discussed in~\cite{WittenNew}.

The plan for this paper is as follows. In section~\ref{sec:GTFD} we discuss the emergence of a type III$_1$ vN algebra in the boundary theory at finite temperature. In section~\ref{sec:locIII} we discuss the emergence of a new type III$_1$ structure for local boundary algebras in the large $N$ limit. In section~\ref{sec:impl} we suggest several physical implications of these emergent type III$_1$ algebras. In section~\ref{sec:half} we review half-sided modular inclusion/translation. In section~\ref{sec:ext} we show that half-sided translations can be uniquely extended to all values of the parameter and that the description of these evolution operators is completely fixed, up to a phase, for algebras generated by generalized free fields. In section~\ref{sec:bdgen} we illustrate our construction of evolution operators in the simple case of generalized free fields on Rindler spacetime. In section~\ref{sec:recon} we review bulk reconstruction in the AdS-Rindler and BTZ spacetimes and then provide new results on the boundary support of such bulk reconstructions. In section~\ref{sec:adsrind} we show how to cross the AdS-Rindler horizon and reconstruct the bulk Poincar\'e time from Rindler patches of the boundary theory.
 In section~\ref{sec:bhh} we discuss boundary descriptions of Kruskal-like time evolution in the BTZ geometry and sharp signatures of the black hole horizons and causal structure in the boundary theory. In section~\ref{sec:largeMass} we show that the emergent bulk evolution becomes a point-wise transformation in the limit with the bulk field having a very large mass. We then conclude in section~\ref{sec:concDisc} with a discussion of our results and we point out many future directions to be explored.

\medskip
\noindent {\bf Conventions and notations:}

In this paper we use $N^2 \sim {1 \ov G_N}$ to denote the number of degrees of freedom of the boundary theory, where $G_N$ is the bulk Newton constant. For two-dimensional CFTs, $N^2$ should be understood as the central charge $c$. 
The $1/N$ perturbative expansion of the CFT is dual to the perturbative $G_N$ expansion around the corresponding classical geometry.  
In this regime, the bulk gravity theory can be described by a weakly coupled quantum field theory in a curved spacetime. 

All operator algebras discussed in this paper should be understood as those of bounded operators.\footnote{This is for mathematical convenience, but this constraint does not sacrifice physical significance as essentially all observables can be made to be bounded by putting restrictions on their spectra.} 

We will consider the boundary theory to be on $\RR \times S^{d-1}$ or $\RR^{1,d-1}$ and the discussion generalizes straightforwardly to other boundary spatial manifolds such as hyperbolic space. 
A boundary point is denoted by $x = (\rt, \vx)$ with $\vx$ denoting points on either $\RR^{d-1}$ or $S^{d-1}$. The corresponding Fourier space will be denoted as $k = (\om, q)$ with $q$ collectively denoting momentum on $\RR^{d-1}$ or spherical harmonic labels on $S^{d-1}$.  A bulk point is denoted by $X = (r, x)$ with $r$ the bulk radial direction (later in the paper we use $w$ as the radial variable). 

$\sA'$ denotes the commutant of an algebra $\sA$, i.e. the algebra of operators commuting with the algebra $\sA$. 
By type III$_1$ algebras we mean a von Neumann (vN) algebra which contains type III$_1$ factor(s). 

We  use $\rt$ to denote the boundary time whose translation is generated by the Hamiltonian $H$,  $\eta$ to denote the boundary time in units where the inverse temperature is $\b =2\pi$, i.e. $\eta = {2 \pi \ov \b} \rt$, and $t$ to denote the modular time. For a detailed description of the notational conventions used in the paper, see Appendix~\ref{app:notation}.

\section{Emergent type III$_1$ algebras at finite temperature} \label{sec:GTFD}

In this section we consider two copies of the boundary CFT in the thermal field double state, which is dual to an eternal black hole in AdS. We argue that there are emergent type III$_1$  vN algebras in the  large $N$ limit. We start with a quick review of the bulk theory to set up the notation.

\subsection{Small excitations around the eternal black hole geometry}

\begin{figure}[h]
\begin{centering}
	\includegraphics[width=2in]{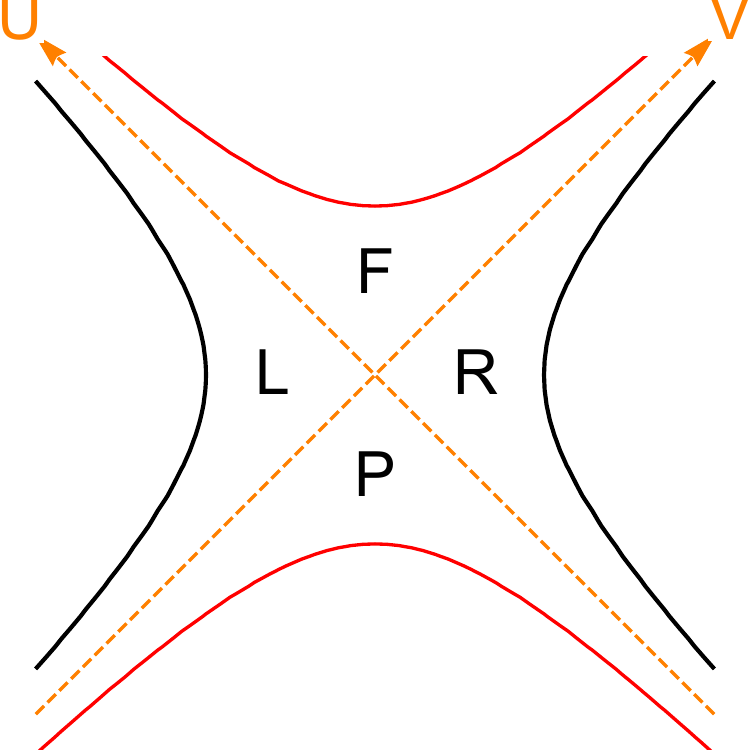}
\par\end{centering}
\caption{Kruskal diagram for an eternal black hole. The dashed lines are event horizons, the solid red lines are the singularities, and the solid black lines are the boundaries.
}
\label{fig:casu}
\end{figure}

Consider an eternal black hole in AdS$_{d+1}$, whose metric can be written in a form 
\bega 
ds^2 = - f d\rt^2 + {1 \ov f} dr^2 + {r^2} d \Sig^2_{d-1} 
\end{gather}
where $d \Sig^2_{d-1}$ is the metric for the boundary spatial manifold $\Sig$ which we will take to be the unit sphere $S^{d-1}$ or $\RR^{d-1}$, and $f$ is a function with a first order zero at event horizon $r= r_0$. 
A bulk point is denoted by $X = (\rt, r, \vx)$ where $\vx$ denotes a point on $\Sig$.  
The Schwarzschild coordinates $(\rt, r)$ can be used to cover any of the four regions of the fully extended black hole geometry of {Fig.~\ref{fig:casu}}, while the Kruskal coordinates $U, V$  cover all the regions.

Small perturbations around the black hole geometry can be described 
using the standard formalism of quantum field theory in a curved spacetime. Their quantization results in 
a Fock space $\sH_{\rm BH}^{\rm (Fock)}$. We will use a real scalar field $\phi$ of mass $m$ as an illustration. 
 The restriction $\phi_R$ of $\phi$ to the $R$ region of the black hole geometry can be expanded 
in terms of a complete set of properly normalized modes $v_{\om q}^{(R)} (X) $ in the $R$ region as
\be \label{moex}
\phi_R (X) = \sum_{ q} \int_0^\infty {d \om \ov 2 \pi}\, \le(v_{\om q}^{(R)} (X)  a_{\om q}^{(R)} + (v_{\om q}^{(R)} (X) )^* (a_{\om
q}^{(R)})^\dagger \ri) \ ,
\ee
where $q$ collectively denotes quantum numbers associated with $\Sig$,\footnote{The sum over $q$ and $\de_{q q'}$ should be understood as integrals and Dirac delta functions if there are continuous quantum numbers.} and 
\be \label{canQ} 
[a_{\om q}^{(R)}, (a_{\om' q'}^{(R)})^\dagger] = 2 \pi \delta(\om-\om') \de_{q q'}  \ .
\ee
Below for notational simplicity we will write~\eqref{moex} as 
\be \label{moex1}
\phi_R (X)=  \sum_{ k} v_{k}^{(R)} (X) a_{k}^{(R)}, \quad k = (\om, q) , \quad v_{-k}^{(R)} = (v_{k}^{(R)})^*, \quad
a_{-k}^{(R)} = (a_{k}^{(R)})^\da \ .
\ee
There is a similar expansion for the restriction of $\phi$ to the $L$ region,
\be  \label{moex2}
\phi_L (X)=  \sum_{ k} v_{k}^{(L)} (X) a_{k}^{(L)}, \quad k = (\om, q) , \quad v_{-k}^{(L)} = (v_{k}^{(L)})^*, \quad
a_{-k}^{(L)} = (a_{k}^{(L)})^\da  \ .
\ee
In the case of a Schwarzschild black hole,  the $R$ and $L$ regions are related by spacetime reflection symmetry
$(U, V, \vx )$ to $(-U , -V, -\vx)$. It is convenient to 
choose $ v_{k}^{(L)}$ to be 
\be 
v_{k}^{(L)} (\rt, r, \vx)  = v_{k}^{(R) *} (\rt, r, \vx) =  v_{-k}^{(R)} (\rt, r, \vx)  
\ee
and the anti-unitary spacetime reflection operator $J$ then acts as
\be 
J \phi_R (\rt, r, \vx)  J = \phi_L (\rt, r, \vx)  , \qquad J a_{k}^{(R)} J = a_{k}^{(L)}  \ .
\ee
Altogether
\be
(a_k^{(\al)})^\da = a_{-k}^{(\al)} , \quad    [a_k^{(\al)} , a_{k'}^{(\b)}] = \ep (\om)   \de_{k+k',0}  \de_{\al \b}  , \quad \al, \b = R, L  \ .
\ee

The behavior of $\phi$ in the $F$ and $P$ regions can be determined from that in the $R$ and $L$ regions by causal evolution or analytic continuation. 


The Hartle-Hawking vacuum can be defined using the standard Unruh procedure by first introducing modes $w_k$ which are analytic in the lower $U$ and $V$ planes for $\om > 0$, 
\bega 
w_k^{(\al)} =  \fb_+ v^{(\al)}_k + \fb_- v^{(\bar \al)}_{-k}, \quad  \bar L = R , \quad \bar R = L \\
\fb_\pm = {e^{\pm {\b |\om| \ov 4} } \ov \sqrt{2 \sinh {\b |\om| \ov 2} }}, \quad \fb^2_+ - \fb_-^2 = 1 \ .
\label{defbb}
\end{gather}
Denoting the oscillators corresponding to the modes $w_k^{(\al)}$ as $c^{(\al)}_k $ we then have on a Cauchy slice 
\be 
\phi = \sum_{\al, k}  v_{k}^{(\al)} a_{k}^{(\al)} =   \sum_{\al, k}   w_{k}^{(\al)} c_{k}^{(\al)} \ ,
\ee
which implies the oscillators $c^{(\al)}_k $ and $a^{(\al)}_k $ are related by 
\bega\label{c111HH}
c^{(\al)}_k = \fb_+ a^{(\al)}_k - \fb_- a^{(\bar \al)}_{-k}, \qquad a^{(\al)}_k = \fb_+ c^{(\al)}_k + \fb_- c^{(\bar \al)}_{-k},  \\
 [c_k^{(\al)} , c_{k'}^{(\b)}] = \ep (\om)   \de_{k+k',0}  \de_{\al \b} , \quad J c_{k}^{(\al)} J = c_{k}^{(\bar \al)} \ .
\end{gather}
The Hartle-Hawking vacuum $\ket{HH}$ is defined to satisfy 
\bega \label{coscHH}
c_k^{(\al)} \ket{HH} = 0  \quad \text{for} \quad \om > 0 \ .
\end{gather} 
The Fock space $\sH_{\rm BH}^{\rm (Fock)}$ is built by acting $c_k^{(\al)}$ with $\om < 0$ on $\ket{HH}$. 
Note that 
\be \label{hheq}
a_{k}^{(R)} \ket{HH} = e^{-{\b \om \ov 2} }  a_{-k}^{(L)} \ket{HH} , \qquad 
 \bra{HH}   a_{k}^{(R)}  = e^{{\b \om \ov 2} } \bra{HH}  a_{-k}^{(L)}  \  .
\ee

We will denote the operator algebra generated by $\phi$ and other matter fields (including metric perturbations) in the $R$ region as $\widetilde \sY_R$ and similarly those generated by 
fields in the $L$ region as $\widetilde \sY_L$.  $\widetilde \sY_R$ and $\widetilde \sY_L$ are commutants of each other, and are expected to be type III$_1$ von Neumann algebras~\cite{Araki1964:2, Longo:1982zz, Fredenhagen:1984dc}. Reflections of the type III$_1$ structure include  the non-existence of the Schwarzschild vacuum state $\ket{0}_R \otimes \ket{0}_L$ (which is defined to be annihilated by $a_k^{(\al)}$ with $\om > 0$) in $\sH_{\rm BH}^{(\rm Fock)}$ and the entanglement entropy between $R$ and $L$ regions being not well defined in the continuum limit.

\subsection{Small excitations around thermal field double state on the boundary} \label{sec:TFD} 


We always consider the boundary CFT at a large but finite $N$ and work to leading order in the $1/N$ expansion.
We denote the Hilbert space of the boundary CFT as $\sH$, its Hamiltonian as $H$, the algebra of bounded operators as $\sB$, and the vector space of all finite products of single-trace operators by $\sA$.
We use $\sO$ to denote the single-trace operator dual to the bulk field $\phi$. 
Now consider two copies of the boundary theory, to which we refer respectively as CFT$_R$ and CFT$_L$. 
Operators or states with subscripts $R, L$ refer 
to those in the respective systems. The doubled system has Hilbert space $\hat \sH = \sH_R \otimes \sH_L$, operator algebra
$\hat \sB = \sB_R \otimes \sB_L$, and single-trace operators $\hat \sA =  \sA_R \otimes \sA_L$. In the large $N$ limit, $\hat \sA$ can be endowed with an algebraic structure defined with respect to the thermal field double state (see~\cite{Leutheusser:2022bgi} for details). We denote the resulting algebra by $\hat \sA_{\rm TFD}.$ The vector space of products of single-trace operators associated to either side of thermal field double are then also endowed with an algebraic structure and become subalgebras of $\hat \sA_{\rm TFD},$ which we denote by $\sA_{R, \rm TFD}$ and $\sA_{R, \rm TFD}$.
Generic operators in $\hat \sA_{\rm TFD}$ will be denoted as $a, b, \cdots$, those in $\hat \sB$ as $u, v, \cdots$, those in 
$\sA_R$ as $A_R, B_R, \cdots$. 

The thermal field double state is defined as 
\bega \label{defd} 
\ktfd = 
{1 \ov \sqrt{Z_\b}} \sum_n e^{- \ha \b E_n}  \ket{\tilde n}_L \ket{n}_R   , \\
\ket{\tilde n} = \th \ket{n} , \quad  \vev{\tilde m |\tilde n} = \vev{n|m} = \de_{mn} , \quad
Z_\b = \sum_n e^{-\b E_n} 
\end{gather} 
where $\ket{n}$ denotes the full set of energy eigenstates of the CFT, with eigenvalues $E_n$. 
$m,n$ here collectively denote all quantum numbers including spatial momenta for the boundary theory on $\RR^{1,d-1}$ or angular quantum numbers for the theory on $\RR \times S^{d-1}$. 
 $\th$ is an anti-unitary operator and will be taken to be the $\sC \sP \sT$ operator of the CFT. 
  When tracing over degrees of freedom of one of the CFTs, we get the thermal density operator 
  at inverse temperature $\b = {1 \ov T}$ for the remaining one 
\be 
\rho_\b = {1 \ov Z_\b} e^{-\b H} \ .
\ee


Perturbatively in the $1/N$ expansion, excitations around $\ktfd$ can be obtained by acting single-trace boundary operators
on it.\footnote{See also~\cite{Witten:2021jzq} for a review of the definition of the thermal field double state in the infinite volume and large $N$ limits.} In fact, the collection of  excitations obtained this way has the structure of a Hilbert space, which can be 
made precise mathematically using the 
Gelfand-Naimark-Segal (GNS) construction.  More explicitly, for each operator $a \in \hat \sA_{\rm TFD} $ we associate a state $\ket{a}$ and define the inner product among them as 
\be \label{inp1}
\vev{a|b} = \lim_{N \to \infty} \vev{\Psi_\b|a^\da b|\Psi_\b} , \quad a, b \in \hat \sA_{\rm TFD} \ .
\ee
In particular, for $A_R, B_R \in \sA_{R, \rm TFD}$, we have 
\be \label{inp0}
\vev{A_R|B_R} =  \lim_{N \to \infty} \vev{\Psi_\b|A_R^\da B_R|\Psi_\b} =  \lim_{N \to \infty} \Tr (\rho_\b A^\da B)  \ .
\ee

Equation~\eqref{inp1} does not yet define a Hilbert space as there can be operators $y \in \hat \sA_{\rm TFD}$ satisfying $\lim_{N \to \infty} \vev{\Psi_\b|y^\da y|\Psi_\b} =0$ 
and the corresponding $\ket{y}$ should be set to zero. Denote the set of  such operators as $\sJ$.
The GNS Hilbert space is the completion of the set of equivalence classes $[a]$ which are defined by the equivalence relations  
\be 
a \sim a + y , \qquad a \in \hat \sA_{\rm TFD}, \quad y \in \sJ  \ .
\ee

The set $\sJ$ is non-empty, as from~\eqref{defd}, for a Hermitian operator $W \in \sB$ 
\bega \label{ehen1}
W_R (\rt,\vx) \ket{\Psi_\b} =   W_L \le(\rt - i {\b \ov 2}, \vx \ri) \ket{\Psi_\b} , \quad 
W_L (\rt,\vx)  \ket{\Psi_\b} = W_R \le(\rt + i {\b \ov 2}, \vx \ri) \ket{\Psi_\b}, 
\end{gather} 
where for simplicity we have assumed that  $\th^\da W (0) \th = W (0)$ and we have chosen the space and time orientations of CFT$_L$ to be the opposite of those of CFT$_R$.\footnote{That is, $W_R (\rt, \vx)= \bid_L \otimes W(\rt, \vx) $ while 
$W_L (\rt, \vx) = W(-\rt, -\vx) \otimes \bid_R$.
We take the single-trace operators $W (\rt, \vx)$ to be analytically continued to ${\rm Im} \, \rt \in [0, {\b \ov 2}]$.  Thus $W_R (\rt, \vx)$ are defined for ${\rm Im} \,\rt \in [0, {\b \ov 2}]$ while $W_L (\rt, \vx)$ for ${\rm Im}\, \rt \in [-{\b \ov 2}, 0]$. 
}
From~\eqref{ehen1} it can be shown that $\sA_{R, \rm TFD}$ or $\sA_{L, \rm TFD}$ alone can be used to generate the full GNS Hilbert space, which we will denote as $\sH_{\rm TFD}^{\rm (GNS)}$. See Appendix~\ref{app:GNS} for details. In other words, any state in $\sH_{\rm TFD}^{\rm (GNS)}$ can be written as $\ket{A_R}$ with $A_R \in \sA_{R, \rm TFD}$ or as a limit of such states.
The state in $\sH_{\rm TFD}^{\rm (GNS)}$ corresponding to the identity operator is denoted as $\ket{\Om_0}$, which we sometimes refer to as the GNS vacuum.

$\sH^{\rm (GNS)}_{\tfd}$ also provides a representation space for $\hat \sA_{\rm TFD}$. The representation $\pi (a)$ of an operator $a \in \hat \sA_{\rm TFD}$ acting on $\sH^{\rm (GNS)}_{\tfd}$ can be defined as 
\be\label{xhni}
\pi (a) [b] = [ab], \qquad a, b \in \hat \sA_{\rm TFD} 
\ee
and as a result the inner product~\eqref{inp1} can also be written as 
\be \label{inp2}
\vev{a|b} = \vev{\Om_0 |(\pi (a))^\da \pi (b)|\Om_0}  \ .
\ee
We denote the representations of $\sA_{R, \rm TFD}$ and $\sA_{L, \rm TFD}$ in $\sH_{\rm TFD}^{\rm (GNS)}$ respectively as $\sY_R$ and $\sY_L$. 
Given that $\sH_{\rm TFD}^{\rm (GNS)}$ can be generated by $\sA_{R, \rm TFD}$ or $\sA_{L, \rm TFD}$ alone, 
the GNS vacuum $\ket{\Om_0}$ is cyclic and separating under both $\sY_R$ and $\sY_L$, and we have $\sY_R' = \sY_L$. 
We denote the operator algebra on $\sH_{\rm TFD}^{\rm (GNS)}$ as $\sY$. 

It can also be shown that $\sH_{\rm TFD}^{\rm (GNS)}$ is isomorphic to $\sH_{\rm \rho_\b}^{\rm (GNS)}$, the GNS Hilbert space 
corresponding to the thermal density operator $\rho_\b$ over the algebra $\sA_{\rho_\beta}$ obtained from the single-trace operators of one copy of the CFT and the thermal state.\footnote{For the construction of latter see Sec.~V.1.4 of~\cite{Haag:1992hx}  and~\cite{Magan:2020iac}.}

To leading order in the $1/N$ expansion, the inner products~\eqref{inp1} and thus~\eqref{inp2}  can be written as sums of products of two-point functions of single-trace operators. We can thus represent single-trace operators by generalized free fields acting on $\sH_{\rm TFD}^{\rm (GNS)}$ and 
the algebras $\sY_R, \sY_L$ are generated by generalized free fields. 
More explicitly, for a single trace scalar operator $\sO$, we can expand its representations in terms of a complete set of functions on the boundary manifold\
\bega \label{eha}
\pi (\sO_R (x)) = \sum_k u_k^{(R)} (x) a_k^{(R)},  \quad  \pi (\sO_L (x) )= \sum_k u_{k}^{(L)} (x) a_k^{(L)}  , \\
u_k^{(R)} (x) = N_k e^{- i \om \rt}  h_q (\vx), \quad u_k^{(L)} (x) = u_{-k}^{(R)} (x) = (u_k^{(R)}(x))^*, \quad x = (\rt, \vx) 
\end{gather} 
where $N_k$  is some function of $k = (\om, q)$,  and $h_{q}(\vx)$ denotes the complete set of functions on the boundary spatial manifold $\Sig$, and $a_k^{(R, L)}$  are operators acting on $\sH^{\rm (GNS)}_{\rm TFD}$, normalized as\footnote{Note that this is purely a boundary discussion. Even though we use the same notation, $a_k^{(\al)}$ as in~\eqref{moex1}, at this stage these operators do not have anything to do with each other.} 
\be\label{eha1} 
(a_k^{(\al)})^\da = a_{-k}^{(\al)} , \quad    [a_k^{(\al)} , a_{k'}^{(\b)}] = \ep (\om)   \de_{k+k',0}  \de_{\al \b}  , \quad \al, \b = R, L \ .
\ee
Using~\eqref{inp0} and~\eqref{inp2}, $N_k$ can be deduced from the condition 
\be \label{inp3}
 \lim_{N \to \infty} \Tr (\rho_\b \sO (x_1) \sO (x_2)) =  \vev{\Om_0 |(\pi (\sO_R (x_1)) \pi( \sO_R (x_2))|\Om_0}  \ .
\ee
Furthermore, applying~\eqref{ehen1} to $\pi (\sO_R)$ and $\pi (\sO_L)$ we have 
\bega
 \label{eqd11}
a_{k}^{(R)} \ket{\Om_0} = e^{-{\b \omega \ov 2}}  a_{-k}^{(L)} \ket{\Om_0} , \qquad 
 \bra{\Om_0}   a_{k}^{(R)}  = e^{{\b \omega \ov 2}} \bra{\Om_0}  a_{-k}^{(L)}  \  .
\end{gather} 
We can introduce an anti-unitary ``swap'' operator $J$ which acts as
\be 
J \ket{\Om_0} = \ket{\Om_0}, \quad J a_k^{(\al)}J = a_k^{(\bar \al)} , \quad J \pi (\sO_\al (x) ) J = \pi (\sO_{\bar \al} (x)) , \quad J^2 =1  \ .
\ee
Equations~\eqref{eqd11} motivate the introduction of ($\fb_\pm$ were introduced in~\eqref{defbb}) 
\bega \label{c11}
c^{(\al)}_k = \fb_+ a^{(\al)}_k - \fb_- a^{(\bar \al)}_{-k}, \qquad a^{(\al)}_k = \fb_+ c^{(\al)}_k + \fb_- c^{(\bar \al)}_{-k},  \quad \bar L = R \end{gather} 
which satisfy 
\bega \label{cosc}
c_k^{(\al)} \ket{\Om_0} = 0  \quad \text{for} \quad \om > 0 , \quad \le(c_k^{(\al)} \ri)^\da =  c_{-k}^{(\al)} ,
\quad  [c_k^{(\al)} , c_{k'}^{(\b)}] = \ep (\om)   \de_{k+k',0}  \de_{\al \b}  \ .
\end{gather}

To conclude this subsection, we make some further general remarks: 

\ben 

\item The algebras $\sA_{R, \rm TFD}, \sA_{L, \rm TFD}$ are defined only in the $1/N$ expansion. 
  The algebras $\sY_R, \sY_L$ act on $\sH_{\rm TFD}^{\rm (GNS)}$, and are von Neumann algebras. 

\item $\pi (\sO)$ is not the same as $\sO$. The former is defined only on $\sH^{\rm (GNS)}_{\rm TFD}$ and 
is state-dependent (i.e. it depends on the state we use to build the GNS representation), while $\sO$ acts on the full CFT Hilbert space and is state-independent. The algebras $\sY_R, \sY_L$ are thus also state-dependent. 
For example,  they depend on $\b$. 


\item The operator algebras $\sB_R, \sB_L$ are type I von Neumann algebras, and $\ktfd$ is cyclic and separating with respect to them. The corresponding modular operator $\mathbf{\De}$ is given by $-\log \mathbf{\De} = \b (H_R - H_L)$. Note that the modular time $t$ defined by modular flow with $\mathbf{\De}^{-i t}$ is related to the usual CFT time $\rt$ by\footnote{Recall that we take the time of CFT$_L$ to run in the opposite direction to that CFT$_R$.}
\be \label{timR} 
t = {\rt \ov \b} \ .
\ee


\item Since $\ket{\Om_0}$ is cyclic and separating for $\sY_R$, there exists a modular operator $\De_0$ which leaves $\ket{\Om_0}$
invariant and generates automorphisms of $\sY_R, \sY_L$. 
The modular flows generated by $\De_0$ again coincide with the time evolution of the respective boundaries.  More explicitly, combining with the previous item, we have 
\be \label{mofl1}
\pi (A_R (\rt=\b t)) = \pi \le(\mathbf{\De}^{- i t} A_R \mathbf{\De}^{it} \ri)  = \De^{- i t}_0 \pi ( A_R)  \De_0^{it}  , \quad A_R \in \sA_{R, \rm TFD}  \ . 
\ee

\een

\subsection{{Complete} spectrum and emergent type III$_1$ structure}  \label{sec:comp}

For the boundary theory on $\RR \times S^{d-1}$, we conjecture that the algebras $\sY_R, \sY_L$ are type I below the Hawking-Page temperature $T_{HP}$, but become type III$_1$ above $T_{HP}$. 
Recall that $T_{HP}$ is the temperature at which the boundary system exhibits a first-order phase transition in the large $N$ limit, with 
$\log Z_\b \sim O(N^0)$ for $T <T_{HP}$ but $\log Z_\b \sim O(N^2)$ for  $T> T_{HP}$. 
Below $T_{HP}$ thermal averages are dominated by contributions from states with energies of $O(N^0)$ while above $T_{HP}$ they are dominated by states with energies of $O(N^2)$. This change of dominance leads to dramatically different behavior for thermal correlation functions. Since the inner products~\eqref{inp1}--\eqref{inp0} of $\sH^{\rm (GNS)}_{\rm TFD}$ are determined by thermal two-point functions of single-trace operators,  the representations of elements of $\sA_{R, \rm TFD}, \sA_{L, \rm TFD}$, and thus the structure of the algebras $\sY_R, \sY_L$ are sensitive to the behavior of these two-point functions. 

Consider thermal Wightman functions of a Hermitian scalar operator $\sO$ of dimension $\De$
\be 
G_+  (x_1 - x_2) = \Tr (\rho_\b \sO(x_1) \sO(x_2)) =  \vev{\Psi_\b|\sO_R (x_1) \sO_R (x_2) |\Psi_\b} \ .
\ee
Its Fourier transform has the Lehmann representation 
\bega \label{ehkl}
G_+ (\om, q) = \sum_{m,n} (2 \pi) \delta (\om -  E_{nm})  e^{-\beta E_m} \rho_{mn}  \equiv {1 \ov 1- e^{-\beta \om}} \rho (\om,q) , \\ 
E_{nm} = E_n - E_m, \quad \rho_{mn}  = |\vev{m |\sO(0)|n}|^2
, 
\end{gather} 
where $\rho (\om, q)$ is the (finite temperature) spectral function. In the large $N$ limit and at strong coupling, $G_+$ and $\rho$ can be computed using the standard procedure from gravity. 
Below $T_{HP}$,  the finite temperature Euclidean correlation function, $G_E,$ of  $\sO$ is determined by the Euclidean function $G_{E0}$  at zero temperature via summation over images in the Euclidean time 
\be
G_E (\tau, \vec x) = \sum_{n} G_{E0} (\tau + n \beta, \vec x)  \ .
\ee
When analytically continued back to the Lorentzian signature, this implies that 
\be
\rho (\om, q) =  \th (\om) \rho_0 (\om, q) - \th (-\om) \rho_0 (-\om,q) 
\ee
where $\rho_0 (\om, q)$  
 is the spectral function at zero temperature\footnote{This can be defined by taking $\b \to \infty$ in~\eqref{ehkl} and can be found from the
 zero temperature momentum space Wightman function as $G_{0+} (\om, q) = \rho_0 (\om, q)$.}
  and has the following form 
\be
 \rho_0 (\om, q) = \sum_{l =0}^\infty c_l  \delta (\om - \Delta - 2l)  \ ,
 \ee
In this case,  $\rho$ is supported only at discrete points on the real $\om$-axis. 

In contrast, for  $T > T_{HP}$, $\rho (\om, q)$ is smooth and supported on the full real $\om$-axis. For $d=2$, i.e. CFT on a circle, from the BTZ black hole~\cite{Banados:1992wn} it can be found that\footnote{Now $q$ is the momentum on the circle and $q_\pm = \ha (\De + i (\om \pm q)),\;  \bar q_\pm = \ha (\De - i (\om \pm q))$. $C$ is a normalization constant. In~\eqref{birm} we have chosen units such that $\b = 2 \pi$.} 
\be \label{birm} 
\rho (\om,q) = C 
\sinh {\pi \om}  \Ga (q_+) \Ga (\bar q_+)  \Ga (q_-) \Ga (\bar q_-)  \ .
\ee
We will refer to such a $\rho$, a smooth function supported on the full real $\om$-axis for any $q$, as having a {\it complete spectrum}. 
For general $d$, the explicit analytic expressions of $\rho (\om,q)$ at strong coupling are not known, but $\rho$ can be shown to always have a complete spectrum due to the presence of the horizon in the black hole geometry (see e.g.~\cite{Festuccia:2005pi}).

It is known that for a generalized free field theory in a thermal field double state if the spectral function has a continuous spectrum, then 
the corresponding subalgebra for a single copy of the theory is type III$_1$~\cite{arakiWoods,Derezinski:2005qv}. This has also been emphasized recently from a different perspective in~\cite{Furuya:2023fei}.\footnote{We thank Eliott Gesteau, Nima Lashkari, and Mudassir Moosa for discussions on these references.}  Here the gravity calculation indicates that $\rho (\om, q)$ has a complete spectrum, which implies that $\sY_R, \sY_L$ are type III$_1$. We thus conjecture that in the large $N$ limit, at $T > T_{HP}$, $\sY_R, \sY_L$ become type III$_1$.
Note also that when there is a continuous spectrum, the ``vacuum'' $\ket{0}_L \otimes \ket{0}_R$ for $a_k^{(\al)}$, which is defined to be annihilated by $a_k^{(\al)}$ with $\om > 0$, does not exist in $\sH_{\tfd}^{\rm (GNS)}$ and thus $\sH_{\tfd}^{\rm (GNS)}$ cannot be tensor factorized.\footnote{From~\eqref{c11}, the normalization of 
$\ket{0}_R \otimes \ket{0}_L$ is proportional to $\prod_\om (1 - e^{- 2 \pi \om})$, which is not well defined.}

We will discuss in the next subsection that the type III$_1$ structure of $\sY_R, \sY_L$ is also required by the duality of $\sY_R, \sY_L$ with the bulk algebras $\tilde \sY_R, \tilde \sY_L$.

The complete spectrum of $\rho (\om, q)$ is stronger than the continuous spectrum required to have a type III$_1$ algebraic structure for $\sY_R, \sY_L$. We believe the complete spectrum is necessary for the half-sided modular inclusion/translation structure to be discussed in Sec.~\ref{sec:ext}  (which will play an important role in later parts of the paper), but we will not attempt a rigorous proof here.

We emphasize that a continuous spectrum is possible only in the large $N$ limit.  CFT on $S^{d-1}$ has a discrete energy spectrum, i.e. the sums $m,n$ in~\eqref{ehkl} are literally discrete. As a result, at finite $N$, the spectral function $\rho$ is supported on only discrete values of $\om = E_{mn}$.  In the large $N$ limit (for $T>T_{HP}$), the dominant contributions to the sums in~\eqref{ehkl} come from states with  energies of $O(N^2)$, where the density of states is $e^{O(N^2)}$. If $\sO$ has nonzero matrix elements between generic states with energy differences $E_{mn} \sim O(N^0)$, a continuous spectrum results in the large $N$ limit. In contrast, for $T < T_{HP}$, the dominant contributions to the sums in~\eqref{ehkl} come from states with  energies of $O(N^0)$, where the density of states is $O(N^0)$, which leads to a discrete spectrum for $\rho$. It is interesting to understand what is responsible for the emergence of the complete spectrum on the gravity side.
From the bulk perspective, the complete spectrum can be attributed to the existence of an event horizon which results in a continuum of modes for both signs of $\om$. The emergent complete spectrum in the large $N$ limit for $T > T_{HP}$ was emphasized before in~\cite{Festuccia:2005pi} as a possible reason for the emergence of a bulk horizon and singularity in holography. 

The complete spectrum of finite temperature spectral functions responsible for the emergent type III$_1$ structure may not be restricted to strong coupling. In~\cite{Festuccia:2006sa} it was argued that a complete spectrum may arise generically for a matrix-type theory in the large $N$ limit even at weak coupling (see also~\cite{Iizuka:2008hg,Iizuka:2008eb}). A complete spectrum may also arise in the SYK model yielding an emergent type III$_1$ algebra in the large $N$ limit.

Our discussion of the emergent type III$_1$ structure is at the generalized free field theory level, which applies at leading order in the large $N$ limit. 
See~\cite{WittenNew} for a discussion on the deformation of this algebra when including $1/N$ corrections.

\subsection{Duality between the bulk and boundary from the algebraic perspective} 

Given that single-trace operators are dual to fundamental fields on the gravity side, we can identify the Hilbert spaces of small excitations on both sides and the corresponding operator algebras, i.e. 
\be \label{ejnn}
\sH^{\rm (GNS)}_{\tfd} = \sH_{\rm BH}^{\rm (Fock)}, \quad
 \ket{\Om_0} = \ket{HH}  , \quad \sY_R = \widetilde \sY_R, \quad \sY_L = \widetilde \sY_L \ .
\ee

More explicitly, for a bulk scalar field $\phi$ dual to a boundary single-trace operator $\sO$, the last two equations of~\eqref{ejnn} imply that we should identify oscillators, $a_k^{(\al)},$ constructed from the generalized free field description of the boundary theory operators~\eqref{eha} with those in the bulk mode expansions~\eqref{moex1}--\eqref{moex2}, which is the reason we have been using the same notation for them. This identification is also reflected in the standard extrapolate dictionary for the bulk and boundary operators ($C$ is a normalization constant)
\bega \label{extr}
\pi (\sO_R (x)) = C \lim_{r \to \infty} r^{-\De} \phi_R (r, x) , \qquad \pi (\sO_L (x))= C \lim_{r \to \infty} r^{-\De} \phi_L (r, x) , \\
u_k^{(R)} (x) = C \lim_{r \to \infty} r^{-\De} v_k^{(R)} (r, x) , \quad u_k^{(L)} (x) = C \lim_{r \to \infty} r^{-\De} v_k^{(L)} (r, x)   \ .
\end{gather} 
We emphasize that it is the representations $\pi (\sO_R), \pi (\sO_L)$ of $\sO_R, \sO_L$ in the GNS Hilbert space that appear in the extrapolation formulas~\eqref{extr}. This makes sense as the mode expansions of $\phi_R, \phi_L$ depend on the bulk geometry, which is reflected in the state-dependence of $\pi (\sO_R), \pi (\sO_L)$. The identification of $\ket{\Om_0}$ with $\ket{HH}$ then follows from~\eqref{hheq} and~\eqref{eqd11}.

With the identifications of $a_k^{(\al)}$ in the boundary and bulk mode expansions, $\phi_R, \phi_L$ of equations~\eqref{moex1}--\eqref{moex2} can now be directly interpreted as boundary operators, which is the statement of bulk reconstruction for the $R$ and $L$ regions of the black hole~\cite{Hamilton:2005ju,Hamilton:2006fh,Papadodimas:2012aq}. We emphasize that the reconstruction formula is in terms of operators in the GNS Hilbert space.

Since the algebras  $\tilde \sY_R, \tilde \sY_L$  of bulk fields restricted to the $R$ and $L$ regions of the black hole are believed to be type III$_1$ von Neumann algebras, the duality can only hold if $\sY_R, \sY_L$ are also type III$_1$. 

\begin{figure}[h]
\begin{centering}
	\includegraphics[width=2.5in]{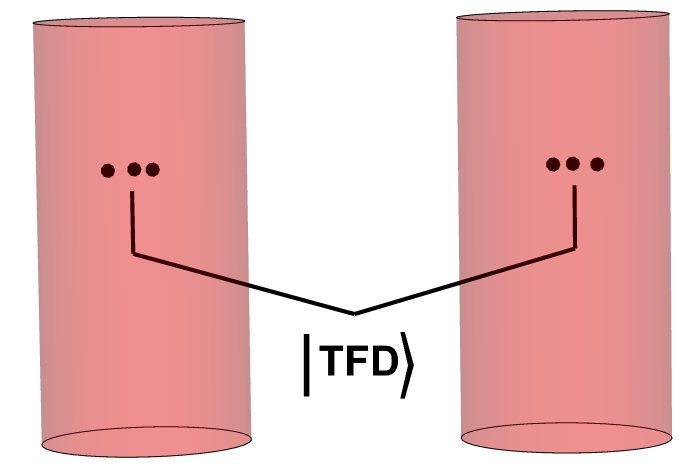} 
\par\end{centering}
\caption{Below $T_{HP}$ the bulk theory is two separate global AdS spacetimes whose small excitations are entangled in the thermal field double state.
}
\label{fig:tads}
\end{figure}

For the boundary theory on $\RR \times S^{d-1}$, the above discussion applies to $T = {1 \ov \b} > T_{HP}$. For $T < T_{HP}$, the bulk dual for~\eqref{defd} is given by two disconnected copies of global AdS whose small excitations are in the thermal field double state, {see Fig.~\ref{fig:tads}.} In this case $\sY_R$ and $\sY_L$ are each dual respectively to the algebra of bulk fields in the global AdS geometry and should be type I.

\section{Emergent type III$_1$ algebras in boundary local regions}  \label{sec:locIII}

The emergent type III$_1$ structure discussed in the previous section concerned the algebras generated by single-trace operators over the entire boundary spacetime. We now would like to argue this phenomenon is more general, applying to spacetime subregions, although in a more subtle way. The operator algebra of a boundary CFT restricted to a subregion should be type III$_1$~\cite{Araki1964:2, Longo:1982zz, Fredenhagen:1984dc}. We argue that there is a further emergent type III$_1$ structure {in the large $N$ limit}, and discuss its manifestation in the bulk gravity dual.



\begin{figure}[h]
\begin{centering}
	\includegraphics[width=2.5in]{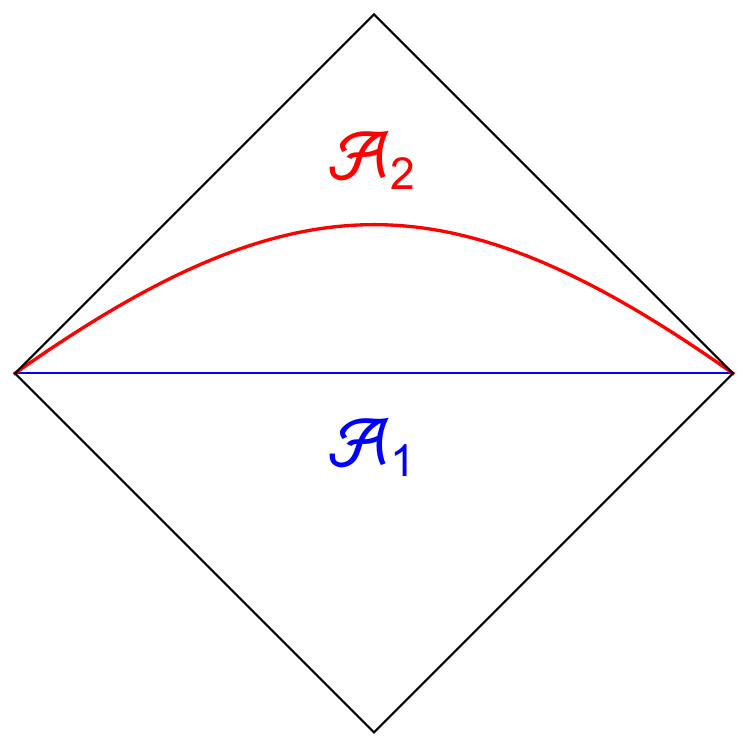}
\par\end{centering}
\caption{The single-trace algebras $\sA_1$ and $\sA_2$ associated with the two different Cauchy slices shown are inequivalent, even though they share a causal diamond, since single-trace operators do not obey any equation of motion among themselves (standard Heisenberg evolution takes a single-trace operator outside of the algebra). The same statements apply to algebras generated by generalized free fields (e.g. subalgebras of $\sY$)  which do not obey any equations of motion. 
}
\label{fig:ineqAlg}
\end{figure}

Our discussion in this section will be for a single copy of the boundary CFT at zero temperature. We again consider a finite but large $N$. 
 For definiteness, we will take the boundary spacetime to be $\RR^{1,d-1}$. 
Recall that the Hilbert space of the boundary CFT is $\sH$, with its full operator algebra $\sB$. The algebra generated by single-trace operators with respect to the vacuum is $\sA_{0}$, which is only defined perturbatively in $1/N$ expansion.  While $\sB$ can be defined on a single time slice, $\sA_{0}$ is defined on the whole spacetime
as single-trace operators do not obey any equations of motion among themselves, {see Fig.~\ref{fig:ineqAlg}.}

 \subsection{GNS Hilbert space and bulk reconstruction}

 For our discussion of the emergent type III$_1$ structure for local boundary subregions, it is again important to introduce the GNS Hilbert space of small excitations, now around the vacuum state of the CFT. The procedure is similar to our discussion of the GNS Hilbert space around the thermal field double state in Sec.~\ref{sec:TFD}, so we will not discuss it in detail.

Consider the GNS Hilbert space $\sH^{\rm (GNS)}_0$ built from the CFT vacuum state $\ket{0}$ over the single-trace operator algebra $\sA_0$. $\sH^{\rm (GNS)}_0$ offers a representation $\pi_0 (A)$ for an operator $A \in \sA_0$ and we denote the algebra $\pi_0 (\sA_0)$ as $\sY$. As the entire operator algebra on $\sH^{\rm (GNS)}_0$, $\sY$ is a type I vN algebra. We will denote the vector corresponding to the identity operator in $\sH^{\rm (GNS)}_0$ as $\ket{\Om_0}_{\rm GNS}$. The definition of $\sY$ is again only sensible perturbatively in the $1/N$ expansion. 

To leading order in the $1/N$ expansion, the algebra $\sY$ is again generated by generalized free fields, with a mode expansion determined by vacuum two-point functions of single-trace operators. 

The boundary theory in the vacuum state $\ket{0}$ is dual to the bulk gravity theory in the empty AdS geometry~(in our case the Poincar\'e patch as we consider the boundary theory on $\RR^{1,d-1}$). We can use the standard procedure to build a Hilbert space of small excitations around the Poincare vacuum $\ket{0}_{\rm bulk}$, which we will denote as $\sH_0^{(\rm Fock)}$. The algebra of bulk fields is denoted as $\tilde \sY$. In terms of the algebraic language we are using, the usual holographic dictionary can be written as 
\be \label{vacid}
\sH^{(\rm Fock)}_0 = \sH^{(\rm GNS)}_0, \quad \ket{0}_{\rm bulk} = \ket{\Om_0}_{\rm GNS}, \quad \tilde \sY  = \sY\ .
\ee
In particular, the last equation in~\eqref{vacid} identifies the bulk and boundary creation/annihilation operators, and is equivalent to the statement of global reconstruction.\footnote{The HKLL global reconstruction~\cite{Hamilton:2006az} is a coordinate space version of the statement.}

 \subsection{Boundary theory in a Rindler region and AdS Rindler duality}


\begin{figure}[h]
\begin{centering}
	\includegraphics[width=2.5in]{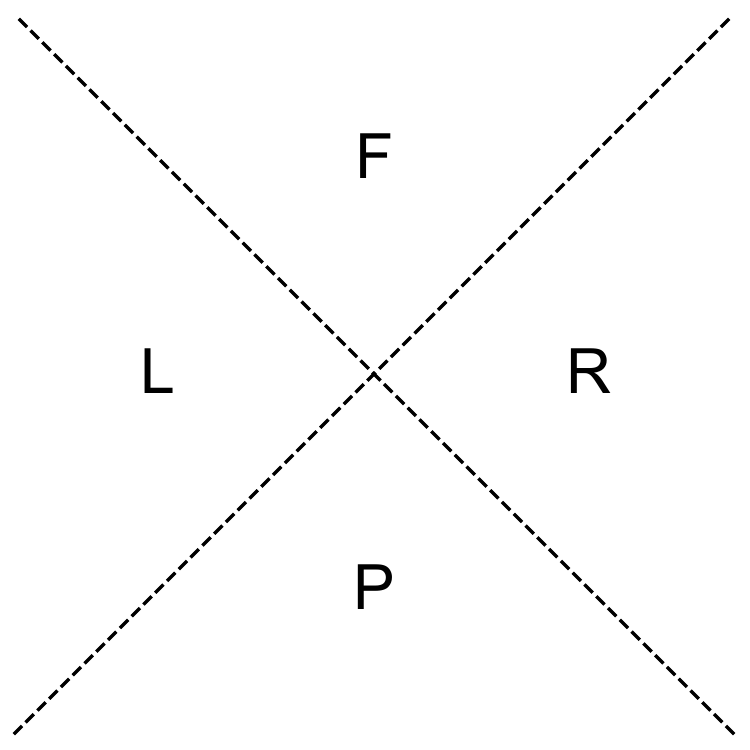}\qquad \;\;
	\includegraphics[width=2in]{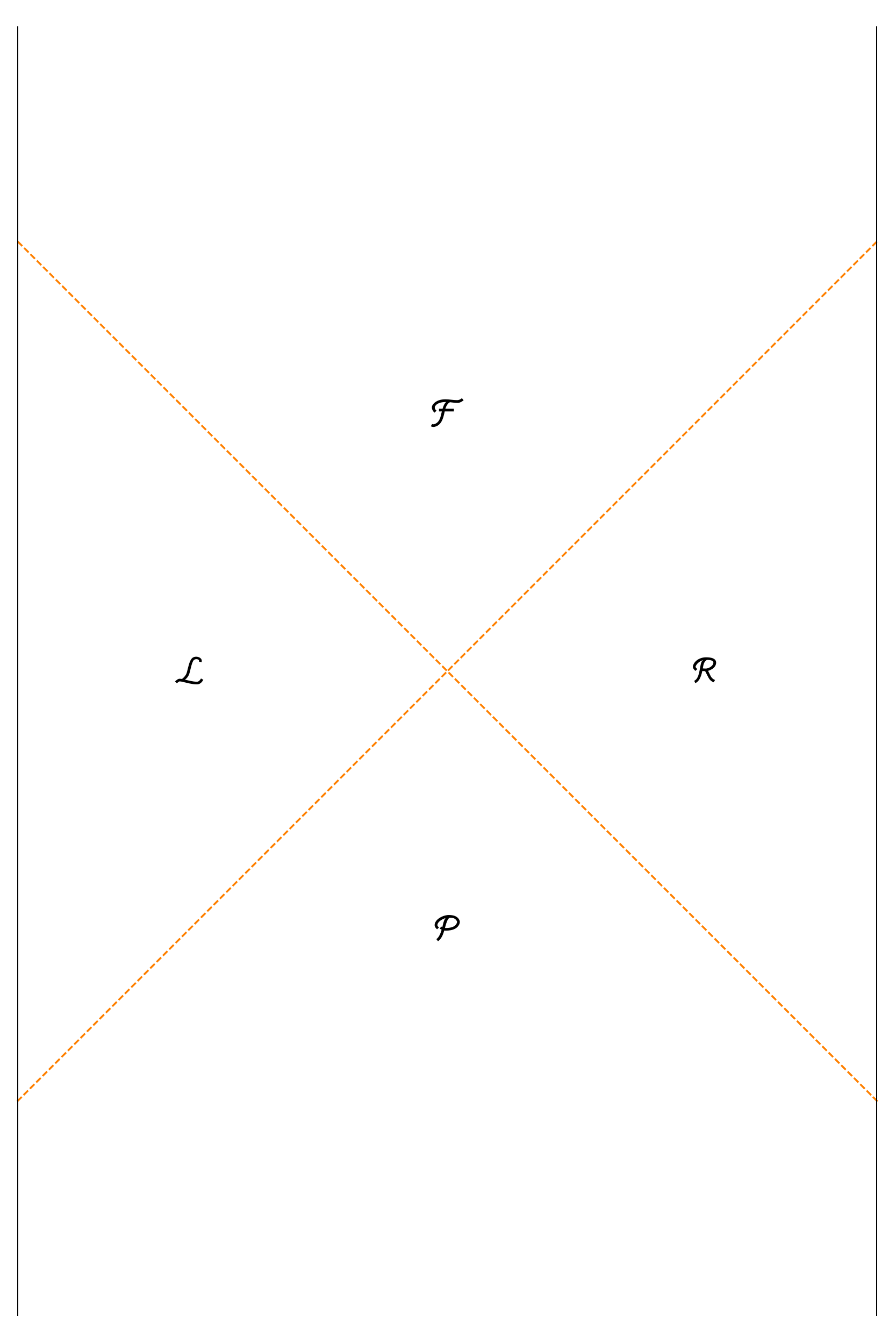} 
\par\end{centering}
\caption{Left: Rindler regions of Minkowski spacetime. Right: AdS Rindler regions of the bulk spacetime. The vertical lines denote the boundary and the dashed lines are Rindler horizons. Each AdS Rindler region has the corresponding Minkowski Rindler region as its boundary.
}
\label{fig:rind}
\end{figure}

Now consider the boundary spacetime separated into Rindler regions, as on the left of {Fig.~\ref{fig:rind}.}
We denote the algebra of operators restricted to the Rindler $R$-region (or $L$-region) as $\sB_R$ (or $\sB_L$). 
These are type III$_1$ vN algebras. 
The single-trace operator algebras restricted to the $R$ and $L$ regions are denoted by $\sA_{R, 0}$ and $\sA_{L, 0}$.  
The restrictions of $\sY$ to the $R$ and $L$ regions are denoted by $\sY_R$ and $\sY_L$. They are von Neumann algebras 
and we have $\sY_R' = \sY_L$.  

 Our proposal is that $\sY_R$ and $\sY_L$ are type III$_1$. This new type III$_1$ structure is only possible perturbatively in $1/N$ expansion, and is mathematically and physically distinct from the type III$_1$ nature of $\sB_R$ and $\sB_L$. The support for our proposal again comes from the complete spectrum of the spectral function of single-trace operators restricted to a Rindler region and the half-sided modular translation structure which we will study in detail in Sec.~\ref{sec:ext} and Sec.~\ref{sec:bdgen}. It is also required by the duality with bulk gravity, which we now elaborate. 
  
 The Poincar\'e patch of AdS can also be separated into four AdS Rindler regions {as on the right of Fig.~\ref{fig:rind}.} 
The standard procedures of the holographic correspondence can be applied to an AdS Rindler region, leading to a duality between the bulk gravity theory in the AdS Rindler $\sR$ ($\sL$) region and the CFT in the boundary $R$ ($L$) region~\cite{Hamilton:2006az,Czech:2012be, Morrison:2014jha}. Denoting the algebras of bulk fields in the AdS $\sR$ and $\sL$ regions 
 as $\tilde \sY_\sR$ and $\tilde \sY_\sL$, we have the identification 
 \be \label{ejnn1}
 \sY_R = \widetilde \sY_\sR, \qquad \sY_L = \widetilde \sY_\sL  \ . 
\ee
As local operator algebras of the bulk  low energy effective theory restricted to a spacetime subregion,  
$\widetilde \sY_\sR$ and $\widetilde \sY_\sL$ are  type III$_1$ vN algebras, thus so are $\sY_R, \sY_L$ due to the identifications~\eqref{ejnn1}. 

The CFT vacuum $\ket{0}$ is cyclic and separating for $\sB_R$ and the corresponding modular Hamiltonian is $- \log \mathbf{\De} = K,$ where $K$ is 
 the boost operator. Similarly, $\ket{\Om_0}_{\rm GNS}$ is cyclic and separating for $\sY_R$, and the flows generated by the corresponding modular operator $\De_0$ should again coincide with boosts. We thus have 
 \be \label{mofl2}
\pi_0 (A_R (\eta = 2 \pi t)) = \pi_0 \le(\mathbf{\De}^{- i t} A_R \mathbf{\De}^{it} \ri)  = \De^{- i t}_0 \pi_0 ( A_R)  \De_0^{it} , \quad A_R \in \sA_{R, 0}  
\ee
 where $\eta$ is the (dimensionless) Rindler time.

With $\sB_R$ and $\sY_R$ being type III$_1$, neither $\mathbf{\De}$ nor $\De_0$ can be factorized into {a product of} operators from the $R$ and $L$ regions, but their non-factorizations are reflected very differently in the bulk.  The non-factorization of $\mathbf{\De}$ implies that the entanglement entropy $S_R$ between the $R$ and $L$ regions in the full CFT can only be defined with a short-distance cutoff $\ep_{\rm b}$ in the boundary, which corresponds to a bulk IR cutoff near the {intersection of the corresponding RT surface with the asymptotic} boundary. 
The non-factorization of $\De_0$ is reflected in the non-factorization of the bulk field theory across the AdS Rindler horizon, which implies  that a bulk UV cutoff $\ep_{\rm UV}$ must be introduced in order to define the bulk entanglement entropy $\sS_\sR$ between the AdS Rindler $\sR$ and $\sL$ regions. {See Fig.~\ref{fig:rind}.}

 The above discussion of a boundary Rindler region can be straightforwardly generalized to ball-shaped regions in the boundary which also have geometric modular flow.



 \subsection{General boundary regions} \label{sec:gbd}

 We now generalize the above discussion of emergent type III$_1$ algebras for Rindler regions to general local boundary subregions. The story is similar, so we will only emphasize those elements which are different. 
 
 We now use $R$ to denote a general {\it spatial} subregion in the boundary. Its causal completion is denoted by $\hat R$. 
 The restriction of $\sB$ to $R$, $\sB_R,$ is the same as  $\sB_{\hat R}$, and is a type III$_1$ vN algebra. 
 Now consider the restriction of $\sY$ to $R$, $\sY_R,$ in the GNS Hilbert space $\sH_0^{\rm (GNS)}$. Note that 
 $\sY_R \neq \sY_{\hat R}$ as $\sY$ is generated by generalized free fields, which do not obey any equations of motion (see Fig.~\ref{fig:ineqAlg}). We now introduce
\be 
\hat \sY_R \equiv (\sY_{\hat R})''  \ .
\ee
From the definition, we have $\hat \sY_R \supseteq \sY_{\hat R}$.  For a half-space (Rindler) or a ball-shaped region,  $\sY_{\hat R} = \hat \sY_R $, as the modular flows are geometric, but for general $R$ it {may be} that $\sY_{\hat R}$ is a proper subset of $\hat \sY_R $.
 We propose that $\hat \sY_R$ is type III$_1$. 
 
 
 We denote the modular operator of $\sB_R$ with respect to the CFT vacuum state $\ket{0}$ as $\mathbf{\De}$ and the modular operator of $\hat \sY_R$ with respect to $\ket{\Om_0}_{\rm GNS}$ as $\De_0$. It is tempting to postulate that modular flows (with $\mathbf{\De}^{-it}$) of elements of $\sA_{\hat R, 0}$ also have a sensible $N \to \infty$ limit, in which case $\De_0$ may be viewed as the representation of $\mathbf{\De}$ in $\sH_{0}^{(\rm GNS)}$, i.e. we should have  
\be \label{yehk}
\pi_0 \le(\mathbf{\De}^{-i s} A \mathbf{\De}^{is} \ri) =  \De_0^{-is} \pi_0 (A)  \De_0^{is} , \quad A \in \sA_{\hat R, 0} 
\ee
which is the statement of the equivalence of bulk and boundary modular flows~\cite{Jafferis:2015del,Faulkner:2017vdd} expressed in our language.
Unlike those in~\eqref{mofl1} and~\eqref{mofl2}, the modular flow parameter $s$ in~\eqref{yehk} does not have any geometric interpretation. 

\begin{figure}[h]
\begin{centering}
	\includegraphics[width=2.5in]{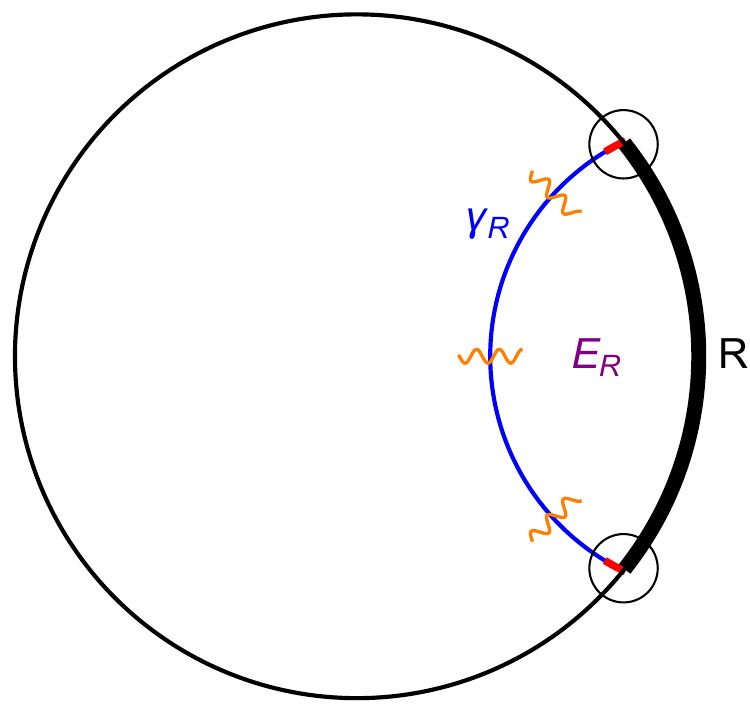} 
\par\end{centering}
\caption{RT surface, $\gamma_R$, for a boundary spatial region $R$. $E_R$ denotes the entanglement wedge. Here we only draw a spatial section of the bulk. {The bulk IR divergence of the area of $\gamma_R$ comes from the part near the boundary (circled red regions) and reflects type III$_1$ nature of boundary algebra $\sB_{R}$.  The type III$_1$ nature of  $\hat \sY_R$  is reflected in the UV divergences of $\sS_{E_R}$, which comes from UV degrees of freedom near $\ga_R$ in the bulk (highlighted by orange wavy lines).}
}
\label{fig:RTs}
\end{figure}

The type III$_1$ nature of $\sB_R$ and $\hat \sY_R$ is again reflected differently in the bulk, with the IR divergence of the area of the Ryu-Takayanagi (RT) surface~\cite{Ryu:2006bv} reflecting the type III$_1$ nature of $\sB_R$, while the divergence in the bulk entanglement entropy $\sS_{E_R}$ for $E_R$ reflects the type III$_1$ nature of $\hat \sY_R$, {see Fig.~\ref{fig:RTs}.} 
 
\section{Physical implications} \label{sec:impl}

The emergent type III$_1$ algebras potentially have many physical implications. One such implication, which will be extensively explored in the rest of the paper, is the emergent half-sided modular inclusion and translation structure, which can be used to generate emergent in-falling flows in the bulk. Here we discuss some other possible implications. Our discussion is somewhat vague, but hopefully offers some pointers for future explorations.

\subsection{Role of the bifurcating horizon and RT surfaces} \label{sec:area}

Consider first the case of the system in the thermal field double state. 
The doubled system has a tensor product structure with $\hat \sH = \sH_R \otimes \sH_L$, $\hat \sB = \sB_R \otimes \sB_L$, and $\hat \sA = \sA_R \otimes \sA_L$. The emergent type III$_1$ nature of $\sY_R, \sY_L$ implies that the GNS Hilbert space $\sH_{\tfd}^{(\rm GNS)}$ does not have a tensor product structure, i.e. it cannot be factorized into Hilbert spaces associated with the $R$ and $L$ theories, and its operator algebra $\sY$ also lacks a tensor product structure in terms of $\sY_R, \sY_L$. This can have important implications for describing the dynamics of low energy excitations around the thermal field double state, including non-factorization of certain objects on the gravity side. 
{An immediate bulk example of such non-factorization comes from operators inserted at the bifurcating horizon (suitably smeared), see the left of Fig.~\ref{fig:wilson}. 
The existence of conserved charges such as energy implies that $\sY_R$ and $\sY_L$ have a nontrivial center at the leading order in the $1/N$ expansion~\cite{WittenNew},  i.e. they are not factors. The presence of diffeomorphisms and other possible gauge symmetries on the gravity side could also lead to a nontrivial center~\cite{Casini:2013rba,Donnelly:2014fua,Donnelly:2015hxa} for $\tilde\sY_R, \tilde \sY_L$ and thus for $\sY_R, \sY_L$. 


 \begin{figure}[h]
\begin{centering}
	\includegraphics[width=2.5in]{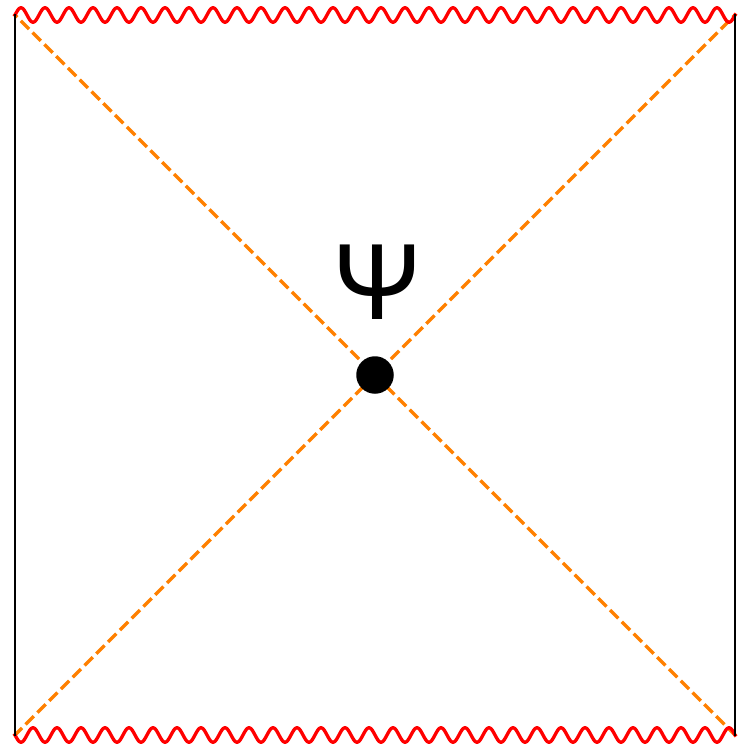}\qquad 
	\includegraphics[width=2.5in]{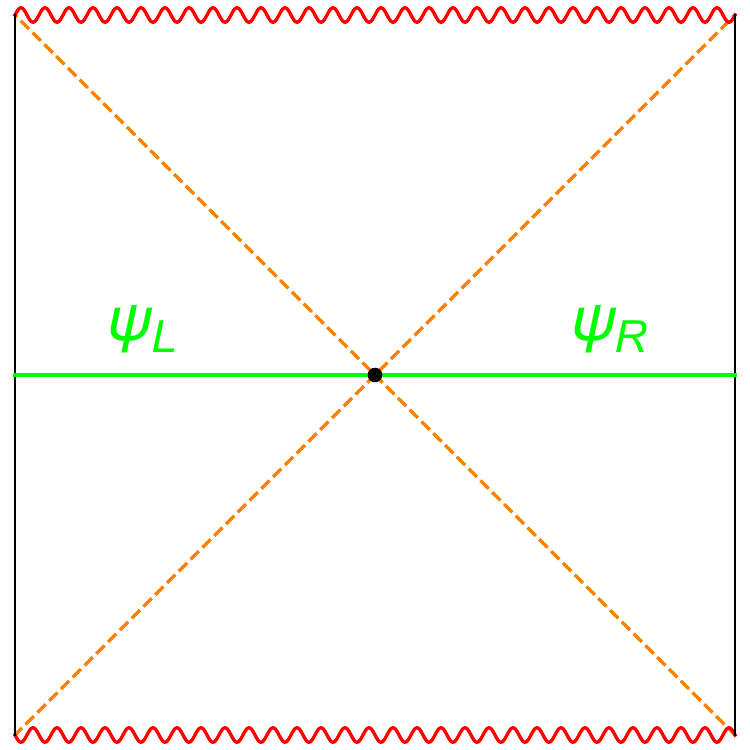} 
\par\end{centering}
\caption{Left: Operators inserted at the bifurcating horizon do not appear to be factorizable. Right: A Wilson line in the eternal black hole geometry and its factorization into left and right operators via the bifurcation surface.
}
\label{fig:wilson}
\end{figure}

{The system can be factorized once we go beyond the $1/N$ expansion. From the bulk perspective this requires going beyond the low energy approximation. Interestingly, there are objects, which naively may be factorized only in the full theory but turn out to be factorizable within the low energy description, with the ``help'' of the bifurcating horizon. }
Here we will briefly comment on two simple examples:


\ben

\item $\sY_R$ is type III$_1$, so its modular operator $\De_0$ with respect to the GNS vacuum $\ket{\Om_0}$ cannot be factorized, which translates to the bulk as the lack of a well-defined entanglement entropy $\sS_R$ between the $R$ and $L$ regions in the continuum limit. Going beyond the $1/N$ expansion, the full theory can be factorized, and there exists a well defined entanglement entropy $S_R$ between the $R$ and $L$ systems. It is a familiar fact that $S_R$ can nevertheless be found using the low energy description on the gravity side by the generalized entropy  
\be \label{ebg}
S_R = {A_{\rm hor} \ov 4 G_N (\ep_{\rm UV})} + \sS_R  (\ep_{\rm UV})
\ee
where $A_{\rm hor}$ is the horizon area and $G_N (\ep_{\rm UV})$ is the (bare) Newton constant at some bulk short-distance cutoff $\ep_{\rm UV}$. The left-hand side is well defined mathematically but 
$\sS_R  (\ep_{\rm UV})$ cannot be defined in the $\ep_{\rm UV} \to 0$ limit, and thus the two terms on the right hand side cannot be individually defined in the continuum limit. 

This emergent type III$_1$ structure also provides a new perspective on the bulk UV divergences and renormalization of the Newton constant $G_N$. Recall that in the usual AdS/CFT dictionary, the bulk UV divergence is understood from the boundary theory as coming from a truncation of operators dual to stringy modes in the bulk. In particular, it is generally expected that the string theory description of a physical quantity should be devoid of UV divergences at each genus order. Here, however, the bulk UV divergences may be understood from the boundary theory as arising from non-factorization of algebra $\sY$. For this reason,  even in string theory  {we expect} that the two terms in~\eqref{ebg} which should come respectively from genus zero (the area term) and from higher genera contributions cannot be individually finite.

\item  Another example was discussed by~\cite{Harlow:2015lma}, as indicated in the right of Fig.~\ref{fig:wilson}, which is a Wilson line $W$ of a bulk gauge field going from the left to the right boundary. To factorize the Wilson line into a product of left and right operators requires breaking it up somewhere in the middle of the black hole geometry, which cannot be done without introducing additional structure. 
But there is an additional structure in the bulk: the bifurcating horizon. 
We can break up the Wilson line in the low energy theory by taking advantage of it, as indicated in Fig.~\ref{fig:wilson}, with
\be 
W = \psi_R - \psi_L , \quad \psi_{L,R} = \int_{\infty}^{r_0} A_r^{(L,R)} dr 
\ee
where $r_0$ is the location of the horizon. From discussions in~\cite{Nickel:2010pr,Glorioso:2018mmw,deBoer:2018qqm}, $\psi_R, \psi_L$ can be identified respectively as effective fields describing diffusion in the right and left theories at a finite temperature. These are collective dynamical variables and cannot be expressed simply in terms of fundamental  degrees of freedom of the boundary theory. The bifurcating horizon can be described in a diffeomorphism invariant way and thus $\psi_R$ and $\psi_L$ are also diffeomorphism invariant and the factorization is well defined.



\een

For both examples above, we see that the horizon plays the role of restoring the factorization in the low energy description. 

{The above discussion can be generalized to the algebra $\hat \sY_R$ associated with a local boundary region $R$. We expect that $\hat \sY_R$ and $(\hat \sY_R)'$ should share a center whose ``size'' is characterized by the area of the RT surface.  
Similarly, the RT surface can be used to restore factorization in the low energy description.}
More explicitly, the entanglement entropy $S_R$ of a region $R$ in the full boundary theory can be obtained from the bulk by~\cite{Faulkner:2013ana}
\be \label{hien}
S_R = {A_{\ga_R} \ov 4 G_N (\ep_{\rm UV})} + \sS_{\sR} (\ep_{\rm UV}) 
\ee
where $A_{\ga_R}$ is the area of the Ryu-Takayanagi (RT) surface $\ga_R$. Recall that in this case $S_R$ is only defined with a UV cutoff in the boundary which translates to a bulk IR cutoff. As remarked earlier, due to the type III$_1$ nature of $\hat \sY_R$, $ \sS_{\sR} (\ep_{\rm UV})$ cannot be defined in the continuum limit, and thus 
 the two terms on the right hand side cannot be defined separately in the limit $\ep_{\rm UV} \to 0$ as was the case in~\eqref{ebg}. 

The parallel with~\eqref{ebg} and the thermal field double case can be made even closer if we put the boundary theory on a lattice. In this regularized theory, at finite $N$ we have a tensor product decomposition of the Hilbert space between the spatial subregion $R$ and its complement $\bar{R},$ $\sH_{{\rm CFT}(N)}^{\rm reg} = \sH_{R} \otimes \sH_{\bar{R}},$ analogous to the tensor product of left and right CFT Hilbert spaces for the thermal field double at finite $N$. In the regulated finite $N$ theory, the entanglement entropy of the subregion $R$ is finite and we can treat the modular operator $\mathbf{\De}$ as being factorizable.
In the strict $N \to \infty$ limit, even the Hilbert space of the boundary theory {\it defined on the lattice} will fail to factorize and the entanglement entropy will diverge. However, the RT formula computes the entropy of the boundary spatial subregion $R$ which is finite in the lattice boundary theory at finite $N$ and thus, just as with the horizon of the black hole in~\eqref{ebg}, the RT surface restores factorization in the low energy bulk description. 

An alternative perspective on factorization in the boundary theory can be obtained if we assume that the boundary theory has the split property, which is believed to be satisfied by general quantum field theories~\cite{Buchholz1,Buchholz2,Fewster}. Consider the situation in Fig.~\ref{fig:split}, 
where we separate the two regions $R$ and $L$ by an infinitesimal distance $\ep_{\rm b}$. 
The split property says that there exists a tensor product decomposition of the global Hilbert space $\sH = \sH_N \otimes \sH_{\bar N}$, giving rise to a type-I factor $\sN_{\rm I}$ corresponding to operators acting on
$\sH_N$, which satisfies 
\be
\sB_R \subseteq \sN_{\rm I} \subseteq \sB_L'   \ .
\ee
The entanglement entropy associated with $\sH_N$ is well defined and in the limit $\ep_{\rm b} \to 0$ it can be used as a regularization of $S_R$.\footnote{Such a regularization was discussed earlier, for example, in~\cite{Dutta:2019gen}.} 
Under this regularization, the type III$_1$ algebra $\hat \sY_R$ can now be viewed as arising from the type I factor $\sN_{\rm I}$. 
In other words, in equation~\eqref{yehk} we can treat the modular operator $\mathbf{\De}$ {in the full theory} as being factorizable. 
Similar to the role played by the event horizon in~\eqref{ebg}, in~\eqref{hien} the RT surface restores the factorization structure of $\sH = \sH_N \otimes \sH_{\bar N}$ in the low energy description.

 \begin{figure}[h]
\begin{centering}
	\includegraphics[width=2.5in]{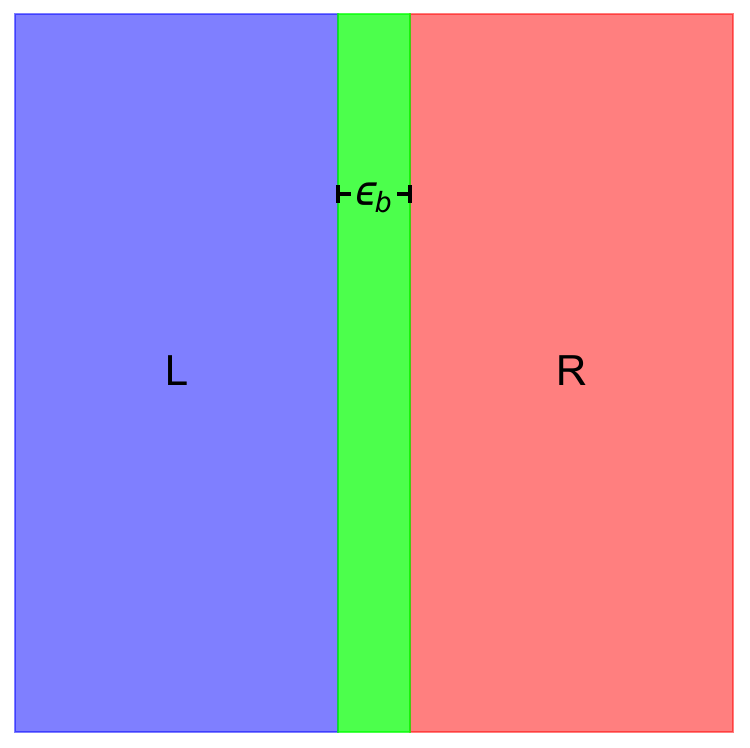} 
\par\end{centering}
\caption{Slightly separated Rindler regions on a spatial slice. The split property implies that there is a tensor factorization of the Hilbert space with respect to an operator algebra $\sN$ contained in the union of the green and red regions above, even though no such tensor factorization exists for the red and blue regions alone when $\ep_{\rm b} = 0$
}
\label{fig:split}
\end{figure}

These discussions also imply that the bifurcating horizon of an eternal black hole can be viewed as a special example of an RT surface from an algebraic perspective. 
For a more general entangled state $\ket{\Psi}$ between the CFT$_R$ and CFT$_L$, the RT surface which 
{provides a signal of the factorization of the full system in the low energy theory} no longer coincides with the horizon. 

The role of the area terms in~\eqref{ebg} and~\eqref{hien} in restoring the tensor product of $\hat \sH$ and $\sH$ also provides a 
 new perspective on their physical origin and their universality. 
  There are other ways to understand the appearance of the area terms from the perspective of quantum error correction~\cite{Harlow:2016vwg, Faulkner:2020hzi} and superselection sectors~\cite{Casini:2019kex}. We believe that all these perspectives can be understood in a unified way, which will be discussed elsewhere. 
  

\subsection{More general dual relations}

 The bulk dual of a boundary subregion $R$ can be defined to be the maximal bulk subregion $E_R$ whose operator algebra can be reconstructed from that of $R$.
 In the static situation and with a {\it spatial} region $R$ which we are considering, the bulk dual $E_R$ for $R$ has been formulated using the RT surface. It is the bulk causal completion of the region between the RT surface $\ga_R$ and $R$.
This definition assumes that the relevant operator algebra in the boundary for $R$ is $\sB_R$ which is equivalent to $\sB_{\hat R}$. 

Our discussion in the previous sections suggests that bulk duals and subregion duality can be formulated more generally, with the definition associated to RT surfaces as a special case. As we emphasized, bulk reconstruction should be more precisely formulated in terms of operators in the GNS Hilbert space, which are built from single-trace CFT operators. For single-trace operators or their representations in the GNS Hilbert space, the algebras associated with different Cauchy slices are inequivalent. 
This leads to new ways of associating algebras with spacetime regions, which in turn leads to new examples of bulk duals that are not related to RT surfaces. We will discuss such examples in Sec.~\ref{sec:bhh}. 
RT surfaces appear in special situations where the operator algebra is associated with the causal completion of a spatial (or null) region. In the more general case, these new notions of bulk duals also raise the interesting question of whether we can define more general notions of entropies that are associated with these bulk subregions. From our discussion of the close connections between bulk area terms and emergent type III$_1$ von Neumann algebras we expect the answer to be yes. We leave this to future investigations. 


\subsection{Emergent symmetries}

There can be emergent symmetries associated with the emergent type III$_1$ structure.
In the example of two copies of CFT in the thermal field double state, it can be shown that there are emergent null translation symmetries along the past and future event horizons of the eternal black hole, which will be discussed in detail in Sec.~\ref{sec:horsy}.  


There are likely many other examples where symmetries in the low energy effective theory of gravity can be understood as being associated with emergent type III$_1$ algebras. Here we mention some possible candidates: 

\ben 

\item In~\cite{Goheer:2003tx} it has been argued that the compactification to $(1+1)$-dimensional Rindler spacetime cannot exist in a quantum gravity, due to incompatibility of an exact $SL(2,R)$ symmetry with a finite number of states. 
It may be possible to understand the Rindler spacetime and associated uncompact symmetries from an emergent type III$_1$ algebra in the  $G_N \to 0$ limit.  

\item An $SL(2, R)$ algebra  in Jackiw-Teitelboim gravity was discussed in~\cite{Maldacena:2018lmt,Lin:2019qwu} (see also~\cite{Harlow:2021dfp}), which implements AdS$_2$ isometries on the matter fields. These symmetries may be  
understood from emergent type III$_1$ algebras in the SYK model.

\item In~\cite{DeBoer:2019kdj} local Poincare symmetry about a RT 
surface was discussed, including its relevance for the modular properties of the boundary theory. 
As with the near-horizon symmetries discussed in item 1 for a black hole, these symmetries should be a consequence of the emergent type III$_1$ structure discussed in Sec.~\ref{sec:gbd}.

\een

\section{Review of half-sided modular translations} \label{sec:half}

In this section we discuss how to generate  new times in the boundary theory. 
Our main tool is half-sided modular inclusion/translation~\cite{Borchers:1991xk, Wiesbrock:1992mg}, and an extension of it. This structure has played a role in proofs of the CPT theorem~\cite{Borchers:1991xk} and the construction of the Poincar\'e group from wedge algebras~\cite{Borchers1996}.
There have also been important applications of the half-sided modular inclusion structure to understanding modular Hamiltonians of regions 
with boundaries on a null plane for a quantum field theory in the vacuum, including average null energy conditions~\cite{Casini:2017roe,Balakrishnan:2017bjg}. See also~\cite{Jefferson:2018ksk} for a discussion concerning black hole interiors.

\subsection{Review of half-sided modular translations}  \label{sec:halfs} 

Suppose $\sM$ is a  von Neumann algebra and the vector $\ket{\Om_0}$ is cyclic and separating for $\sM$. The associated modular and conjugate operators are $\De_\sM$ and $J_\sM$. The commutant of $\sM$ is denoted as $\sM'$. 
$\De_\sM$ leaves $\ket{\Om_0}$ invariant and 
can be used to generate flows within $\sM$ or $\sM'$, 
\be 
\De_\sM^{-it} A \De_{\sM}^{it} = e^{i K_\sM t} A e^{- i K_\sM t} \in \sM, \quad A \in \sM , \quad K_\sM =- \log \De_\sM   \ ,
\ee
while the anti-unitary operator $J_\sM$ takes $\sM$ to $\sM'$ and vice versa
\be 
\sM' =  J_\sM \sM J_\sM , \quad J_\sM^2 = 1 \ .
\ee
$\De_\sM$ acts on both $\sM$ and $\sM'$, and in general cannot be factorized into operators which act only 
on $\sM$ or $\sM'$. 

Now suppose there exists a von Neumann subalgebra $\sN$ of $\sM$ with the half-sided modular inclusion properties:

\ben 
\item $\ket{\Om_0}$ is cyclic for $\sN$ (it is automatically separating for $\sN$ as $\sN \subset \sM$). 

\item The half-sided modular flow of $\sN$ under $\De_\sM$ lies within $\sN$, i.e. 
\be \label{ghb}
\De^{-it}_\sM \sN \De^{it}_\sM   \subset \sN , \quad t \leq 0 \ .
\ee
\een
We will denote the modular operator of $\sN$ with respect to $\ket{\Om_0}$ as $\De_\sN$ with 
$K_\sN = - \log \De_\sN$.

With these assumptions there are the following theorems~\cite{Borchers:1991xk, Wiesbrock:1992mg,Borchers:1998ye,Borchers:1998}.

{\bf Theorem 1:} There exists a unitary group $U(s), \, s \in \RR$ with the following properties:

\ben 

\item $U(s)$ has a positive generator, i.e. 
\be\label{pGen}
U(s) = e^{-i G s}, \qquad G \geq 0
\ee

\item It leaves $\ket{\Om_0}$ invariant
\be 
U(s)\ket{\Om_0} = \ket{\Om_0}, \quad \forall s \in \RR
\ee

\item Half-sided inclusion 
\be \label{hsi}
U^\da (s) \sM U (s) \subseteq \sM , \quad \forall s \leq 0 \ .
\ee

\item $\sN$ can be obtained from $\sM$ with an action of $U$
\be \label{mo}
\sN = U^\da (-1) \sM U(-1)  \ .
\ee

\een

{\bf Theorem 2:} Suppose $U(s) = e^{- i G s}$ is a  continuous unitary group satisfying~\eqref{hsi}, 
then any of the two conditions below imply the third: 
\bea
1. &\quad & G \geq 0 \ . \\
2. &\quad & U(s)\ket{\Omega_0} = \ket{\Omega_0}, \quad s \in \RR  \ . \\
3. & \quad & \Delta^{-it}_\sM U(s) \Delta^{it}_\sM = U(e^{-2\pi t} s), \;\; \text{and} \;\; J_\sM U(s) J_\sM = U(-s) \  .
\label{eju0}
\eea

{\bf Theorem 3:} 
Introducing 
 \be
 \sN_t \equiv \Delta_{\sM}^{-it} \sN \Delta_{\sM}^{it}
 \ee
 we then have 

\ben 

\item The family of algebras $\sN_t$ with $t \in \RR$ is nested, i.e. $\sN_{t_1} \subset \sN_{t_2}$ for $t_1 < t_2$, with 
$\sN_{\infty} = \sM$ and $\sN_0 = \sN$. 

\item The half-sided modular flow of any member of this family gives another algebra in the family. 
In particular, 
\be \label{p19}
\De_\sN^{-i s} \sN_t \De_{\sN}^{is} = \sN_{f_0 (s,t)} , \qquad  f_0 (s,t)= - {1 \ov 2 \pi} \log \le(1+ e^{-2 \pi s} (e^{-2 \pi t} -1) \ri)
\ee 
valid for all $s, t$ such that the argument of the logarithm is positive. Note:

\ben 

\item For $t< 0$, which means $\sN_t \subset \sN$, we always have $f_0 < 0$ for any $s$ and $f_0 < t$ for $s< 0$. $f_0$ increases as $s$ increases and $f_0 \to 0$ as $s \to \infty$. 

\item For $t > 0$, which means that $\sN \subset \sN_t$, the logarithm is defined only for $s \geq s_t \equiv {1 \ov 2 \pi} \log (1 - e^{-2 \pi t}) < 0$, and $f_0 < t$ for $s > 0$.  As $s \to s_t$, $f_0 \to +\infty$, while as $s \to +\infty$, $f_0 \to 0$.  This can be intuitively understood as that the part of $\sN_t$ which is outside $\sN$ is pushed further away from (closer to) $\sN$ for $s< 0$ ($s> 0$).

\een 

\item The action of $U(s)$ on $\sN_t$ has the structure 
\be \label{bdrytshift}
e^{i G s} \sN_t e^{- i G s}  = \sN_{f_1 (s,t)}, \quad f_1 (s, t) = - {1 \ov 2 \pi} \log ( e^{-2 \pi t} - s) 
 \ee
valid for all $s, t$  such that the argument of the logarithm is positive. Note that $f_1 \to -\infty$ as $s \to -\infty$ and 
$f_1 \to +\infty$ as $s \to e^{-2 \pi t}$. 

\item Modular operators of $\sM$ and $\sN$ satisfy the algebra  
\be 
[K_\sM , K_\sN] = - 2 \pi i (K_\sM - K_\sN)  \ . 
\ee


\item $U(s)$ can also be expressed in terms of modular flow operators of $\sM$ and $\sN$ as 
\be \label{p21}
\De_\sM^{-it} \De_{\sN}^{ it} = U (\lam(t)) , \qquad \lam(t) = 
 e^{-2 \pi t} -1  \ .
\ee
Expanding both sides to linear order in $t$ we have 
\be \label{motr0} 
K_\sM - K_\sN = 2 \pi G  
\ee
which gives the explicit form of $G$ in terms of modular Hamiltonians of $\sM$ and $\sN$.\footnote{{Note that the positivity of $G$ is manifest in this expression since $\sK_\sM - \sK_\sN = \log \De_\sN - \log \De_\sM \geq 0$}} 

\een

{\bf Theorem 4}~\cite{Borchers:specCond}: Suppose we have (i) nested von Neumann algebras $\sN_a, a \in \RR, \; \sN_a \subset \sN_b ,$ for $ a < b$
with common cyclic and separating vector $\ket{\Om_0}$; (ii) a one-parameter unitary group $T(a)$ with a positive generator
and $T(a) \ket{\Om_0} = \ket{\Om_0}$; (iii) $T(a)$ translates the algebras
\be 
\sN_a = T(a) \sN_0 T(-a) \ .
\ee
Then $T(a)$ is unique. Theorem 4 then says that given $\sM, \sN$ and $\ket{\Om_0}$, $U(s)$ is unique.

The above structure is called half-sided modular translation and exists only if $\sM$ is a type III$_1$ von Neumann algebra~\cite{Borchers:1998}.  

Similarly, we can define half-sided modular inclusion~\eqref{ghb} for $t \geq 0$ with the corresponding half-sided modular translation for $s \geq 0$. All the statements are parallel except with the following sign changes for equations~\eqref{eju0},~\eqref{p19},~\eqref{bdrytshift},~\eqref{p21}  
\bega 
\Delta^{-it}_\sM U(s) \Delta^{it}_\sM = U(e^{2\pi t} s) , \\
f_0 (s,t)=  {1 \ov 2 \pi} \log \le(1+ e^{2 \pi s} (e^{2 \pi t} -1) \ri), \\
f_1 (s, t) =  {1 \ov 2 \pi} \log ( e^{2 \pi t} + s),  \\
 \lam(t) = 
 e^{2 \pi t} -1  \ .
\end{gather} 


\subsection{Example I: null translations in Rindler spacetime} \label{sec:rind}

Consider a quantum field theory in $(1+1)$-dimensional Minkowski spacetime $\RR^{1.1}$ with coordinates $x^\mu = (x^0, x^1)$ and 
momentum operators $P^\mu = (P^0, P)$. 
Suppose the system is in the vacuum state $\ket{0}$ with respect to the Minkowski time $x^0$. 

Consider the half space $A$ given by
\be 
A = \{ x^\mu \in \RR^{1,1}|x^0 =0, x^1 > 0 \}
\ee
whose domain of dependence is the Rindler $R$-region (see Fig.~\ref{fig:rind} Left). We take  $\sM$ to be the operator algebra in the $R$-region, so $\ket{0}$ is cyclic and separating under $\sM$. The corresponding modular Hamiltonian $K_\sM$ in this case  is proportional to the boost operator $K$
\be 
K_\sM =  - \log \De_\sM 
= 2 \pi K \ ,  \quad \De_\sM^{-it} = e^{i K_\sM t} 
\ee
and $J_\sM$ is the $\sC \sP \sT$ operator. 

It is convenient to use light-cone coordinates 
\be 
x^\pm = x^0 \pm x^1, \quad P^\pm = \ha (H \pm P) = - P_\mp , \quad H = P^0 \ ,
\ee
where the translation operator by a vector $a^\mu$ is given by
\be 
e^{- i P_\mu \cdot a^\mu} = e^{i H a^0 - i P a^1} 
= e^{i P^- a^+ + i P^+ a^-} \ .
\ee
 Note that 
\bega 
[K, P^\pm] = \pm i P^\pm , \qquad e^{i K_\sM s} P^\pm e^{- i K_\sM s} = e^{\mp 2 \pi s} P^\pm , \\
e^{i K_\sM s}  \phi (x^\mu)e^{-i K_\sM s}  = \phi (x'^\mu (s)) , \qquad x'^\pm (s) = e^{\pm 2 \pi s} x^\pm  
\end{gather}
where $\phi (x^\mu)$ is a scalar operator. 

$\sM$ is the operator algebra in the region $x^+ > 0, x^- < 0$. Below for simplicity we will simply use the spacetime region
to denote the operator algebra in that region. 
We take $\sN$ to be the region $\{x^+ > 0, x^- < - 1\}$ (see {Fig~\ref{fig:shiftedWedge} Left}), and then
\be 
\sN_t \equiv  e^{i K_\sM t}  \sN  e^{-i K_\sM t}= \{x^+ > 0, x^- < -  e^{- 2 \pi t}\}   
\ee
with $\sN_t \subset  \sN$  for $t < 0$. We thus have the half-sided modular inclusion structure~\eqref{ghb}. 
In this case the modular operator of $\sN$ can be found explicitly and existence of the positive generator $G$ can be directly verified. 
More explicitly, flows generated by the modular operator of $\sN$ correspond to boosts which which leave the point $a^\mu = (a^+, a^-) = (0, -1)$ invariant.  
Thus  $K_\sN$ should be given by 
\be
e^{i t K_\sN} = e^{-i a^\mu P_\mu} e^{i t K_\sM} e^{i a^\mu P_\mu} = e^{-i  P^+} e^{i t K_\sM} e^{i  P^+} , 
\ee
which gives 
\be 
K_\sN = K_\sM - 2 \pi  P^+ \ .
\ee
From~\eqref{motr0} we conclude that the corresponding $G$ is given by 
\be 
G =  P^+ 
\ee
and thus 
\be 
U(s) = e^{- i  s P^+}  \ .
\ee

We can now verify explicitly the statements of various theorems of last subsection. For example, 
\bega 
U^\da (s) \sM U (s)  = \{x^+ > 0, x^- < s \} \subset \sM , \quad s < 0 , \\ 
\sN = U^\da (-1) \sM U (-1)  , \quad \sN_t = U^\da (- e^{-2 \pi t}) \sM  U (- e^{-2 \pi t}) , \\
\De_\sM^{-it} U(s) \De_\sM^{it} = e^{i K_\sM t} e^{- i P^+ s} e^{-i K_\sM t} = U(e^{-2 \pi t} s)  \ . 
\end{gather} 

By taking $\sN$ to be the operator algebra associated with the region in Fig.~\ref{fig:shiftedWedge} (b), there is a half-sided modular inclusion structure with $t \geq 0$, and the corresponding modular translation operator is given by $G =  P^-$.

\begin{figure}[h]
\begin{centering}
	\includegraphics[width=2.5in]{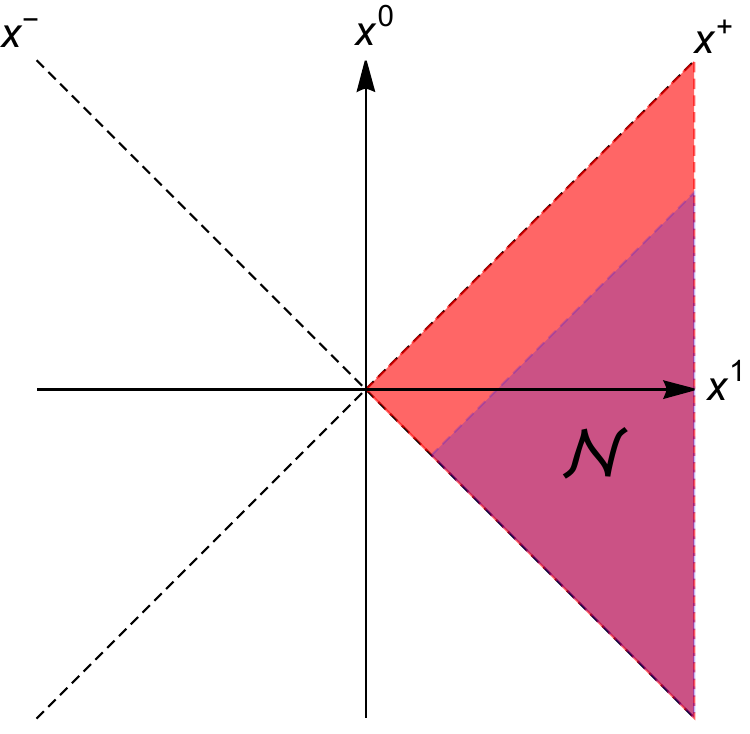} \qquad  \includegraphics[width=2.5in]{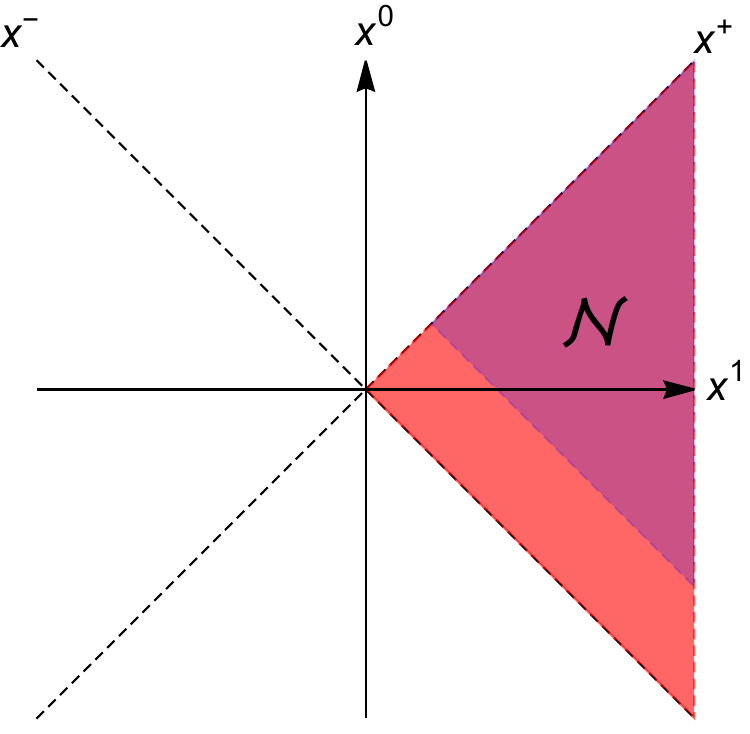}
\par\end{centering}
\caption{Left: The algebra of the subregion $\sN$ that leads to the half-sided modular inclusion structure for $x^-$ translation. Right: The algebra of the subregion $\sN$ that leads to the half-sided modular inclusion structure for $x^+$ translation. 
}
\label{fig:shiftedWedge}
\end{figure}

\subsection{Example II: Two copies of a large $N$ theory in the thermal field double state} \label{sec:htfd}

\begin{figure}[h]
\begin{centering}
	\includegraphics[width=2in]{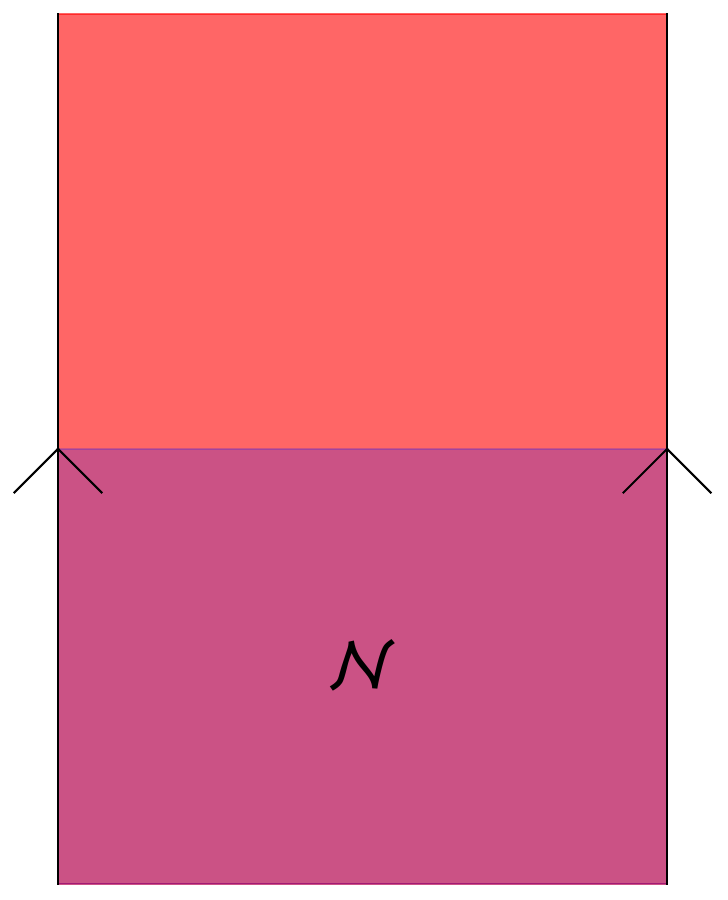} \qquad \includegraphics[width=2in]{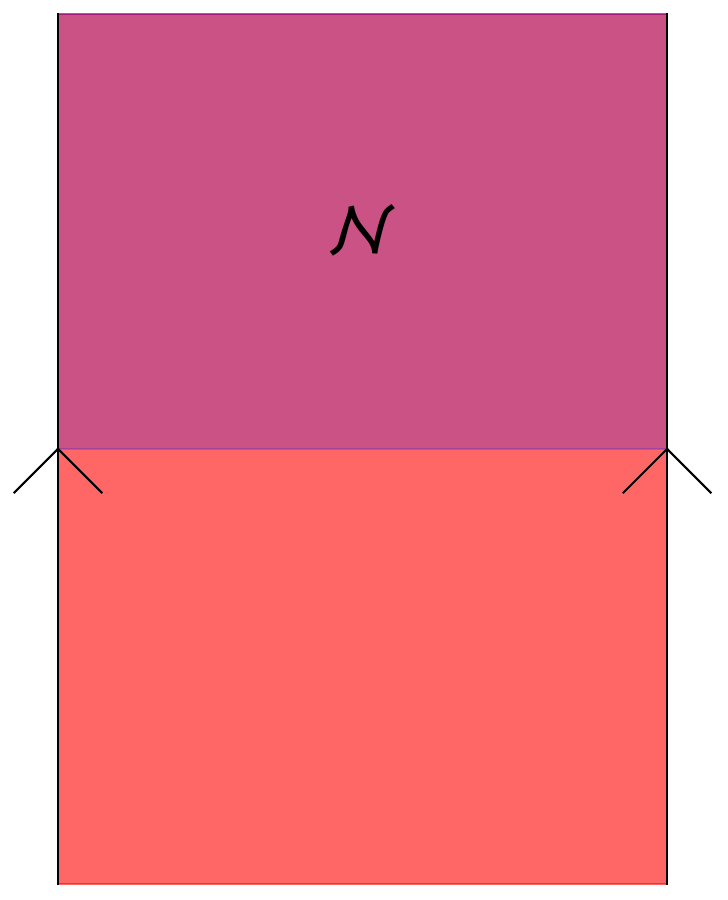}
\par\end{centering}
\caption{Left: $\sN$ denotes the spacetime subregion with $\rt \leq 0$. The vertical axis is time and for simplicity we have taken the spatial manifold to be a circle (vertical boundaries in the figure are identified). Right: $\sN$ denotes the spacetime subregion with $\rt \geq 0$.
}
\label{fig:tShiftBdryBH}
\end{figure}

Now consider two CFTs in the thermal field double state in the large $N$ limit, as discussed in Sec.~\ref{sec:TFD}. 
We now take $\sM = \sY_R$, which is the representation of the single-trace algebra $\sA_{R, TFD}$ in the GNS Hilbert space $\sH_{\rm TFD}^{(\rm GNS)}$. The associated modular operator is $\De_\sM = \De_0$ with corresponding modular time $t$ related to the usual time $\rt$ by $t = {\rt \ov \b}$.

By choosing different subalgebras $\sN$ we can construct different generators $G$ whose spectra are bounded from below and thus generate new ``times".
As the simplest possibility we take $\sN$ to be the representation of the single-trace operator algebra associated with the region indicated in the left plot of Fig.~\ref{fig:tShiftBdryBH}. 
Since generalized free fields do not satisfy any Heisenberg equations, $\sN$ is inequivalent to $\sM$ (recall Fig.~\ref{fig:ineqAlg}). 

The GNS vacuum $\ket{\Om_0}$ is separating with respect to $\sN$. While we do not have a rigorous mathematical proof, we will assume that it is also cyclic with respect to $\sN$. 
Since $\De_0^{-i t}$ generates a time translation, clearly 
\be 
\sN_t \equiv \De_0^{-i t} \sN \De_0^{i t}  \subset \sN, \;\; \text{for} \; t < 0 \ .
\ee
We thus have the half-sided modular inclusion structure of~\eqref{ghb}. In this case $\De_{\sN}$ and $G$ are not explicitly known. The theorems in Sec.~\ref{sec:halfs} can be used to anticipate the action of $U(s) = e^{- i Gs}$, for example as in~\eqref{bdrytshift}. 
In Sec.~\ref{sec:bhh} we will give the explicit action of $U(s)$ by proposing the gravity description of it, which can be explicitly worked out. Equation~\eqref{bdrytshift} then provides a nontrivial check of the proposal. 

We can also consider choosing $\sN$ to be associated with the region in the right plot of Fig.~\ref{fig:tShiftBdryBH}, which gives a half-sided modular inclusion structure for $t > 0$. 

For both plots in Fig.~\ref{fig:tShiftBdryBH}, instead of letting the region describing $\sN$ be bounded by the $\rt =0$ slice, we can choose an arbitrary Cauchy slice (not necessarily with constant $\rt$), {see Fig.~\ref{fig:arbT}}. There is still a half-sided modular inclusion structure and the associated modular translations. Thus there are an infinite number of emergent ``times'' in the large $N$ limit.

\begin{figure}[h]
\begin{centering}
	\includegraphics[width=2.2in]{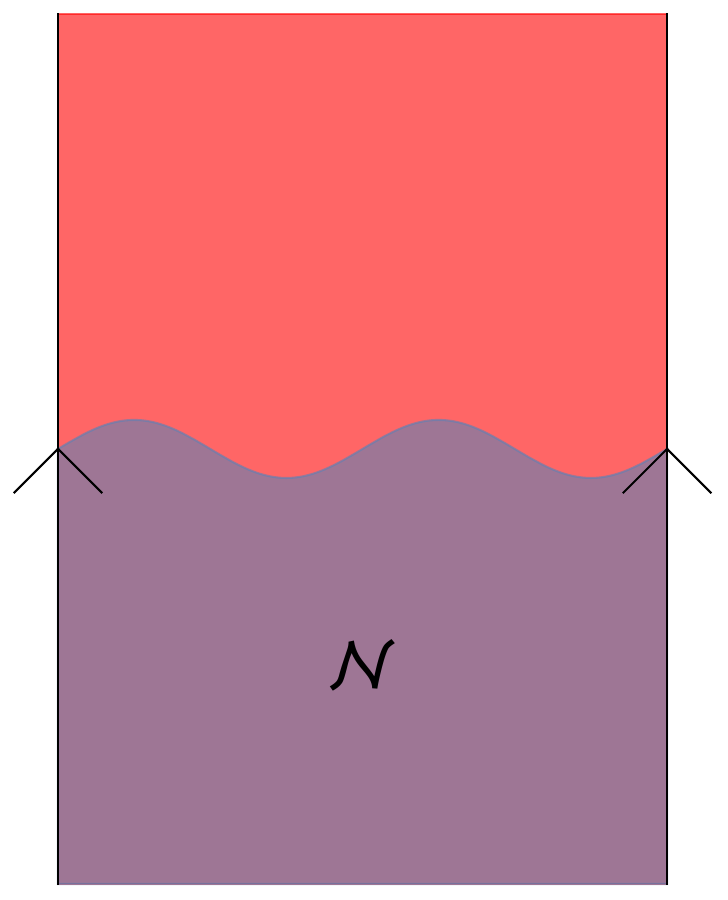} \qquad
	\includegraphics[width=2.2in]{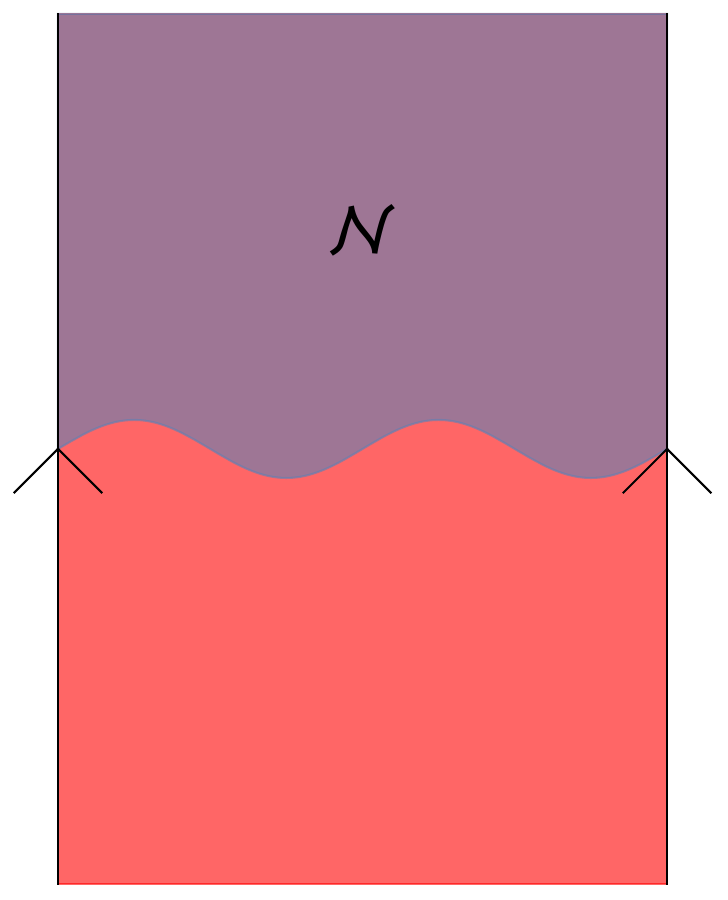}
\par\end{centering}
\caption{For both plots in Fig.~\ref{fig:tShiftBdryBH} instead of letting the region describing $\sN$ be bounded by the $\rt =0$ slice, we can choose a slice $\rt = f (\chi)$ where $\chi$ is the boundary spatial coordinate and $f$ an arbitrary periodic function.
}
\label{fig:arbT}
\end{figure}


\section{Positive extension of half-sided translations for generalized free fields} \label{sec:ext}

We now consider the general structure of half-sided modular translations for a generalized free field theory.
We will show that in this context it is possible to determine the general structure of the action of $U(s)$ for all values of $s$ 
without the need of specifying $\sN$ or $\De_\sM$. 

As an illustration , we will use two CFTs in the thermal field double state in the large $N$ limit, as discussed in Sec.~\ref{sec:TFD} and Sec.~\ref{sec:htfd}, with $\sM = \sY_R, \; \sM' = \sY_L$, and $\De_\sM = \De_0$. The generalized free fields that generate the algebras are defined by~\eqref{eha}--\eqref{eha1}. Below and for the rest of the paper for notational simplicity we will write $\pi (\sO_R (x) )$ simply as $\sO_R (x)$, but it should be kept in mind that they are operators in the GNS Hilbert space. 
Also for convenience for the rest of the paper we will rescale the CFT time such that $\b =2 \pi$. 
The rescaled time will be denoted as $\eta$. From~\eqref{timR} we thus have the relation between modular time $t$ and $\eta$
\be
 t =   {\eta \ov 2 \pi}  \ .
 \ee
 From now on $\om$ will be conjugate to $\eta$. 

The discussion in this section also applies to $\sY_R$ for a local subregion. For a Rindler region, $\eta$ is simply the Rindler time.


\subsection{Unitary automorphism of the algebra} 

For convenience we first copy here some relevant equations of Sec.~\ref{sec:TFD}.  
In the GNS Hilbert space $\sH^{\rm (GNS)}_{\rm TFD}$, single-trace operators $\sO_L, \sO_R$  
can be represented by generalized free fields with mode expansions 
\bega \label{mes}
\sO_\al (x)  =  \sum_k u_k^{(\al)} (x) a_k^{(\al)} = \sum_k  ( \fb_+   u_k^{(\al)} \, c_k^{(\al)} +  \fb_- u_{-k}^{(\al)} \, c_k^{(\bar \al)}), \quad \al = R, L , \quad \bar R = L \ ,
\end{gather} 
where $k = (\om, q)$ denotes all quantum numbers including $\om \in (-\infty, \infty)$ with $\sum_\om = \int {d \om \ov 2 \pi}$. 
The $R$ and $L$ systems are assumed to be symmetric with 
\be
u_{-k}^{(R)} (x) = u_k^{(R)*}  (x) = u_k^{(L)}  (x)  \ .
\ee
The various oscillators satisfy the equations 
\bega  
(a_k^{(\al)})^\da = a_{-k}^{(\al)} , \quad    [a_k^{(\al)} , a_{k'}^{(\b)}] = \ep (\om)   \de_{k+k',0}  \de_{\al \b}  , 
\quad a_{k}^{(\al)} \ket{\Om_0} = e^{-\pi\omega}  a_{-k}^{(\bar \al)} \ket{\Om_0} , \\
\label{cosc1}
\le(c_k^{(\al)} \ri)^\da =  c_{-k}^{(\al)} ,
\quad  [c_k^{(\al)} , c_{k'}^{(\b)}] = \ep (\om)   \de_{k+k',0}  \de_{\al \b}, \quad c_k^{(\al)} \ket{\Om_0} = 0  \;\;
 \text{for} \;\; \om > 0 , \\
 c^{(\al)}_k = \fb_+ a^{(\al)}_k - \fb_- a^{(\bar \al)}_{-k}, \qquad a^{(\al)}_k = \fb_+ c^{(\al)}_k + \fb_- c^{(\bar \al)}_{-k},  \quad
 \fb_\pm = {e^{\pm {\pi |\om| \ov 2} } \ov \sqrt{2 \sinh {\pi |\om| }}}\ .
 \label{c111}
\end{gather} 
The anti-unitary modular conjugation operator $J$ 
 takes $\sO_R$ to $\sO_L$ and vice versa, i.e. 
\be 
J \sO_\al (x) J = \sO_{\bar \al}  (x), \quad  J a_k^{(\al)} J =  a_k^{(\bar \al)}, \quad  J c_k^{(\al)} J =  c_k^{(\bar \al)} \ .
\ee

Now suppose there is a one-parameter unitary automorphism group  $U(s), s \in \RR$, 
\be \label{utr}
\sO_R (x; s) \equiv U(s)^\da \sO_R (x) U(s)  , \qquad \sO_L (x; s) \equiv U(s)^\da \sO_L (x) U(s) 
\ee
which we require to satisfy  the following properties: 
\ben 

\item Half-sided inclusion, i.e. 
\be \label{oneI}
\sO_R (x; s) \in \sY_R , \quad \text{for }  s < 0  \ . 
\ee

\item It leaves the state $\ket{\Om_0}$ invariant 
\be \label{invV} 
U(s) \ket{\Om_0} = \ket{\Om_0} , \quad \forall s \ .
\ee

\item $J$ acts on $U$ as 
\be
J U(s) J = U(-s) \ .
\ee
Acting on both sides of the first equation of~\eqref{utr} by $J$ we find 
\be 
J \sO_R (x; s)  J = J U(s)^\da  J J \sO_R (x) J J U(s) J = \sO_L (x; -s) \ .
\ee

 \item  Under modular flow we require 
\be \label{modF}
\De^{-it}_0 U(s) \De_0^{it}  = U(e^{-2 \pi t} s)    \ .
\ee

\item $U(1)$ group property
\be \label{jen1}
U(s)^\da = U(-s) , \qquad U(s_1) U(s_2) = U(s_1+ s_2) , \quad s, s_1, s_2 \in \RR  \ .
\ee

\een
From Theorem 2 of Sec.~\ref{sec:halfs}, $U(s)$ satisfying the above conditions 
is generated by a Hermitian operator $G$ that is bounded from below.

We will now use the above properties to deduce the explicit transformations of the oscillators under $U(s)$. 
For this purpose we 
 denote 
\be \label{utr1}
\sig_s (a_k^{(\al)} ) \equiv U^\da (s)  a_k^{(\al)} U(s) = \Lam^{\al \b}_{kk'} (s)  a_{k'}^{(\b)}   , \qquad \sig_s (c_k^{(\al)} ) \equiv U^\da(s)  c_k^{(\al)} U(s) 
= \Sig^{\al \b}_{kk'}  (s) c_{k'}^{(\b)} \ .
\ee
 In the above equations and  also subsequent discussions, repeated indices $k'$ and $\b$ should be understood as being summed. 
The transformation matrices $\Sig$ and $\Lam$ can be related to each other using the basis changes~\eqref{c111}. 
More explicitly, we have ($\fb'_\pm$ denotes the corresponding expression as in \eqref{c111} for $\om'$)
\bega \label{rr1}
\Lam^{\al \b}_{kk'} (s)  = \fb_+ \Sig^{\al \b}_{kk'} (s) \fb'_+   + \fb_-  \Sig^{\bar \al \b}_{-kk'} (s) \fb'_+ 
- \fb_+ \Sig^{\al \bar \b}_{k-k'} (s) \fb'_-
- \fb_- \Sig^{\bar \al \bar \b}_{-k-k'} (s) \fb'_-  , \\
\Sig^{\al \b}_{kk'} (s)  = \fb_+ \Lam^{\al \b}_{kk'} (s) \fb'_+   - \fb_-  \Lam^{\bar \al \b}_{-kk'} (s) \fb'_+ 
+ \fb_+ \Lam^{\al \bar \b}_{k-k'} (s) \fb'_-
- \fb_- \Lam^{\bar \al \bar \b}_{-k-k'} (s) \fb'_-  \ .
\label{rr2}
\end{gather}  
Introducing 
\bega \label{leh}
u^{(\al \b)}_k (x; s) = u^{(\al)}_{k'} (x) \Lam^{\al \b}_{k' k} (s) , \quad
 w_k^{(\al \b)} (x; s)  = \fb_+'   u_{k'}^{(\al)} (x) \Sig^{\al \b}_{k' k}(s) +  \fb_-' u_{-k'}^{(\al)} (x) \Sig^{\bar \al \b}_{k' k}(s), 
 \end{gather} 
($\alpha$ is not summed) we have 
\bega
\sO_\al (x; s)  =  u^{(\al \b)}_{k'} (x; s) a_{k'}^{(\b)}  
 =  w_{k'}^{(\al \b)} (x; s) \, c_{k'}^{(\b)}  ,  \\
 w_k^{(\al \b)} (x; s) = u^{(\al \b)}_k (x; s)  \fb_+ + u^{(\al \bar \b)}_{-k}  (x; s)  \fb_-    \ .
\label{leh1} 
 \end{gather}


We now work out the conditions $ \Sig^{\al \b}_{kk'} $ and $ \Lam^{\al \b}_{kk'} $ should satisfy. 
Equation~\eqref{oneI} implies that 
\be 
\Lam^{RL}_{kk'}  (s) =0 , \quad \text{for } s < 0
\ee
and we denote 
 \be 
\label{ehb11}
 \Lam^{RR}_{kk'} (s < 0) = C_{kk'} (s)  \ .
  \ee
Taking the Hermitian conjugate of~\eqref{utr1}  
\be 
 (\sig_s (c_k^{(\al)}))^\da = \sig_s (c_{-k}^{(\al)})   \quad \to \quad 
 \Sig^{\al \b*}_{k k'}(s)  c_{-k'}^{(\b)}  = \Sig_{-k k'}^{\al \b}(s)  c_{k'}^{(\b)}  
\ee
which requires that 
\be \label{heo11}
 \Sig^{\al \b*}_{k k'} (s)  =  \Sig_{-k ,-k'}^{\al \b} (s)  \ .
 \ee
 Similarly we have 
\be  \label{hjk}
 \Lam^{\al \b*}_{k k'} (s)  =  \Lam_{-k ,-k'}^{\al \b} (s), \qquad    C_{kk'}^* (s) = C_{-k -k'} (s)   \ .
 \ee
Acting $J$ on~\eqref{utr1} we have 
\be \label{ue31}
J (\sig_s (c_k^{(\al)})) J =J U^\da (s) c_k^{(\al)} U(s)  J = \sig_{-s} (c_k^{(\bar \al)}) \quad \to \quad
\Sig^{\al \b*}_{k k'}  (s) c_{k'}^{(\bar \b)}  = \Sig^{\bar \al \bar \b}_{k k'}  (-s) c_{k'}^{(\bar \b)} 
\ee
 which implies that 
 \be \label{heo01}
 \Sig^{\al \b*}_{k k'}  (s) =  \Sig^{\bar \al \bar \b}_{k k'}  (-s)  
 \ee
and similarly 
\be 
 \Lam^{\al \b*}_{k k'}  (s) =  \Lam^{\bar \al \bar \b}_{k k'}  (-s)   \ .
 \ee 
From~\eqref{invV} and~\eqref{cosc1} we have 
\be 
\sig_s (c_k^{(\al)} ) \ket{\Om_0} =0 \quad \text{for } \om > 0, \quad  \bra{\Om_0} \sig_s (c_k^{(\al)} )  =0  \quad \text{for } \om <  0 , 
\ee
i.e. 
the action of $\sig_s$ does not mix $c-$type creation and annihilation operators, which implies that $\Sig$ should have the structure 
\be \label{ehni}
\Sig_{kk'}^{\al \b} (s)  = \th (\om) \th(\om') A_{kk'}^{ \al \b} (s)  +  \th(-\om) \th(-\om') B_{kk'}^{\al \b} (s)  \ . 
\ee
Equations~\eqref{heo01} and~\eqref{heo11} imply that 
\be
 A_{kk'}^{ (\al \b)*} (s)   = A_{kk'}^{ (\bar \al \bar \b)} (-s)  = B_{-k-k'}^{ (\al \b)} (s) , \quad 
 B_{kk'}^{ (\al \b)*} (s)   = B_{kk'}^{ (\bar \al \bar \b)} (-s)  = A_{-k-k'}^{ (\al \b)} (s) \ .
 \ee

We will now show that~\eqref{ehni} further implies that the full $\Lam^{\al \b}_{kk'} (s)$ and $\Sig^{\al \b}_{kk'} (s)$ 
can be expressed in terms of $C_{kk'} (s)$ with $s<0$. 
From~\eqref{leh} and~\eqref{leh1} we have for $s < 0$
  \bega
u_k^{(RR)} (x; s)  \fb_+ = \fb_+' u_{k'}^{(R)} (x) \Sig^{RR}_{k' k} +  \fb_-' u_{-k'}^{(R)} (x)
\Sig^{LR}_{k' k}, 
 \\
u_{-k}^{(RR)} (x; s) \fb_- =
 \fb_+' u_{k'}^{(R)} (x) \Sig^{RL}_{k' k} +  \fb_-' u_{-k'}^{(R)} (x)  \Sig^{L L}_{k' k} \ .
 \end{gather} 

From~\eqref{ehni} we then have 
\bega 
u_{k'}^{(R)} (x) C_{k' k}  (s)   \fb_+ =\fb_+' u_{k'}^{(R)} (x) \le( \th(\om) \th(\om') A_{k'k}^{RR} (s)  +  \th(-\om) \th(-\om') B_{k'k}^{RR} (s) \ri) \cr
+  \fb_-' u_{k'}^{(R)} (x)  \le( \th(\om) \th(-\om') A_{-k'k}^{LR} (s)  +  \th(-\om) \th(\om') B_{-k'k}^{LR} (s) \ri) \\
u_{k'}^{(R)} (x) C_{k' -k}  (s)   \fb_- =\fb_+' u_{k'}^{(R)} (x) \le( \th(\om) \th(\om') A_{k'k}^{RL} (s)  +  \th(-\om) \th(-\om') B_{k'k}^{RL} (s) \ri) \cr
+  \fb_-' u_{k'}^{(R)} (x)  \le( \th(\om) \th(-\om') A_{-k'k}^{LL} (s)  +  \th(-\om) \th(\om') B_{-k'k}^{LL} (s) \ri)
\ .
 \end{gather} 
Considering respectively $\om > 0$ and $\om < 0$ on both sides we find that  
\bega 
A_{k'k}^{RR} (s) =\th (\om)  \th (\om') {\fb_+ \ov \fb_+'}  C_{k' k}  (s), \quad 
A_{k'k}^{LR} (s) =\th (\om)  \th (\om') {\fb_+ \ov \fb_-'}  C_{-k' k}  (s), \\
A_{k'k}^{RL} (s) =\th (\om)  \th (\om') {\fb_- \ov \fb_+'}  C_{k' -k}  (s), \quad 
A_{k'k}^{LL} (s) =\th (\om)  \th (\om') {\fb_- \ov \fb_-'}  C_{-k' -k}  (s), \\
B_{k'k}^{RR} (s) =\th (-\om)  \th (-\om') {\fb_+ \ov \fb_+'}  C_{k' k}  (s), \quad 
B_{k'k}^{LR} (s) =\th (-\om)  \th (-\om') {\fb_+ \ov \fb_-'}  C_{-k' k}  (s), \\
B_{k'k}^{RL} (s) =\th (-\om)  \th (-\om') {\fb_- \ov \fb_+'}  C_{k' -k}  (s), \quad 
B_{k'k}^{LL} (s) =\th (-\om)  \th (-\om') {\fb_- \ov \fb_-'}  C_{-k' -k}  (s) \ .
  \end{gather} 
The above equations can be written more compactly as 
\be
A_{k'k}^{\al \b}(s) = {\fb_\b \ov \fb_\al'} \th (\om) \th (\om') C_{\al k' \b k} (s), \quad 
B_{k'k}^{\al \b}(s) = {\fb_\b \ov \fb_\al'} \th (-\om) \th (-\om') C_{\al k' \b k} (s)
\ee
where in subscripts for $\fb$ and $C$ it should be understood that $R =+$ and $L=-$. 
From the above we also have the relations 
\bega
  A_{k'k}^{ (\al \b)*} (s)  = B_{-k'-k}^{ (\al \b)} (s) , \quad {\fb_\al' \ov \fb_\b} A_{k'k}^{\al \b} (s) = {\fb_{\bar \al }' \ov \fb_{\bar \b}} B_{-k'-k}^{\bar \al \bar \b} (s), 
   \end{gather} 
 where the first relation follows from the second equation of~\eqref{hjk}.  The above expressions apply to $s < 0$.  
 We can now find the expressions for $A$ and $B$ for $s> 0$  from~\eqref{heo01} 
\bega
 A_{k'k}^{ (\al \b)} (s)   = A_{k'k}^{ (\bar \al \bar \b)*} (-s) 
 = {\fb_{\bar \b } \ov \fb'_{\bar \al}} \th (\om) \th (\om') C_{\al k' \b k} (-s)
  , \\
B_{k'k}^{ (\al \b)} (s)   = B_{k'k}^{ (\bar \al \bar \b)*} (-s)   
 =  {\fb_{\bar \b } \ov \fb'_{\bar \al}}  \th (-\om) \th (-\om') C_{\al k' \b k} (-s)
 \ .
\end{gather}
 
Collecting everything together we thus have 
 \bega \label{sig1}
 \Sig_{k'k}^{\al \b} (s)  = \bca {\fb_{\bar \b } \ov \fb'_{\bar \al}}   \th(\om\om')   
 C_{\al k' \b k} (-s) , & s > 0 \cr
 {\fb_{ \b } \ov \fb'_{ \al}}   \th(\om\om')  
 C_{\al k' \b k} (s), & s < 0 
 \eca
\end{gather}
and more explicitly 
 \bega
 \Sig_{k' k}^{RR}(s) = {\sqrt{\sinh \pi |\om'|} \ov \sqrt{\sinh \pi |\om|}} 
 \bca 
  \le[ e^{{\pi \ov 2} (\om - \om')} \th(\om) \th(\om')   + e^{- {\pi \ov 2} (\om - \om')}  \th(-\om) \th(-\om')  \ri]  C_{k'k} (s) & s < 0 \cr
   \le[ e^{-{\pi \ov 2} (\om - \om')} \th(\om) \th(\om')   + e^{ {\pi \ov 2} (\om - \om')}  \th(-\om) \th(-\om')  \ri]  C_{k'k} (- s) & s > 0 
   \eca , \\
   \Sig_{k' k}^{RL}(s) = {\sqrt{\sinh \pi |\om'|} \ov \sqrt{\sinh \pi |\om|}} 
 \bca 
  \le[ e^{-{\pi \ov 2} (\om + \om')} \th(\om) \th(\om')   + e^{ {\pi \ov 2} (\om + \om')}  \th(-\om) \th(-\om')  \ri]  C_{k'-k} (s) & s < 0 \cr
   \le[ e^{{\pi \ov 2} (\om + \om')} \th(\om) \th(\om')   + e^{- {\pi \ov 2} (\om + \om')}  \th(-\om) \th(-\om')  \ri]  C_{k'-k} (- s) & s > 0 
   \eca , 
    \end{gather}
and similarly for $\Sig^{LR}$ and $\Sig^{LL}$. From~\eqref{rr1} we then have 
\bega \label{lam1}
\Lam^{RR}_{k'k}(s) = \bca  C_{k'k} (s)& s < 0 \cr
{\sinh \pi \om' \ov \sinh \pi \om}   C_{k' k} (-s)  & s > 0 \quad
\eca, \quad
\Lam^{RL}_{k'k}(s) = \bca  0 & s < 0 \cr
{\sinh \pi (\om + \om') \ov \sinh \pi \om} C_{k' -k} (-s)  & s > 0
\eca , \\
\label{lam2}
\Lam^{LL}_{k'k}(s) = \bca  {\sinh \pi \om' \ov \sinh \pi \om}   C_{-k' - k} (s) & s < 0 \cr
C_{-k' -k} (-s)  & s > 0 \quad
\eca, \quad
\Lam^{LR}_{k'k}(s) = \bca   {\sinh \pi (\om + \om') \ov \sinh \pi \om} C_{- k' k} (s)  & s < 0 \cr
0  & s > 0
\eca \ .
\end{gather}

\subsection{Determining the structure of $C_{kk'} (s)$}

We first collect the conditions $C_{k k'} (s)$ should satisfy and then show that under certain assumptions
it can be completely determined up to a phase. 

The $U(1)$ property implies that for $s_1, s_2 < 0$
 \be \label{dgb}
 C_{k k'} (s_1) C_{k' k''} (s_2) = C_{k k''} (s_1 + s_2) \ . 
 \ee
Since the modular operator $\De_0$ generates a translation in time $t$, it acts on $a_k^{(\al)}$ as\footnote{Recall that $t = {\eta \ov 2 \pi}$ and $\om$ is the frequency for $\eta$.}
\be 
\De_0^{-it}  a^{(\al)}_k \De_0^{it}  = e^{-2 \pi i \al \om t} a^{(\al)}_k 
\ee
which also implies 
\be 
\De^{-it}_0  c^{(\al)}_k \De^{it}_0  = e^{-2 \pi i \al \om t} c^{(\al)}_k \ .
\ee
Acting $\De^{-it}_0$ on~\eqref{utr1} we find from~\eqref{modF} 
\be 
\De^{-it}_0 U(s)^\da c^{(\al)}_k U(s) \De_0^{it}  =e^{-2 \pi i \al \om t}  \Sig^{\al \b}_{k k'} (e^{-2 \pi t} s) c^{(\b)}_{k'}
= \Sig^{\al \b}_{k k'} (s) c^{(\b)}_{k'}  e^{-2 \pi i \b \om' t}
\ee
which implies that 
\be \label{mod0} 
e^{-2 \pi i \al \om t}  \Sig^{\al \b}_{k k'} (e^{-2 \pi t} s)  = \Sig^{\al \b}_{k k'} (s)   e^{-2 \pi i \b \om' t} \ .
\ee
The above equation implies that the $s$-dependence of $\Sig$ must have the form 
\be \label{mod1}
\Sig^{\al \b}_{k k'}(s) \propto s^{- i (\al \om - \b \om')}   
\ee
which in turn implies that 
  \be \label{mod2}
C_{k k'} (s)  \propto (-s)^{- i (\om - \om')}  \ .
\ee
We will make a further assumption that $C_{kk'}$ is diagonal in other quantum numbers, i.e. 
\be 
C_{kk'} (s) = (-s)^{-i (\om-\om')} g  (k, k') \de_{q, q'}  
\ee
From~\eqref{hjk} $g$ should satisfy 
\be\label{newj}
  g^* (k, k') =  g  (-k, -k')  \ . 
 \ee

Equation~\eqref{dgb} requires  
\bega \label{helk0}
\int {d \om' \ov 2 \pi} (-s_1)^{-i (\om-\om')}  g(k, k')
(-s_2)^{-i (\om'-\om'')}   g (k', k'') 
= (-s_1- s_2)^{-i (\om-\om'')}  g (k, k'')  \ . 
\end{gather} 
Without loss of generality,  we take $|s_2| > |s_1|$. The above equation can then be written as 
\bega \label{helk}
\int {d \om' \ov 2 \pi} z^{-i (\om-\om')}  g(k, k')  g (k', k'') = (1+z)^{-i (\om-\om'')}   g(k, k''), \quad z = {|s_1| \ov |s_2|} < 1 \ .
\end{gather} 
To compare with the LHS, let us expand the RHS in powers of $z$ 
\be 
(1+z)^{-i (\om-\om'')} = {1 \ov \Ga (i (\om-\om''))} \sum_{n=0}^\infty {(-1)^n z^n \ov n!}  \Ga ( i (\om-\om'')+n)  
\ee
and equation~\eqref{helk} can follow if the integral on the LHS can be evaluated using Cauchy's theorem with poles with 
$\om - \om' =  i n$.  This motivates us to consider the function 
\be \label{i-func}
I_{\om \om'} (x) = x^{- i (\om -\om')} \Ga (i (\om-\om') +\ep) 
= \int_0^\infty {dp \ov p} p^{i (\om - \om')+\ep} e^{- p x} ,
 \quad x > 0, 
\ee
which satisfies 
\bega 
\int {d \om' \ov 2 \pi} I_{\om \om'} (x_1) I_{\om' \om''} (x_2) 
= I_{\om \om''} (x_1 + x_2) , \quad I_{\om \om'} (0) = 2 \pi \de(\om - \om') \equiv \de_{\om  \om'}   \ .
\end{gather} 

Equation~\eqref{helk0} can then be satisfied if $g (k, k')$ has the form 
\be
g (k, k') = {\lam (k) \ov \lam (k')} \Ga (i (\om-\om') +\ep) 
\ee
which gives $C$ of the form 
\be \label{tenta1} 
C_{kk'} (s) = 
 {\lam (k) \ov \lam (k')}  I_{\om \om' } (-s) \de_{q q'}  \ .
\ee
Equation~\eqref{newj} requires 
\be\label{jbn}
{\lam^* (k) \ov \lam^*(k')}  = {\lam (-k) \ov \lam(-k')} \ .
\ee

We still need to consider the full consequences of~\eqref{invV}. The invariance of $\ket{\Om_0}$ under $U(s)$ 
requires that 
\be
\vev{\Om_0| \sig_{s} (c_{k_1}^\al)  \sig_{s} ( c_{k_2}^\b)  |\Om_0} 
=  \Sig^{\al \ga}_{k_1 k_1'} (s) \Sig^{\b\ga}_{k_2, -k_1'} (s)  \th (\om_1') 
\ee
 is independent of $s$, which leads to\footnote{Note  
\be 
 \vev{\Om_0| c_k^\al c_{k'}^\b |\Om_0} = \de_{\al \b} \de_{k, -k'} \th (\om)  \ .
\ee}
\be \label{hez}
  \Sig^{\al \ga}_{k_1 k_1'} (s) \Sig^{\b\ga}_{k_2, -k_1'} (s)  \th (\om_1')  = \de_{\al \b} \de_{k_1, -k_2} \th (\om_1) \ .
  \ee
  The above equations are in turn  equivalent to 
  \be
  C_{k_1 k'} (s) C_{k_2 -k'} (s) ( \fb'^2_+ \th (\om') + \fb'^2_- \th (-\om') ) 
  = (\fb_{1+}^2 \th (\om_1) + \fb_{1-}^2 \th (-\om_1)  ) \de_{k_1,-k_2}
  \ee
  which can be written more explicitly as 
  \be \label{dgb1}
\sum_{k'}  C_{k_1 k'} (s) C_{k_2 -k'} (s) {e^{\pi \om'} \ov 2 \sinh \pi |\om'|} = 
(\fb_{1+}^2 \th (\om_1) + \fb_{1-}^2 \th (-\om_1)  ) \de_{k_1,-k_2} \ .
\ee  
Inserting~\eqref{tenta1} into~\eqref{dgb1} leads to 
\be \label{kjl}
\sum_{k'} I_{\om_1 \om'} (s) {\lam (k_1) \ov \lam (k')}  \de_{q_1 q'} I_{\om_2 -\om'} (s) {\lam (k_2) \ov \lam (-k')}\de_{q_2, -q'} 
{e^{\pi \om'} \ov 2 \sinh \pi |\om'|} = {e^{\pi \om_1} \ov 2 \sinh \pi |\om_1|} \de_{k_1, -k_2}  \ .
\ee
It can be checked that 
\be 
\sum_{\om'} I_{\om_1 \om'} (s) I_{\om_2 -\om'} (s) e^{\pi \om'}  = \de_{\om_1 , -\om_2} e^{\pi \om_1} 
\ee 
and thus~\eqref{kjl} is satisfied  if 
\be 
\lam (k') \lam (-k') = {1 \ov 2 \sinh \pi |\om'|}  \ .
\ee
From~\eqref{jbn} we then have 
\be 
\lam (k) ={e^{i \ga_k} \ov \sqrt{2 \sinh \pi |\om|}}, \quad \ga_{-k} = - \ga_k
\ee
and 
\be \label{genSt} 
C_{kk'} (s) = \sqrt{{\sinh \pi |\om'| \ov \sinh \pi |\om|}} e^{i \ga_k - i \ga_{k'}}   I_{\om \om' } (s) \de_{q q'}  \ .
\ee
With the above form $C_{kk'}$ has the following ``transpose'' property  
\be \label{ctrans}
C_{k k'}  (s )= C_{k' k}^* (s) {\sinh \pi |\om'| \ov \sinh \pi |\om|} \ .
\ee
With~\eqref{ctrans},~\eqref{dgb1} can be rewritten as 
\be \label{dgb2}
\sum_{k'}  C_{k_1 k'} (s) C_{k' -k_2} (s) e^{\pi \om'}  = 
e^{\pi \om_1} \de_{k_1,-k_2} \ .
\ee

  From~\eqref{jen1} we have 
\be 
\sig_{s_1} (\sig_{s_2} (a_k^{(\al)})) = \sig_{s_1+ s_2} (a_k^{(\al)})  
\ee
which implies that 
\be  \label{heo21}
\Lam_{kk'}^{\al \b} (s_1)  \Lam_{k' k''}^{\b \ga} (s_2) = \Lam_{k k''}^{\al \ga} (s_1 + s_2) \ .
\ee
For $s_1, s_2$ with the same sign, it can be shown that the group properties~\eqref{heo21} follow from~\eqref{dgb}. 
For $s_1, s_2$ of opposite signs,~\eqref{dgb} is not enough, but~\eqref{heo21} can be shown to  follow from the more explicit form~\eqref{genSt}. {See Appendix~\ref{app:u1} for details.} 

To conclude, equations~\eqref{genSt} and~\eqref{lam1}--\eqref{lam2} give the explicit transformation of $a_k^{(\al)}$ and thus $\sO(x)$ under $U(s)$ (for all $s \in \RR$), which satisfies all the desired properties~\eqref{oneI}--\eqref{jen1} for half-sided modular translations. Without needing any explicit information about $\sN$ we have determined the action of $U(s)$ up to a phase factor $e^{i \ga_k}$. Information about different choices of $\sN$ as well as the nature of emergent time $s$ is encoded in this phase factor. 

\section{An example: generalized free fields in Rindler spacetime} \label{sec:bdgen}

In this section we use a simple example to illustrate the formalism developed in Sec.~\ref{sec:ext}.  Consider a generalized free field $\sO (x)$ in Minkowski spacetime and the following question: given the restrictions $\sO_R, \sO_L$ of $\sO$ to the $R$ and $L$ Rindler regions (see Fig.~\ref{fig:rind} Left), is it possible to recover the behavior of the field in the full spacetime (i.e. also in the $F$ and $P$ regions)? Intuitively the answer is no, since a generalized free field does not satisfy any equation of motion, so we cannot obtain the behavior of $\sO$ in the $F$ and $P$ regions by evolving $\sO_R$ and $\sO_L$ from a Cauchy slice as in a standard quantum field theory. 
Here we show that by using the procedure of Sec.~\ref{sec:ext} we in fact can 
express $\sO$ in the $F$ and $P$ regions in terms of those in $R$ and $L$ regions.


\subsection{The transformation}

Consider $(1+1)$-dimensional Minkowski spacetime 
\bega \label{henx}
ds^2 = - (dx^0)^2 + (dx^1)^2 = - dx^+ dx^- = e^{2 \chi} \le(-  d \eta^2  + d \chi^2 \ri) 
=-  e^{2 \chi} d\xi^+ d\xi^- , \\
x^0 = e^{\chi}   \sinh { \eta }, \quad x^1 =e^{\chi}   \cosh{\eta } , 
\quad x^\pm = x^0  \pm x^1 
= \pm e^{\pm \xi^\pm}, \quad \xi^\pm = \eta \pm \chi 
\end{gather}
where the coordinates $\xi^\pm$ cover the $R$ patch. There is a similar description for the $L$ patch. 

For simplicity and also for the later connection with the AdS Rindler discussion we take $\sO$ to be given 
by an operator of dimension $\De$ in a CFT. The expression for the two-point function of $\sO$ is fixed by conformal symmetry 
including the restriction to the $R$ region. Accordingly the mode expansion for $\sO_R (\xi)$ in the $R$ region 
can be written as ({see Appendix~\ref{app:bdlim}})
\bega 
\sO_R (\xi) = \int \frac{d^2 k}{(2\pi)^2}  \, 
 u_k^{(R)} (\xi) a_{k}^{(R)}  \equiv \sum_k  u_k^{(R)} (\xi) a_{k}^{(R)}  \\
 \label{nelk}
 u^{(R)}_k =  N_k e^{{\De \ov 2} (\xi^--\xi^+)}  e^{- i \om \eta + i q \chi  } = N_k e^{\bar q_+ \xi^- - q_- \xi^+}   ,\quad N_k =  {\sqrt{\sinh \pi |\om| }\ov \sqrt{2 \pi} \Ga (\De)} \le|\Gamma\left(q_+ \right) \Gamma\left(q_- \right) \ri|\  \\
 k = (\om, q) , \quad k^\pm = \ha (\om \pm q) , \quad 
  q_\pm = \ha (\De + i (\om \pm q)), \quad  \bar q_\pm = \ha (\De - i (\om \pm q)) \ .
\end{gather} 
There is an analogous mode expansion for $\sO_L$ with $u^{(L)}_k = u^{(R)}_{-k}$. 

Taking $\sM$ to be the algebra generated by $\sO_R$ and 
$\sN$ the subalgebra associated with the region $\xi^- < 0$ (Fig.~\ref{fig:shiftedWedge} Left),  as discussed in Sec.~\ref{sec:rind} 
the generator 
$G$ for half-sided modular translation simply corresponds to a null translation 
$x^- \to x^- +s$, or in terms of $\xi^\pm$
\be \label{nut0}
\xi^- \to \xi_s^- =  \xi^- - \log \le(1- s e^{\xi^-} \ri), \quad \xi^+ \to \xi^+_s = \xi^+ , \quad s < 0  \ .
\ee
We now show how to use the formalism developed in Sec.~\ref{sec:ext} to extend the action of $U(s)$ 
to positive $s$ and thereby extend $\sO_R, \sO_L$ to the $F$ and $P$ regions.

For $s < 0$, we have 
\be 
\sO_R (\xi; s) \equiv U^\da (s) \sO_R (\xi) U(s) = \sO_R (\xi_s) 
\ee
which implies that 
\be 
 u_k^{(R)} (\xi_ s)  = \ u_{k'}^{(R)} (\xi)  C_{k' k} (s), \quad s < 0 \ .
\ee
We then find that 
 \bea 
 C_{k' k}  (s) &=&  {N_k \ov  N_{k'}}  \int d^2 \xi \, e^{- i k' \cdot \xi + i k \cdot \xi_s+ {\De \ov 2} (\xi_s^- - \xi^-)}  
 =   \ha {N_k \ov  N_{k'}} \de_{k'^-, k^-} {\Ga (\bar q'_+)  \ov \Ga (\bar q_+)}
 I_{\om'\om}(-s) \\
& = &  \ha \de_{k'^-, k^-}  {\sqrt{\sinh \pi |\om| } \le|\Gamma\left(q_+ \right) \ri|
 \ov  \sqrt{\sinh \pi |\om'| } \le|\Gamma\left(q'_+ \right) \ri|}  {\Ga (\bar q'_+)  \ov \Ga (\bar q_+)}
  I_{\om'\om}(-s)
  \label{rindcc}
\eea
where function $I$ was introduced previously in~\eqref{i-func} and the last expression has the form of~\eqref{genSt}
with
\be
e^{i \ga_k} ={\Ga (\bar q_+) \ov |\Ga (\bar q_+)|} \ .
\ee

\subsection{Crossing the Rindler horizon}

We now consider $\sO_R (\xi; s)$ with $s > 0$, 
\bega \label{oie}
\sO_R (\xi; s)  =  \sum_k  u_k^{(R)} (\xi) \sig_s (a_k^{(R)} )  
=\sum_k  u_k^{(R\b)} (\xi; s) a_k^{(\b)},  \quad 
  u_k^{(R\b)} (\xi; s)  
 \equiv u_{k'}^{(R)} (\xi) \Lam_{k' k}^{R \b} (s)   \ .
\end{gather} 
From~\eqref{lam1} we have 
\bega 
 u^{(RR)}_{k'}  (\xi; s) =  N_{k'} { e^{ \bar q'_+ \xi^-- q'_- \xi^+}   \ov   \Ga (\bar q'_+) \sinh(\pi\om')} J_1 , \quad
 u^{(RL)}_{k'}  (\xi; s) =  N_{k'} { e^{q'_+ \xi^- - \bar q'_- \xi^+} \ov \sinh(\pi \om')  \Ga (q'_+)}  J_2  ,  \\
\label{j1}
J_1 = \int {d\om  \ov 2 \pi}  \,  \sinh \pi (\om +\om') \,   e^{- i \om (\xi^- + \log s) } 
 \Ga (i \om +\ep) \Ga (- i \om + \bar q_+')
,  \\
 J_2 =\int {d\om  \ov 2 \pi}  \, e^{- i \om (\xi^- + \log s)}   \Gamma\left(- i \om + q'_+ \right)   
 \Ga (i\om + \ep) 
  \sinh \pi \om   \ .
  \label{j2} 
\end{gather} 
The integrals~\eqref{j1}--\eqref{j2} can be evaluated using contour integration. We have the following situations: 

\ben 

\item $\xi^- + \log s < 0$, i.e. $s  < s_0 \equiv e^{-\xi^-}$. In this case we can close the contour in the upper half complex-$\om$ plane picking up poles at 
\be
\om = i \le(n + \ep \ri), \quad n=0,1,\cdots  \
 \ee
which results in
\bega
J_1 = \Ga (\bar q_+') \sinh \pi \om' \le(1 - s e^{\xi^-} \ri)^{-\bar q_+'} , \qquad J_2 =0  
\end{gather} 
and 
\be
 u^{(RR)}_{k'}  (\xi; s) = u^{(R)}_{k'} (\xi^-_s, \xi^+) , \quad \hat u^{(RL)}_{k'}  (\xi; s) =0 , \quad \to \quad
\sO_R (\xi; s) = \sO_R (\xi^-_s, \xi^+)  \ .
\ee
This is the situation where the transformation~\eqref{nut0} is still well defined and $\sO_R (\xi; s)$ remains in the $R$ region.

\item $\xi^- + \log s > 0$, i.e. $s > s_0$ or equivalently $s e^{\xi^-} > 1$. In this case we can close the contour in the lower half complex-$\om$ plane picking up poles  for~\eqref{j1}--\eqref{j2} respectively at 
\be
\om = - i \le(n + \bar q_+' \ri), \quad \om  = - i \le(n + q_+' \ri), \quad  n=0,1,\cdots  \
 \ee
which results in
\bega
J_1 =  (-i) \sin \pi q'_- \Ga (\bar q'_+)  \le( s e^{\xi^-} - 1 \ri)^{-\bar q'_+}, \\
J_2 = (-i) \sin \pi q_+'  \Ga ( q'_+) \le( s e^{\xi^-} - 1 \ri)^{- q'_+}, 
\end{gather} 
and 
\bega \label{bewq} 
 u^{(RR)}_{k}  (\xi; s)  
= -i {\sin \pi q_-  \ov  \sinh(\pi\om)} N_{k} (x^-_s)^{-\bar q_+}  (x^+)^{ -q_-}  , \quad x_s^- = x^- + s,  \\
 u^{(RL)}_{k}  (\xi; s) 
=  -i {\sin \pi q_+  \ov  \sinh(\pi\om)}  N_{k} (x^-_s)^{- q_+}  (x^+)^{ -\bar q_-}
 \label{bewq1} 
\end{gather} 
In this range of $s$,~\eqref{nut0} becomes complex and is no longer well-defined. But
the action of $U(s)$ leads to a well-defined new transformation described by~\eqref{bewq}--\eqref{bewq1} if we use Minkowski $x^\pm$ coordinates of the initial point. Note $x^-_s = x^- +s > 0$, i.e. we are now in the future region. It can be checked that the expressions~\eqref{bewq}--\eqref{bewq1} precisely agree with behavior of the CFT in the $F$ region, 
see the second line of~\eqref{contBdryR} and~\eqref{contBdryL}.

Thus we see that $s_0$ is the ``critical'' value for the half-sided modular translation to take $\sO_R (\xi)$ beyond the Rindler horizon and into the $F$ region. Crossing the Rindler is signaled by the appearance of $a_k^{(L)}$ in $\sO_R (\xi; s)$. 

\een
By taking $\sN$ to be given by the region indicated on the right of Fig.~\ref{fig:shiftedWedge}, we can take $\sO_R$ beyond the past Rindler horizon and into the $P$ region. 

 \section{Bulk reconstruction for AdS Rindler and BTZ revisited} \label{sec:recon}

We will now use the formalism developed in earlier sections to study emergent in-falling times in a black hole geometry. 
In particular, we will give an explicit construction in the boundary theory of an evolution operator for a family of bulk in-falling observers, making manifest the boundary emergence of the black hole horizons, the interiors, and the associated causal structure.  
As an illustration, we will work with the BTZ black hole in AdS$_3$. For contrast, it is also instructive to see how the AdS-Rindler horizon emerges in the boundary theory in this framework. 

In this section we first review the metrics of the BTZ black hole and an AdS$_3$ Rindler region, 
as well as the mode expansions of a bulk scalar field in these geometries.
We then discuss the boundary support of a bulk field in the BTZ black hole or AdS-Rindler spacetime. This part is new and will provide an important preparation for our discussion in Sec.~\ref{sec:bhh}. 

We will set the AdS radius to be unity throughout the paper.

\subsection{AdS Rindler and BTZ geometries} \label{sec:adsgeo} 

Consider the Poincar\'e patch of AdS$_3$
\be \label{poinads}
ds^2 =  {1 \ov z^2} (- (dx^0)^2 + (dx^1)^2 + dz^2 )  =  {1 \ov z^2} (- dx^+ dx^- + dz^2 ), \quad x^\pm = x^0 \pm x^1
\ee
which can be separated into four different AdS Rindler regions, labeled by $\sR, \sL, \sF, \sP$ on the right of Fig.~\ref{fig:rind},
corresponding respectively to regions with $(x^+ , x^-)$ having signs $(+,-)$, $(-,+)$, $(+,+)$, $(-,-)$. They have respectively $R, L,F,P$ Rindler regions of Minkowski spacetime $\RR^{1,1}$
(depicted on the left of Fig.~\ref{fig:rind}) as their boundaries (i.e. as $z \to 0$). It is also convenient to introduce the so-called BTZ coordinates  $(\eta,w,\chi)$, which for the $\sR$ region have the form 
\be 
        z = w  e^{\chi} , \quad x^+  = e^{\xi^+} \sqrt{1-w^2} , \quad x^-  = - e^{- \xi^-} \sqrt{1-w^2}, \quad 
         \xi^\pm = \eta \pm \chi   , 
          \label{RindtoPC1}
    \ee
and in terms of which the metric has a ``black hole'' form 
\be \label{Rindc1}
    ds^2 =  \frac{1}{w^2}\left[-\left(1 - w^2  \right)d\eta^2 + \left(1 - w^2  \right)^{-1}dw^2 + d\chi^2 \right]  \ .
      \ee
The AdS Rindler horizon is at $w =1$ and the boundary is at $w=0$. When $w > 1$, the metric~\eqref{Rindc1} covers the part of the $\sF$ or $\sP$ regions with $z^2 - x^+ x^- > 0$. $w=\infty$ is a coordinate singularity beyond which we have $z^2   - x^+ x^- < 0$ and the BTZ coordinates $(\eta, w, \chi)$ no longer apply. We will refer to the parts of $\sF$ with $z^2 - x^+ x^- > 0$ and $z^2   - x^+ x^- < 0$
respectively as the $\sF_1$ and $\sF_2$ regions. Similarly the $\sP$ region is split into $\sP_1$ and $\sP_2$. 

The BTZ black hole can be obtained by making $\chi$ compact~\cite{Banados:1992wn}, in which case $w=\infty$ becomes a genuine singularity where the spacetime ends, and
 $w =1$ becomes an event horizon. The black hole has inverse temperature $\b =2 \pi$ corresponding to the time $\eta$. 
For compact $\chi$, the Poincar\'e coordinates~\eqref{RindtoPC1} can no longer be used to connect different regions. Instead, we can introduce the Kruskal coordinates in the $\sR$ region 
\bega\label{krubtz1}
 U = -e^{\ze-\eta} =
- \sqrt{1-w \ov 1+w} e^{- \eta} , \quad
 V = e^{ \ze+\eta} =
\sqrt{1-w \ov 1+w} e^{ \eta} \ , 
\end{gather} 
where $\ze$ is the tortoise coordinate 
\be \label{torC1}
\ze = - \int {dw \ov 1-w^2} = \ha \log {1 - w \ov 1+w}   \ .
\ee
The event horizons lie at $U, V=0$ and the boundary lies at $UV =-1$. See Fig.~\ref{fig:casu}. 

Note that the Kruskal coordinates $U, V$ can also be used for AdS-Rindler, with $U, V=0$ corresponding to the AdS-Rindler horizons and $UV =1$ now a coordinate singularity.\footnote{Note that in terms of the range of Kruskal coordinates $U,V$, the AdS-Rindler $\sR$ and $\sL$ regions coincide respectively with the $R$ and $L$ regions of the BTZ black hole, but the $F$ and $P$ regions of the BTZ black hole only 
cover the $\sF_1$ and $\sP_1$ regions of AdS-Rindler.}

 For more extensive discussion of the AdS-Rindler and BTZ spacetimes, see Appendix~\ref{app:adsR}.

\subsection{Mode expansion in AdS-Rindler and BTZ}

Consider a bulk scalar field $\phi$ dual to a boundary operator $\sO$ of dimension $\De$. The restriction $\phi_R (X)$ of $\phi$ to the AdS Rindler $\sR$ region (with $X = (\eta, w, \chi)$) or to the $R$ region of the BTZ black hole has the same mode expansion except that the momentum $q$ along the $\chi$ direction is continuous for AdS-Rindler and discrete for BTZ. Below we will use the same notation for both cases.

$\phi_R$ can be expanded in modes as 
\bega \label{ebs}
 \phi_R (X) = 
 \sum_k  v_{k}^{(R)} (X) a_{k}^{(R)}  , \quad  v_k^{(R)}  (X) = N_{k}  f_{k} (w)  e^{i k \cdot x }  \\
\label{qdef}
 k = (\om, q), \quad 
 k \cdot x = -  \om \eta + q \chi  \quad  q_\pm = \ha (\De + i (\om \pm q)), \quad  \bar q_\pm = \ha (\De - i (\om \pm q)) \\
N_k 
 = {\sqrt{\sinh \pi |\om| }\ov \sqrt{2 \pi} \Ga (\De)} \le|\Gamma\left(q_+ \right) \Gamma\left(q_- \right) \ri| , \quad
   f_{k} (w) =  w^{\Delta} (1-w^2)^{-i\omega/2} {}_2F_1\left(\bar q_-  , \bar q_+ ; \Delta;  w^2\right)
    \ .
\end{gather} 
The $a_k^{(R)}$ are creation {(for $\omega < 0$)} and annihilation {(for $\omega > 0$)} operators of the boundary generalized free field theory in the $R$ region, and thus 
$\phi_R (X)$ can be interpreted as an operator in the boundary theory. 
There is a similar ``bulk reconstruction'' equation for $\phi_L$  in terms of $a_k^{(L)}$. 
Note that $f_k (w)$ is normalizable at infinity 
\be 
\lim_{w \to 0} f_{k} (w)  = w^\De + \cdots  , 
\ee
and satisfies  
\be 
f_k (w) = f^*_k (w) = f_{-k} (w) = f_{-\om, q} (w) = f_{\om, -q} (w)  \ . 
\ee
Near the horizon, $w \to 1$, we have 
\be
v^{(R)}_k (X) 
= {1 \ov \sqrt{2 |\om|}} e^{i k \cdot x}  \le(e^{- i \om \xi + i \de_k} + e^{ i \om \xi - i \de_k} \ri) \ ,
\ee
where the phase shift $\de_k$ is given by 
\be \label{btzph}
e^{i \de_k} =  {\Ga (i \om)  | \Ga (q_- ) \Ga (q_+)| \ov |\Ga (i \om) | \Ga (q_- ) \Ga (q_+) } e^{- i \om \log 2}   \ .
\ee

We also note the asymptotic behavior  for $|\om| \to \infty$ 
\be  \label{hyas}
{}_2F_1\left(\bar q_-, \bar q_+ ; \Delta ; w^2 \right) \approx   w^{\ha - \De}   {\Ga (\De) \ov  2 \sqrt{\pi}} \bca (1-w)^{i \om}
 \le(- {i \om \ov 2} \ri)^{\ha-\De}  & {\rm Im} \om> 0 \cr
 (1+w)^{i \om } 
 \le({i \om \ov 2} \ri)^{\ha-\De} & {\rm Im} \om < 0 
 \eca 
 \ee
 which can be obtained from the discussion of Appendix~\ref{app:hyper}. 
 As $|q| \to \infty$ we have~\cite{Morrison:2014jha}
\bega \label{asy0}
 f_{k} (w) =  |q|^{\ha-\Delta} w^\ha (1-w)^{-{ 1\ov 4}}
 \left(e^{q \arcsin{w}} + e^{-q \arcsin{w}}\right) , \\
|N_{k}|^2 =  {2^{1-2d} \ov (\Ga (\De))^2} \sinh (\pi |\om| )  |q|^{2 (\De-1)} e^{- \pi |q|} \le(1 + O(|q|^{-2}) \ri) , \\
N_{k}  f_{k} (w) \sim \left(e^{- |q| ({\pi \ov 2} - \arcsin{w})} + e^{-|q| ({\pi \ov 2} + \arcsin{w}) }\right) 
 \ ,
\end{gather}
and similarly for $\phi_L$. The expressions in the $F$ and $P$ regions of BTZ, and in the $\sF$ and $\sP$ regions of Poincare AdS can be obtained from analytic continuation which we discuss in Appendix~\ref{app:cont}.

The corresponding boundary operators are obtained by the extrapolate dictionary, i.e. removing $w^{\De}$ and taking $w \to 0$. We find 
\be \label{bdryRindBTZ}
	\sO_R(x) = \sum_k u^{(R)}_k(x) a^{(R)}_k , \qquad u^{(R)}_k(x) = N_k e^{ik \cdot x} \ ,
\ee
and similarly for $\sO_L(x)$ with $u^{(L)}_k(x) = (u^{(R)}_k(x))^*$. For the AdS Rindler the boundary limit is taken by removing $z^\De$, so the corresponding boundary operator in the Rindler patch has an extra factor $e^{-\De \chi}$ as in~\eqref{nelk}.  


\subsection{Boundary support of a bulk operator} \label{sec:bdsu}


The identification of bulk and boundary oscillators $a_k^{(\al)}$ implies that $\phi_R$ of~\eqref{ebs} can be regarded as a boundary operator. This is the statement of bulk reconstruction. We will now examine the support of $\phi_R (X)$ on the boundary. We will use the notation for the BTZ spacetime and exactly the same conclusion applies to AdS-Rindler.

Consider the smearing function $K(X, y)$ defined by~\cite{Hamilton:2005ju,Hamilton:2006fh} 
\bega 
\phi_R (X) =  
 \sum_k N_{k}  e^{i k \cdot x } f_{k} (w) a_{k}^{(R)}
= \int d^2 y  \, K(X, y) \sO_R (y) \\
\label{defK0}
\sO_R (x) = \sum_k N_k e^{i k \cdot x }  a_{k}^{(R)}, \quad K (X, y) = 
\sum_k e^{i k \cdot (x-y)}  f_{k} (w) 
\end{gather} 
where $\sO_R (x)$ is obtained by taking the boundary limit of $\phi_R (X)$ (see \eqref{bdryRindBTZ} or Appendix~\ref{app:bdlim}). 
From the large $q$ behavior of $f_k (w)$, see~\eqref{asy0},  the $q$-integral in~\eqref{defK0} is divergent and thus $K(X,y)$ cannot be consistently defined as a function~\cite{Morrison:2014jha}. The origin of the divergence can be traced to  the complete spectrum feature emphasized in Sec.~\ref{sec:comp}: for any $\om$, $\sO$ has nonzero support for arbitrary large values of $|q|$, but this support decays exponentially at large $|q|$.\footnote{We expect the amplitude for creating a mode of large $q$ with a finite $\om$ should be proportional to $e^{-c \b |q|}$ with $c$ an $O(1)$ number.} The same statements apply to 
all AdS-Rindler and black hole systems in all dimensions (see~\cite{Bousso:2012mh} for other arguments from the bulk). 

The divergences can be avoided if we smear $\phi_R (X)$ in the $\chi$ direction by a function with sufficiently soft large $q$ behavior~\cite{Morrison:2014jha}. 
Alternatively, instead of $\phi_R (X)$, we can consider $\phi_q^{(R)} (\eta, w)$ with a fixed momentum $q$ in the $\chi$-direction. This gives
\bega
 \phi_q^{(R)} (\eta,w)  =  \int \frac{d \om}{2\pi} \,  N_{\om q}  e^{-i \om \eta } f_{\om q} (w) a_{\om q}^{(R)}
= \int d\eta' \, K_q (\eta,w; \eta') \,  \sO_q^{(R)} (\eta') \\
\sO_q^{(R)} (\eta) = \int d \chi \, e^{- i q \chi} \sO_R (\eta, \chi) = \int \frac{d \om}{2\pi} \, N_{\om q}  e^{-i \om \eta } a_{\om q}^{(R)} , \\
\quad  K_q (\eta,w; \eta')= \int \frac{d \om }{2\pi} \, e^{- i \om (\eta-\eta')}  f_{\om q} (w) \ . 
 \label{hex}
\end{gather}
The kernel $ K_q (\eta,w; \eta')$ is now well-defined and we can study its support in $\eta'$.

From~\eqref{hyas}, we have the asymptotic behavior 
\be 
f_k (w) =  w^{\ha}   {\Ga (\De) \ov  2 \sqrt{\pi}} \bca ({1-w \ov 1+ w})^{i \om \ov 2}
 \le(- {i \om \ov 2} \ri)^{\ha-\De}  & {\rm Im } \om > 0 \cr
 ({1+w \ov 1- w})^{i \om \ov 2}
 \le({i \om \ov 2} \ri)^{\ha-\De} & {\rm Im } \om < 0 \cr
 (({1-w \ov 1+ w})^{i \om \ov 2} - i\ep(\om) e^{-i\pi\ep(\om)(\De -1)} ({1+w \ov 1- w})^{i \om \ov 2} )
 \le(- {i \om \ov 2} \ri)^{\ha-\De} & {\rm Im } \om = 0
 \eca , \quad |\om| \to \infty \ .
\ee
This behavior implies that we can close the contour of~\eqref{hex} in the upper half $\om$-plane for
\be \label{ineq1}
\eta- \eta' - \ha \log {1-w \ov 1+w} < 0 \quad \to \quad \eta' > \eta - \ha \log {1-w \ov 1+w}  \quad \to \quad 
U (\eta', w=0) > U(\eta, w) 
\ee
and can close the contour in the lower half $\om$-plane for
\be \label{ineq2}
\eta- \eta' - \ha \log {1+w \ov 1-w} > 0 \quad \to \quad  \eta' < \eta - \ha \log {1+w \ov 1-w} 
 \quad \to \quad V (\eta', w=0) < V (\eta, w)  , 
\ee
where in the last equations of~\eqref{ineq1}--\eqref{ineq2} we have expressed the conditions in terms of the Kruskal coordinates~\eqref{krubtz1}.  

Since $f_k (w)$ is an entire function in the complex $\om$-plane, when we can close the contour either in the upper half or the lower half planes, $K_q (\eta,w;\eta')$ is zero and thus it is only supported in the region 
\be \label{bdra}
\eta_{\rm max} \equiv \eta - \ha \log {1-w \ov 1+w} >  \eta' > \eta - \ha \log {1+w \ov 1-w} \equiv \eta_{\rm min} \ .
\ee
Using the last expressions of~\eqref{ineq1}--\eqref{ineq2}, the above equation corresponds to the region 
on the boundary which is spacelike connected to the bulk point $(\eta,w)$ on the Penrose diagram, see Fig.~\ref{fig:bdsurp}.
Since the range~\eqref{bdra} is $q$-independent, we conclude that any bulk field smeared in the $\chi$ direction 
is also supported in the same window of boundary time. 

A bulk field operator in the $F$ region must be supported on both the $R$ and $L$ boundaries. Using the 
expression of $\phi$ in the $F$ region it can be shown that it is supported on the right boundary for $V (\eta', w=0) > V(\eta, w)$ and on the left boundary $U (\eta', w=0) > U(\eta, w)$. {See Fig.~\ref{fig:bdsurp}.}


\begin{figure}[h]
\begin{centering}
	\includegraphics[width=2in]{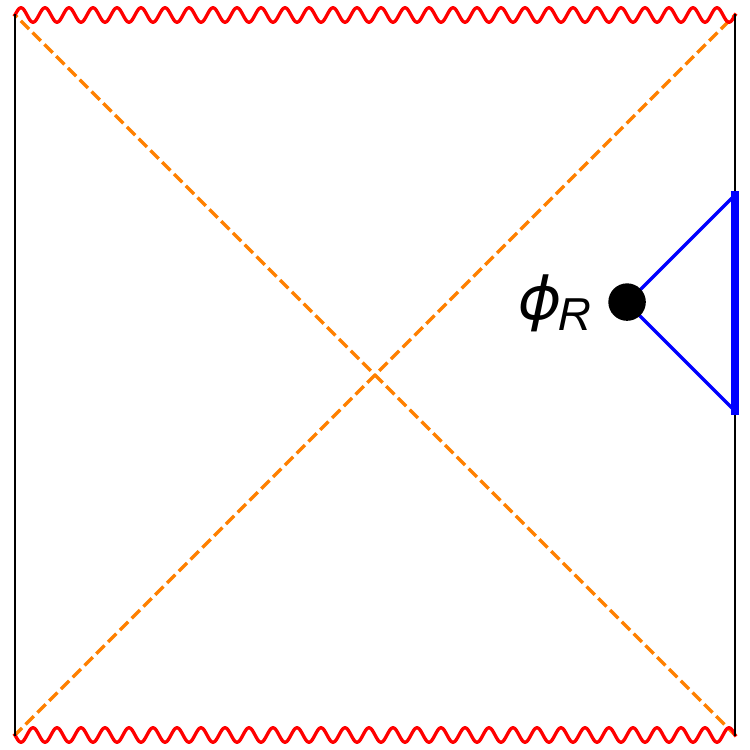} \quad 
	 \includegraphics[width=2in]{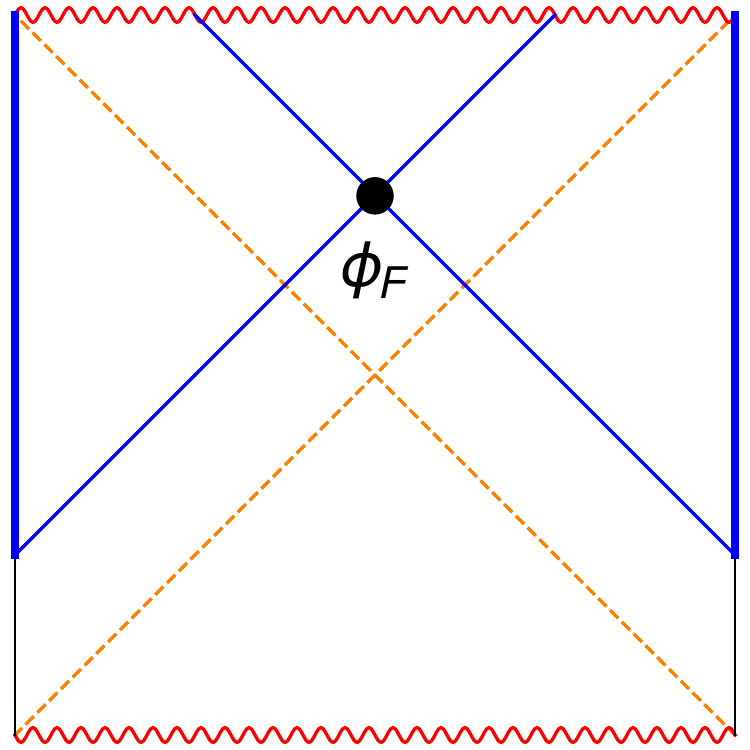}
\par\end{centering}
\caption{Left: Support of a bulk field operator $\phi_R (X)$ in the right region on the boundary in the Penrose diagram. 
The supported region is highlighted with blue color. Right: Boundary support of a bulk field operator $\phi_F$ in the future region.
}
\label{fig:bdsurp}
\end{figure}

\section{Emergence of AdS Rindler horizons} \label{sec:adsrind}

As a warmup for the black hole story we consider the emergence of the bulk AdS-Rindler horizon from the boundary system using 
the unitary group $U(s)$ constructed in Sec.~\ref{sec:bdgen}. Recall that under the duality, $a_k^{(\al)}$ for the bulk mode expansion in the AdS-Rindler regions are identified with those of the generalized free theory in the corresponding boundary Rindler regions. 
We show that the same transformation on $a_k^{(\al)}$ that took a boundary CFT operator across the boundary Rindler horizon also takes a bulk field operator in the $\sR$ region of AdS-Rindler across the bulk horizon. 
In this case going beyond the AdS-Rindler horizon is dictated by symmetries\footnote{See~\cite{Magan:2020iac} for a discussion. Going behind the horizon of a black hole in Jackiw-Teitelboim gravity~\cite{Maldacena:2018lmt,Lin:2019qwu} is also similar to the AdS-Rindler case, {as it can be done using a symmetry operator.}}
 as the null shifts discussed in Sec.~\ref{sec:bdgen} become part of the AdS isometry group. This approach does not apply to a general black hole for which such isometries do not exist. 
 In Sec.~\ref{sec:bhh} we will consider an alternative approach which also applies to black holes. 
 


The discussion is parallel to that of Sec.~\ref{sec:bdgen} except that the wave functions in the AdS-Rindler case are more complicated than the Rindler case. 
Consider the evolution of a bulk field operator initially at a point $X  = (\eta, w, \chi) = (x^+, x^-, z) \in \sR$, 
  \bega 
  \Phi(X; s) = U(s)^{\da} \phi_R (X) U(s) = \sum_{\b} \int \frac{d^2 k'}{(2\pi)^2} \,  v_{k'}^{(R\b)} (X; s) a_{k'}^{(\b)},  \\
  v_{k'}^{(R\b)} (X; s)  = \int \frac{d^2 k}{(2\pi)^2} \, v^{(R)}_k(X) \Lam_{k k'}^{R\b} (s)   
= \int \frac{d^2 k}{(2\pi)^2} N_{k}  e^{i k \cdot x } f_{k} (w)  \Lam_{k k'}^{R \b} (s) \ .
\end{gather} 
where $\Lam_{k k'}^{R \b}$ are given by~\eqref{lam1} with $C_{kk'}$ given by~\eqref{rindcc}.   We then have 
\bega
v_{k'}^{(RR)} (s)  = {N_{k'}  e^{i k' \cdot \xi} \ov \Ga (\bar q'_+)}  w^{\Delta} (1-w^2)^{-i\omega'/2} \bca  J_1 & s < 0 \cr
{1 \ov \sinh \pi \om'}   J_2 & s > 0 
\eca , \\
v_{k'}^{(RL)} (s)  
= {N_{k'}  e^{-i k' \cdot \xi} \ov \sinh \pi \om' \Ga ( q'_+)}  w^{\Delta} (1-w^2)^{i\omega'/2} \bca  0 & s < 0 \cr
 J_3 & s > 0 
\eca ,
\end{gather} 
with 
\bega 
\label{y11}
J_1 =
 \int {d\om \ov 2 \pi}  e^{- i \om \eta_s  }  F \left(\bar q_-'  , -i \om + \bar q_+' ; \Delta;  w^2\right)   \Ga (-i \om + \bar q_+')  \Ga (i\om+ \ep) , \\
 \label{y12}
J_2 =
 \int {d\om \ov 2 \pi}  e^{- i \om \eta_s  }  \sinh \pi (\om + \om')  F \left(\bar q_-'  , -i \om + \bar q_+' ; \Delta;  w^2\right)   \Ga (-i \om + \bar q_+')  \Ga (i\om+ \ep) , \\
 \label{y13} 
 J_3 =  \int {d\om \ov 2 \pi}  e^{- i \om \eta_s  } \sinh \pi \om  F \left( q_-'  , -i \om +  q_+' ; \Delta;  w^2\right)   \Ga (-i \om + q_+')  \Ga (i\om+ \ep) , \\
\eta_s = \xi^- + \log |s| + \ha \log (1-w^2) \ .
 \end{gather} 

We can evaluate the above integrals by closing the contours in the upper or lower half planes. For this purpose we need to know the asymptotic behavior (as $|\om| \to \infty$) of the hypergeometric functions that appear in the integrands. It is convenient to use the identity 
 \ie 
F \left(\bar q_-'  , -i \om + \bar q_+' ; \Delta;  w^2\right) = {\Ga (\De) \Ga (i \om+ i \om') \ov \Ga (q_-') \Ga (q_+'+i \om)} F (\bar q_-' , \bar q'_+ - i\om ; 1 -i \om'-i \om ; 1-w^2) \cr
+(1-w^2)^{i \om' + i \om} {\Ga (\De) \Ga (- i\om'-i \om) \ov \Ga (\bar q_-') \Ga (\bar q_+'-i \om)}  F (q_-', q'_+ + i\om ;
1+ i \om'+i \om; 1-w^2)  \ .
\label{hypergeomXfm}
  \fe 
From equation~\eqref{hypAsmpForSplitRind}} of Appendix~\ref{app:hyper}, the hypergeometric functions in the above equations are of order $O(1)$ as $|\om| \to \infty$  for $ 1-w^2 > -1$ (as is the case for all possible initial points in $\sR$). 
  Each integral in $J_{1,2,3}$ can then be separated into two terms, 
 \be 
 J_{i} = J_{i a} + J_{i b}  , \quad i =1,2,3 
 \ee
 where $J_{ia}$ and $J_{ib}$ are obtained by respectively inserting the first and second term of~\eqref{hypergeomXfm} into the expression for $J_i$. 
Denoting the corresponding integrands as $j_{ia}$ and $j_{ib}$, we have\footnote{$\om$ being pure imaginary gives the most stringent conditions.}
\be 
j_{ia} \sim e^{- i \om \eta_s} , \quad j_{ib} \sim  e^{- i \om \eta_s} (1-w^2)^{i \om} , \quad \om \to \pm i \infty \ .
\ee

We then conclude that for $J_{ia}$ we can close the contour in the upper half plane for $\eta_s < 0$ 
and in the lower half plane for $\eta_s > 0$. While for $J_{ib}$ we can close the contour 
in the upper half plane for $\eta_s - \log (1-w^2) < 0$ and in the lower half plane for $\eta_s - \log (1-w^2) > 0$. 
Denoting the values of $s$ for $\eta_s =0$ and $\eta_s - \log (1-w^2) = 0$ respectively as $s_0$ and $s_1$, we then have 
\be \label{tyeb}
J_i = \bca J_{ia}^{(+)} +  J_{ib}^{(+)} = J_i^{(+)}  &  |s| < s_0 \cr
 J_{ia}^{(+)} +  J_{ib}^{(-)} & s_0 < |s| < s_1 \cr
 J_{ia}^{(-)} +  J_{ib}^{(-)}= J_i^{(-)} &  |s| >  s_1 
 \eca
 \ee
where $J^{(\pm)}_{ia}$ denotes the expression for $J_{ia}$ obtained by closing the contour in the upper (lower) half plane. Note that, for $s > 0$, using~\eqref{RindtoPC1} we have
\be 
 s = s_0  \quad \to \quad \xi^- + \log s - \ha \log (1-w^2) = 0 
 \; \to \;  s_0 = \sqrt{1-w^2}  e^{-\xi^-} = - x^- \ ,
 \ee
so $s= s_0$ is the coordinate distance from $X$ to the horizon along the $x^-$ direction. 
  Also note 
  \be 
 s=  s_1 \quad \to \quad   \xi^- + \log s + \ha \log (1-w^2) = 0 \quad \to \quad 
 s_1 = {1 \ov \sqrt{1-w^2}} e^{-\xi^-}  = {z^2 \ov x^+} - x^- \ .
 \ee
 $s_1$ is then the coordinate distance along the $x^-$ direction from $X$ to the hypersurface separating the $\sF_1$ and $\sF_2$ regions. 
 
 Consider first $J_1$ which applies to $s< 0$. For $|s| < s_0$ from~\eqref{tyeb} we can close the contour of~\eqref{y11} in the upper half plane which gives 
\ie \label{rindJ1}
J_1 
 & = \sum_{n=0}^\infty {(-1)^n \ov n!} e^{n \eta_s}  F \left(\bar q_-'  , n + \bar q_+' ; \Delta;  w^2\right) \Ga (\bar q_+' + n) \\
 & = \Ga( \bar q'_+) F_2 \le(\bar q'_+ ; \bar q'_-, 1 ; \De, 1 ; w^2, -e^{\eta_s} \ri) \cr
&  =  \Ga (\bar q_+')  (1+ e^{\eta_s} )^{-\bar q_+'} F \left(\bar q_-'  ,  \bar q_+' ; \Delta;  w_1^2 \right) ,
 \qquad w_1^2 = {w^2 \ov 1+ e^{\eta_s}}
\fe
where $F_2$ is the second Appell hypergeometric function, and we have used~\eqref{prud4th671} and~\eqref{schlossForm}.
We then find 
  \bega \label{rindWFXfmBeforeHor}
 v_{k'}^{(RR)} (X; s)  
    = N_{k'}  e^{i k' \cdot \xi_s}  w^{\Delta}_s (1-w^2_s)^{-i\omega'/2} 
     F \left(\bar q_-'  ,  \bar q_+' ; \Delta;  w_s^2 \right)  = v_{k'}^{(R)} (X_s)
      \end{gather} 
  where $X_s = (w_s, \xi^+_s, \xi^-_s)$ is given by    
   \bega\label{nsh11}
w_s = {w \ov \sqrt{1-a_s}} , \quad a_s \equiv  s e^{\xi^-} \sqrt{1-w^2} , \\
e^{\xi^-_s} = {e^{\xi^-} \sqrt{1-w^2} \ov \sqrt{1 - a_s}\sqrt{1 - a_s - w^2}} ,
\quad  e^{\xi^+_s} = {e^{\xi^+} \sqrt{1-w^2} \sqrt{1 - a_s} \ov \sqrt{1 - a_s - w^2}}   \ .
\label{nsh12}
\end{gather}
Comparing~\eqref{nsh11}--\eqref{nsh12} with~\eqref{nsh} we conclude that  
\be \label{kds}
\Phi (X;s) = \phi_R (X_s) , \quad   X_s = (x^+, x^- +s , z) \in \sR  
\ee
It can be checked that the expression~\eqref{rindWFXfmBeforeHor}
is in fact valid for all $s <0$. 

Now consider $s > 0$ and $J_2, J_3$.  For $s < s_0$, we can close the contours of~\eqref{y12} and~\eqref{y13}
in the upper half plane. Since the integrand for $J_3$ has no poles in the upper half plane,  $J_3 =0$, while $J_2$ can be evaluated in a similar manner to~\eqref{rindJ1}, giving 
\bega \label{rindJ2}
J_2 
= \sinh \pi \om'  \Ga (\bar q_+')  (1- e^{\eta_s} )^{-\bar q_+'} F \left(\bar q_-'  ,  \bar q_+' ; \Delta;  w_2^2 \right)
\qquad w_2^2 = {w^2 \ov 1- e^{\eta_s}} \ . 
  \end{gather} 
We then find that for $s \in (0, s_0)$, $v_{k'}^{(RR)} (X; s)$ is still given by~\eqref{rindWFXfmBeforeHor} while $v_{k'}^{(RL)} (X; s) =0$, and~\eqref{kds} results.  
 
For $s \in (s_0, s_1)$, from~\eqref{tyeb}, we find\footnote{The evaluation is again similar to~\eqref{rindJ1}. The poles of the integrand in the upper half $\om$ plane are at $\om = i(n + \ep)$ and we have used~\eqref{prud8th671}.}
\bega
	J_{2a} 
	= \sinh\pi\om' {\Ga(\De) \Ga(i\om') \Ga(\bar q'_+) \ov \Ga(q'_+) \Ga(q'_-)} \le(1 - e^{\eta_s} \ri)^{-\bar q'_+} F \le(\bar q_-' , \bar q'_+ ; 1 -i \om'; 1- w_2^2\ri)
	\label{rindJ2Interior}
\end{gather} 
while $J_{2b} =0$ since the integrand has no poles in the lower half plane (the potential poles of $\Ga (-i \om- i \om')$ are canceled by $\sinh\pi(\om + \om')$). We then have  
  \bega \label{rindWFXfmInterior}
  v_{k'}^{(RR)} (X; s)  
    = N_{k'}  e^{i k' \cdot \xi_s}  w^{\Delta}_s (w^2_s-1)^{-i\omega'/2} {\Ga(\De) \Ga(i\om') \ov \Ga(q'_+) \Ga(q'_-)}
     F \le(\bar q_-' , \bar q'_+ ; 1 -i \om'; 1- w_s^2 \ri)  
     \end{gather}  
 where $w_s$ is given by~\eqref{nsh11} but $\xi_s^\pm$ are now given by     
 \be\label{newxi}
e^{\xi^-_s} = {e^{\xi^-} \sqrt{1-w^2} \ov \sqrt{1 - a_s}\sqrt{ a_s -1 + w^2}} ,
\quad  e^{\xi^+_s} = {e^{\xi^+} \sqrt{1-w^2} \sqrt{1 - a_s} \ov \sqrt{a_s -1 + w^2}}    \ .
\ee
We can similarly evaluate $J_3$ 
\be
	J_{3} = \sinh\pi\om' {\Ga(\De) \Ga(i\om') \ov \Ga(q'_-)} (1-e^{\eta_s})^{-\bar q'_-} (e^{\eta_s} -1 +w^2)^{- i\om'} F\le(\bar q'_+, \bar q'_- ; 1 - i\om' ; 1-w_2^2 \ri)
	\label{rindJ3Interior}
\ee
which results in 
  \bega \label{rindLeftWFXfmInterior}
 v_{k'}^{(RL)} (X; s)      = N_{k'}  e^{-i k' \cdot \xi_s}  w^{\Delta}_s (w_s^2 - 1)^{-i\omega'/2} {\Ga(\De) \Ga(i\om') \ov \Ga(q'_+) \Ga(q'_-)}
     F \le(\bar q_-' , \bar q'_+ ; 1 -i \om'; 1- w_s^2 \ri) \ ,
    \end{gather}  
where $\xi^\pm_s$ are given by~\eqref{newxi}. 
Collecting~\eqref{rindWFXfmInterior} and~\eqref{rindLeftWFXfmInterior}, and comparing them with the expressions for the wave functions in the $\sF_1$ region~\eqref{contRindvR}--\eqref{contRindvL} (note~\eqref{newxi} is exactly~\eqref{nsh1}) we conclude 
\be 
\Phi (X; s) = \phi_F (X_s) , \quad X_s = (x^+, x^- +s , z) \in \sF_1 \ .
\ee
From the boundary perspective $s_0$ is the ``critical'' value of $U(s)$ evolution after which $\Phi (X;s)$ now also involves $a^{(L)}_k$. This is the signature of the emergence of the AdS-Rindler horizon from the boundary perspective. 
    
For $s > s_1$, the integrals $J_2$ and $J_3$ can be closed in the lower half plane. We find 
\bega\label{rindJ2FF}
	J_{2} 
	= \sinh\pi(-iq'_-) \Ga(\bar q'_+) (e^{\eta_s} - 1)^{-\bar q'_+} F \le(\bar q_-' , \bar q'_+; \De ; {w^2 \ov 1 - e^{\eta_s}} \ri) , \\
		J_{3} 
	= \sinh\pi(-iq'_+) \Ga( q'_+) (e^{\eta_s} - 1)^{- q'_+} F \le( q_-' , q'_+; \De ; {w^2 \ov 1 - e^{\eta_s}} \ri) , 
	\label{rindJ3FF}
\end{gather}  
leading to 
  \bega \label{rindWFXfmFF}
  v_{k'}^{(R)} (X; s) 
    = N_{k'}  e^{i k' \cdot \xi_s}  w^{\Delta}_s (1 + w^2_s)^{-i\omega'/2} {\sinh\pi(-iq'_-) \ov \sinh\pi\om'}
     F \le(\bar q_-' , \bar q'_+; \De ; - w_s^2 \ri)  \\
      \label{rindLeftWFXfmFF}
  v_{k'}^{(L)} (X; s) 
    = N_{k'}  e^{-i k' \cdot \xi_s}  w^{\Delta}_s (1 + w^2_s)^{i\omega'/2} {\sinh\pi(-iq'_+) \ov \sinh\pi\om'}
     F \le(q_-' , q'_+; \De ; - w_2^s \ri) \ ,
       \end{gather}  
       where now $w_s, \xi^\pm_s$ are given by  (for this range of $s$, $a_s > 1$)
  \be\label{newxi1}
  w_s = {w \ov \sqrt{a_s - 1}} ,  \quad e^{\xi^-_s} = {e^{\xi^-} \sqrt{1-w^2} \ov \sqrt{a_s-1}\sqrt{ a_s -1 + w^2}} ,
\quad  e^{\xi^+_s} = {e^{\xi^+} \sqrt{1-w^2} \sqrt{a_s-1} \ov \sqrt{a_s -1 + w^2}}    \ .
\ee      
Comparing with~\eqref{nsh2} and~\eqref{contRindvR}--\eqref{contRindvL}  we then find that 
\be 
\Phi (X;s) = \phi_{F} (X_s) , \quad X_s = (x^+, x^- +s , z) \in \sF_2 \ .
\ee

We have demonstrated that the transformation $U(s)$, determined from half-sided modular translations in the boundary theory in Sec.~\ref{sec:bdgen}, implements a null shift isometry in AdS, and can be used to generate the full Poincar\'e AdS from its $\sR$ and $\sL$ AdS-Rindler regions. The transformation is well-defined and point-wise for all real values of $s$ and for any choice of initial location of the bulk operator.

\section{Boundary emergence of an in-falling time in a black hole geometry} 
 \label{sec:bhh}

We now consider the generation of an in-falling time in a black hole geometry from the boundary. Our goal is to identify $U(s)$ in the boundary theory which can ``globally evolve'' a Cauchy slice of a black hole geometry across the horizon. 
We will show that the half-sided modular translations discussed in Sec.~\ref{sec:htfd} can be used to for this purpose. 
That is, here we take $\sM = \sY_R$ and $\sN$ to be the algebra of single-trace operators in the GNS Hilbert space $\sH_{\tfd}^{(\rm GNS)}$ associated with the subregion $\eta \leq 0$ in CFT$_R$ (see Fig.~\ref{fig:tShiftBdryBH}).  
Recall the identifications~\eqref{ejnn} and that the modular operator for $\sM$ is $\De_0$ with modular time $t = 2 \pi \eta$. 
  
\subsection{Expressions for the transformations}   
  
As discussed in Sec.~\ref{sec:htfd}, finding the explicit modular operator and the associated half-sided modular translation generator $G$ for the subalgebra $\sN$ indicated in Fig.~\ref{fig:tShiftBdryBH} directly in the boundary theory appears to be difficult. Here we will find it by proposing the bulk dual for $\sN$. 
  
\begin{figure}[h]
\begin{centering}
	\includegraphics[width=2.5in]{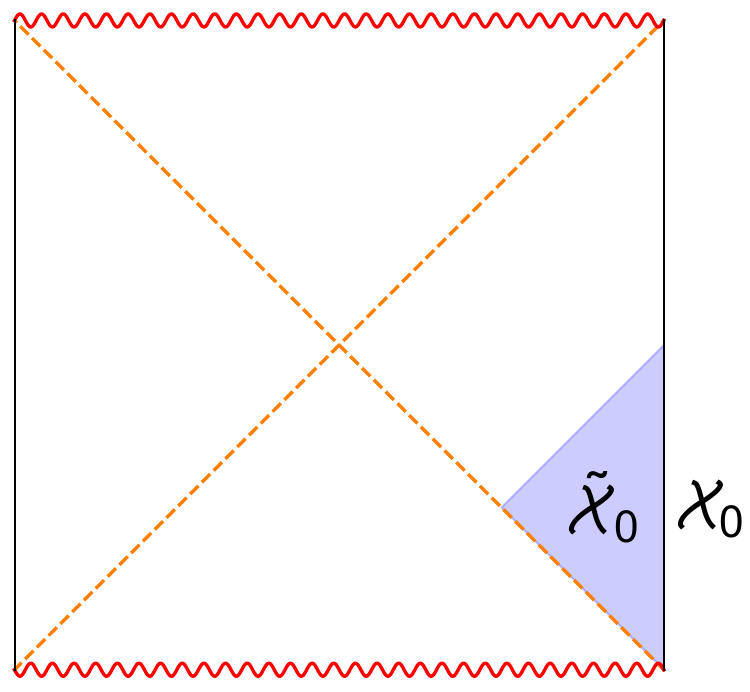} \qquad \includegraphics[width=2.5in]{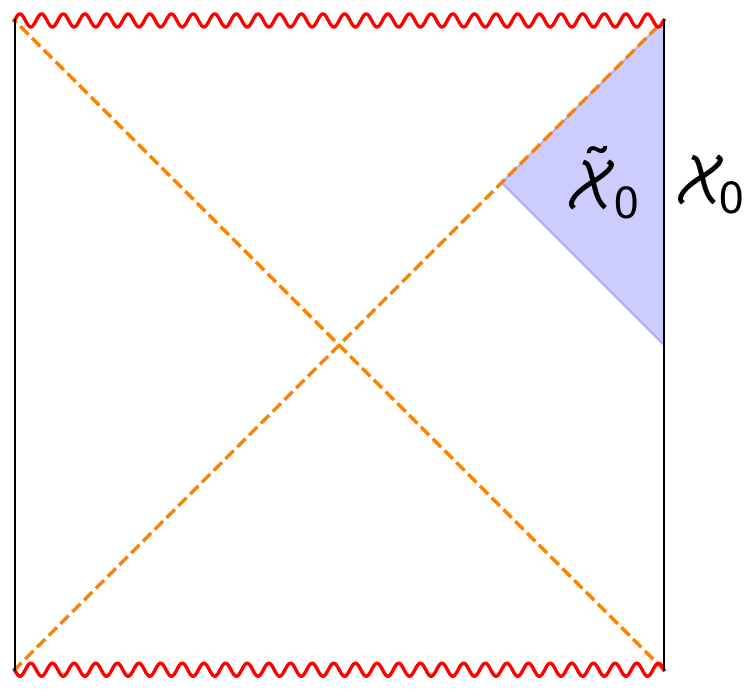}
\par\end{centering}
\caption{The respective proposed bulk bulk duals for the boundary subregions 
indicated in Fig.~\ref{fig:tShiftBdryBH}. 
}
\label{fig:propEW}
\end{figure}

Consider a boundary subregion defined by $\eta \leq \eta_0$ and the corresponding algebra $\sX_{\eta_0}$ of single-trace operators  in the GNS Hilbert space $\sH_{\tfd}^{(\rm GNS)}$. We denote the algebra of bulk fields in the bulk subregion defined by $U \leq U_0 = - e^{-\eta_0}$ (see Fig.~\ref{fig:propEW}), as $\tilde \sX_{\eta_0}$. We propose that 
\be 
\sX_{\eta_0} = \tilde \sX_{\eta_0} \ .
\ee
In other words, the bulk dual of the boundary subregion $\sX_{\eta_0}$  is  given by the bulk region $\tilde \sX_{\eta_0}$.
 This proposal is natural from various perspectives. Firstly, from our discussion of Sec.~\ref{sec:bdsu}, a bulk field operator in $\tilde \sX_{\eta_0}$ has boundary support only in the region $\sX_{\eta_0}$.  Secondly, under modular flow of $\De_0$,  $\sX_{\eta_1} = \De_0^{-it} \sX_{\eta_0} \De_0^{it}$ is the region $\eta \leq \eta_1 = \eta_0 + 2 \pi t$. Under such a flow, the bulk region $\tilde \sX_{\eta_0}$ is taken to 
$\tilde \sX_{\eta_1}$ defined with $U \leq U_1 = - e^{-\eta_1}$. So the identification is consistent with this flow. 
In Sec.~\ref{sec:bdtr} below we will show that~\eqref{bdrytshift} is recovered from this identification, 
providing further nontrivial support. Under this identification we then have $\sN = \sX_0 = \tilde \sX_0$. 

Near the horizon, the black hole spacetime is approximately given by Rindler. In the bulk field theory, the half-sided modular flows associated with $\sM = \tilde \sY_R$ and its subalgebra $\sN = \tilde \sX_{0}$ should then given by that in the Rindler spacetime near the horizon. 
From the discussion of Sec.~\ref{sec:rind}, $G$ should then correspond to a null shift in the Kruskal coordinate $U$.  
We can thus determine the matrix $C_{k k'}$ from the transformation of $a_k^{(\al)}$ near the horizon, from which we obtain the full $\Lam_{k k'}^{\al \b}$. The discussion in this section applies to a general black hole, not restricted to the BTZ. We will use the more general notation of Sec.~\ref{sec:GTFD} except that the boundary time is now taken to be $\eta$ in whose units $\b =2 \pi$.  That is, a bulk point is $X=(\eta, r, \vx)$ with $r$ the radial direction.  We will switch to the BTZ coordinates when restricting to that case.

Consider a bulk operator in the $R$ region
\be 
\phi_R (X) = \sum_k v_k^{(R)} (X) a_k^{(R)}, \quad   v_k^{(R)} (X)  = e^{- i \om \eta} f_k (r) h_q (\vx) 
\ee
where $k = (\omega, q)$ and $h_{q}(\vx)$ denotes the wavefunction in the boundary spatial directions. 
Near the horizon, the bulk wave function can be written as
\bega 
v_k^{(R)} (X) = {h_{q}(\vx) \ov \sqrt{2 |\om|}} e^{-i \om \eta}  \le(e^{- i \om \ze + i \de_k} + e^{ i \om \ze - i \de_k} \ri)
= {h_{q}(\vx) \ov \sqrt{2 |\om|}} \le(  e^{i \de_k} V^{- i \om} + e^{-i \de_k} (-U)^{i \om}  \ri) \\
U = - e^{-\xi^-} , \quad V = e^{ \xi^+} , \quad \xi^\pm = \eta \pm \ze  \ ,
\end{gather} 
where $\ze$ is the tortoise coordinate. 

Consider the evolution with $s < 0$
  \bega 
  \Phi(X; s) = U(s)^{\da} \phi_R (X) U(s) = \sum_{k}   v_{k}^{(RR)} (X; s) a_{k}^{(R)},  \\
   v_{k}^{(RR)} (X; s)  = \sum_{k'} v_{k'}^{(R)} (X) C_{k' k} (s)   \ .
  \label{tena}
\end{gather} 
Note that $C_{k'k}$ is independent of $X$. Now consider $X$ to be close to the past horizon, i.e. $V \to 0$,  where, as discussed above, we expect $ v_{k}^{(RR)} (X; s)  = v_{k}^{(R)} (X_s)$ with $X_s = (U+s, V, \vx)$. 
Then~\eqref{tena} can be written as 
\bega \label{efc1}
 {h_{q}(\vx) \ov \sqrt{2 |\om|}} \le( e^{i \de_k} V^{- i \om} + e^{-i \de_k} (-U-s)^{i \om} \ri) 
=
\sum_{k'} {h_{q'}(\vx) \ov \sqrt{2 |\om'|}} \le( e^{i \de_{k'}} V^{- i \om'} + e^{-i \de_{k'}} (-U)^{i \om'} \ri) C_{k' k} (s) \ .
\end{gather} 
By equating $U$-dependent terms on both sides of~\eqref{efc1}, we find 
\ie  \label{abv1}
C_{k' k} (s) 
 &=\de_{qq'} {\sqrt{|\om'|} \ov \sqrt{|\om|}} e^{i \de_{k'} - i \de_{k}}   {\Ga (-i (\om'+ i \ep)) \ov \Ga (-i (\om + i \ep))} 
  I_{\om' \om} (-s) \cr
& = \de_{qq'}  \sqrt{\sinh \pi |\om| \ov \sinh \pi |\om'|} 
e^{i \ga_{k'} - i \ga_{k}} I_{\om' \om} (-s) 
\fe
where $e^{i \ga_k}$ is defined by 
\bega 
e^{i \ga_k} \equiv e^{i \de_k} {|\Ga (i \om)| \ov \Ga (i \om)}  \ . 
\end{gather} 
We still need to check that $V$-dependent terms~\eqref{efc1}  are also equal when we use~\eqref{abv1}.  
More explicitly, the $V$-dependent term on the right hand side  can be written as 
\be 
R_1 = {h_{q}(\vx)   e^{-i \de_{k}} \ov \sqrt{2 |\om|} \Ga (-i \om)}  
  F(V)  
 \ee
 where 
 \ie \label{fvef}
 F(V) &= \int {d \om' \ov 2 \pi} 
 e^{2 i \de_{k'}} V^{-i \om'}  \Ga (-i (\om'+ i \ep))  
  I_{\om' \om} (-s)  \cr
  & =  (-s)^{i \om} \int {d \om' \ov 2 \pi}\,  e^{2 i \de_{k'}}
  \Ga (-i \om'+\ep)   \Ga (-i (\om -\om' + i \ep)) 
e^{-i  \om' \log (- s V) }  \ .
 \fe
In the near horizon limit $V \to 0$,  we can close the contour in the upper half $\om'$-plane.\footnote{Since $\log V \to -\infty$ this statement is independent of possible $e^{- i \om' \ga}$ (with $\ga$ a constant) type dependence in $e^{2 i \de_{k'}}$.}
Any pole in the upper half plane (include those coming from $e^{2 i \de_{k'}}$) that is a finite distance away from the real axis will lead to a contribution that vanishes as $V \to 0$. Thus the only relevant contribution comes from the pole at $\om' = \om + i \ep$,\footnote{We do not expect the phase shift $e^{i \de_k}$ to have poles on the real axis.} leading to 
\be
F(V) = e^{2 i \de_k} V^{-i \om} \Ga (-i \om) 
\ee
which reproduces the $V$-dependent term on the left hand side of~\eqref{efc1}.

For the case of a BTZ black hole we have from~\eqref{btzph}
\be 
 e^{i \de_k} = {\Ga (i \om)  | \Ga (q_- ) \Ga (q_+)| \ov |\Ga (i \om) | \Ga (q_- ) \Ga (q_+) } e^{- i \om \log 2} 
\ee
which then leads to 
\ie\label{ccbtz}
C_{k' k} (s) & = \de_{qq'} \sqrt{\sinh \pi |\om| \ov \sinh \pi |\om'|} 
 e^{i (\om -\om') \log 2}  {  \Ga (q_- ) \Ga (q_+) \ov  |\Ga (q_- ) \Ga (q_+)| }  { \Ga (\bar q_-' ) \Ga (\bar q_+') \ov  |\Ga (q_-' ) \Ga (q_+') |} I_{\om' \om} (-s) \cr
 & = \de_{qq'}  {N_k \ov N_{k'}}  e^{i (\om -\om') \log 2} { \Ga \le(\bar q'_+\ri) \Ga \le(\bar q'_-\ri)  \ov  \Ga \le(\bar q_+\ri) \Ga \le(\bar q_-\ri) }  I_{\om' \om} (-s) \ .
\fe

With $C_{k'k}$ in hand, we can write down the transformation for general $s$, 
 \bega \label{tena2}
  \Phi(X; s) = U(s)^{\da} \phi_R (X) U(s) = \sum_{k'}  v_{k'}^{(R\b)} (X; s) a_{k'}^{(\b)},  \\
  v_{k'}^{(R\b)} (X; s)  = \sum_k v_k^{(R)} (X) \Lam_{k k'}^{R \b}  (s)   \ .
  \label{tena1}
\end{gather} 
with $\Lam_{k k'}^{R \b}$ given by~\eqref{lam1}. Using the explicit expression~\eqref{ccbtz} for the BTZ black hole we have 
\bega \label{rFromJ}
 v_k^{(RR)} (X; s) = w^{\Delta}   e^{iq\chi} {N_k e^{i \om \log 2} \ov  \Ga \le(\bar q_+\ri) \Ga \le(\bar q_-\ri)}      \bca (-s)^{i\omega} J_1 & s < 0 \cr
s^{i\omega} {J_2 \ov \sinh \pi \om} & s > 0 
\eca,  \\
\label{lFromJ}
 v_k^{(RL)} (X; s) = w^{\Delta}   e^{iq\chi} {N_k e^{-i \om \log 2} \ov  \sinh \pi \om  \Ga \le(q_+\ri) \Ga \le(q_-\ri)}   s^{-i\omega} J_3, \quad s > 0 ,\\
J_1 = \int \frac{d\omega'}{2\pi} \Gamma(\bar q'_+) \Gamma(\bar q'_-) \Gamma(i(\omega'-\omega - i \ep)) a^{-i\omega'} F \left(\bar q'_+, \bar q'_- ; \Delta ; w^2 \right) ,
\\
J_2 = \int \frac{d\omega'}{2\pi}  \sinh \pi \om' \Gamma(\bar q'_+) \Gamma(\bar q'_-) \Gamma(i(\omega'-\omega - i \ep)) a^{-i\omega'} F \left(\bar q'_+, \bar q'_- ; \Delta ; w^2 \right) ,
\\
J_3 = \int \frac{d\omega'}{2\pi}  \sinh \pi (\om' +\om) \Gamma(\bar q'_+) \Gamma(\bar q'_-) \Gamma(i(\omega'+\omega - i \ep)) a^{-i\omega'} F \left(\bar q'_+, \bar q'_- ; \Delta ; w^2 \right) , \\
a \equiv 2 |s| e^{\eta} \sqrt{1-w^2} \ .
\label{Adef}
\end{gather} 
We can evaluate $J_1,~J_2,$ and $J_3$ by contour integration. The discussion is very similar to that of Sec.~\ref{sec:adsrind}, except in this case the integrals $J_{1,2,3}$ can no longer be evaluated explicitly. Here we mention some general features, and in the rest of this section we discuss more specifically various aspects of the transformation. 

Recall that the Gauss hypergeometric function $F$ is an entire function of its first two parameters, and the asymptotic behavior of the hypergeometric functions in $J_{1,2,3}$ for 
$|\om'| \rightarrow \infty$ was given by~\eqref{hyas}. 
Note that the integrand of $J_1$ decays exponentially at large $|\om'|$ for real $\om'$, while the integrands of $J_2$ and $J_3$ (denoted by $j_2,~j_3$ respectively) have power law decay in $|\om'|$ times a Fourier phase factor:
\bega
	j_2 = A |\om'|^{-1-i\om} \le(\le({1-w \ov 1+ w}\ri)^{i \om' \ov 2} - i\ep(\om') e^{-i\pi\ep(\om')(\De -1)} \le({1+w \ov 1- w}\ri)^{i \om' \ov 2} \ri)  (1+ O(|\om'|^{-1})), \\
	j_3 = A |\om'|^{-1+i\om} \le(\le({1-w \ov 1+ w}\ri)^{i \om' \ov 2} - i\ep(\om') e^{-i\pi\ep(\om')(\De -1)} \le({1+w \ov 1- w}\ri)^{i \om' \ov 2} \ri)  (1+ O(|\om'|^{-1})) , \\
	A  = \ha w^{\ha - \De}\Ga(\De)e^{{\pi \ov 2}\ep(\om')\om} e^{-i\om'\log(|s|e^{\eta})}  
	 \ .
\end{gather}

For all $J_{1,2,3}$, we can close the $\om'$ integral in the upper half plane for 
\be \label{opw1}
{a \ov 2 (1-w)} =  |s| e^{\eta} \sqrt{1+w \ov 1-w} = {|s| \ov s_0} < 1 , \quad s_0 \equiv e^{-\eta} \sqrt{1-w \ov 1+w} = - U \ ,
\ee
while for
\be \label{opw2} 
{a \ov 2(1+w)} = |s| e^{\eta} \sqrt{1-w \ov 1+w} = {|s| \ov s_2} > 1 , \qquad s_2 \equiv  e^{-\eta} \sqrt{1+w \ov 1-w} = {1 \ov V} \ ,
\ee 
we can close the contour in the lower half $\om'$ plane. Note that $U$ and $V$ are the values of Kruskal coordinates for the initial point $X$. Also notice that $J_3 =0$ for $s < s_0$, i.e. $\Phi (X;s)$ only involves $a_k^{(R)}$ for $s < s_0$. Thus $s_0$ can be interpreted as the
``critical value'' for crossing the future event horizon.

For  $s \in (s_0, s_2)$ as in the discussion of Sec.~\ref{sec:adsrind} we can split the integrals by using the transformation 
on the hypergeometric function in $J_{1,2,3}$
\be \label{hypGeoXfmSplitNS}
\begin{aligned}
	&F \left(\bar q'_+  , \bar q'_-  ; \Delta ; w^2 \right)  = {\Ga(\De) \Ga(i\om') \ov \Ga\le(q'_+ \ri)\Ga\le(q'_- \ri)} F \left(\bar q'_+  , \bar q'_-   ; 1-i\om' ; 1- w^2 \right) \\
	&+ (1-w^2)^{i\om'}{\Ga(\De) \Ga(-i\om') \ov \Ga\le(\bar q'_+ \ri)\Ga\le(\bar q'_- \ri)} F \left( q'_+  , q'_-   ; 1+i\om' ; 1- w^2 \right) \ .
\end{aligned}
\ee
Then one of the terms can be evaluated by closing the contour in the lower half plane and the other in the upper half plane.

\subsection{Near horizon transformation} \label{sec:horsy}

We now examine~\eqref{tena2} near the horizon ($V \to 0$) for $s> 0$, which we will show to be
\be \label{tranJ}
\Phi (X; s) = \bca \phi_R (X_s)  & s < s_0 \equiv -U \cr
\phi_F (X_s) & s > s_0  
\eca, \quad X_s = (U+s, V, \vx) \ ,
\ee
with $\phi_F (X)$ the expression in the $F$ region described by~\eqref{NearcontBTZvR}--\eqref{phaseShiftApp}. $s_0$ is the critical value for crossing the future event horizon.  Thus the action of $U(s)$ reveals an emergent translational symmetry near the horizon. 
The discussion applies to a general black hole without knowledge of the details of the phase shift $e^{i \de_k}$. 

More explicitly, for $s > 0$, we have 
\bega \label{efc2}
 v^{(RR)}_k (X; s) 
={h_{q}(\vx) \ov \sinh \pi \om} \int  {d \om' \ov 2\pi}  {\sinh \pi \om'  \ov \sqrt{2 |\om'|}} \le( e^{i \de_{k'}} V^{- i \om'} + e^{-i \de_{k'}} (-U)^{i \om'} \ri)    C_{k' k} (-s) , \\
 v^{(RL)}_k (X; s) 
={h_{q}(\vx) \ov \sinh \pi \om} \int  {d \om' \ov 2\pi}  {\sinh \pi (\om'+\om) \ov \sqrt{2 |\om'|}} \le( e^{i \de_{k'}} V^{- i \om'} + e^{-i \de_{k'}} (-U)^{i \om'} \ri)     C_{k' -k} (-s)  \ .
\end{gather} 
From $C_{k'k}$ given in \eqref{abv1}, 
\bega
 v^{(RR)}_k (X; s)  ={\ep(\om) h_{q}(\vx)  e^{-i \ga_{k}} \ov \sqrt{2 \pi \sinh\pi|\om|} } (A_1 + A_2) , \quad
 v^{(RL)}_k (X; s)  ={\ep(\om) h_{-q}(\vx)  e^{i \ga_{k}} \ov \sqrt{2 \pi \sinh |\om|} } (B_1 + B_2) 
\\
  A_1 =  s^{i \om} \int  {d \om' \ov 2\pi}  \sinh \pi \om'  
  e^{2i \ga_{k'}}
  \Ga (i (\om'+ i \ep))   \Ga (-i (\om -\om' + i \ep)) 
e^{-i  \om' \log ( s V) } ,  \\
A_2 = s^{i \om} \int  {d \om' \ov 2\pi}  \sinh \pi \om'  
 \Ga (-i (\om'+ i \ep))  \Ga (-i (\om -\om' + i \ep)) 
e^{i  \om' \log (-U/s)  } , \\
   B_1 =  s^{-i \om} \int  {d \om' \ov 2\pi}  \sinh \pi (\om'  +\om)
  e^{2i\ga_{k'}}
  \Ga (i (\om'+ i \ep))   \Ga (i (\om +\om' - i \ep)) 
e^{-i  \om' \log ( s V) }  \\
B_2 = s^{-i \om} \int  {d \om' \ov 2\pi}  \sinh \pi (\om'  +\om) 
 \Ga (-i (\om'+ i \ep))  \Ga (i (\om +\om' - i \ep)) 
e^{i  \om' \log (-U/s)  } \ .
\end{gather} 

The evaluation of $A_1$ and $B_1$ is similar to~\eqref{fvef}: with $V \to 0$ we can always close the integration contour in the upper half  $\om '$-plane to find 
\be 
A_1 =  e^{2 i \ga_k} V^{-i \om} \Ga (i \om) \sinh \pi \om   +\cdots , \quad B_1 = 0 + \cdots 
\ee
where $\cdots$ denotes contributions that vanish as $V \to 0$. For $A_2, B_2$, we can close the contour in the upper (lower) half plane for $s < s_0 \equiv -U$ ($s > s_0$), leading to 
\bega 
A_2 = \bca   \sinh \pi \om \Ga (-i \om) (-U-s)^{i \om}  & s < - U \cr
 0 & s > - U
\eca , \\
B_2 =\bca 0  & s < -U \cr
\sinh \pi \om \Ga (i \om) (U + s)^{-i \om}  & s > -U 
\eca \ , 
\end{gather} 
where for $A_2$ ($B_2$) there is no pole in the lower (upper) half plane for $s > -U$ ($s < -U$). 
Putting these results all together we find
\bega 
 v^{(RR)}_k (X;s) =\bca  {h_{q}(\vx) \ov \sqrt{2 |\om|}} \le( e^{i \de_k} V^{- i \om} + e^{-i \de_k} (-U-s)^{i \om} \ri) 
& s < - U \cr
{h_{q}(\vx) \ov \sqrt{2 |\om|}} e^{i \de_k} V^{- i \om} & s > - U 
\eca , \\
 v^{(RL)}_k (X;s) =\bca 0
& s < - U \cr
 {h_{-q}(\vx) \ov \sqrt{2 |\om|}}  e^{i \de_k} (U+s)^{-i \om}  & s > - U 
\eca   \ .
\end{gather}
which indeed give~\eqref{tranJ} by comparing with~\eqref{NearcontBTZvR}--\eqref{phaseShiftApp}.

Since the above discussion fixes $s$ and assumes $V \to 0$, the valid range of $s$ is for 
$|s| < s_2 = {1 \ov V}$, which is infinite in this limit.

\subsection{Left-right commutators and causal structure}

For the action of $U(s)$ on a bulk field at a generic point in the $R$ region,  
the details of the phase shift $e^{i \de_k}$ will matter and from now on we will specialize to the BTZ black hole with the corresponding expressions given by~\eqref{tena2}--\eqref{Adef}. The transformation is complicated and is nonlocal. 
Below~\eqref{Adef} we already commented that $v^{(RL)}_k(X; s) \neq 0$ only for $s > s_0$, which can be interpreted as the boundary emergence of the event horizon. This is further confirmed by the near-horizon analysis of the last subsection. Here we show that despite the transformation being nonlocal it respects  sharp causal structure. 
 
Consider the commutator of an evolved bulk field from the right exterior and a bulk field operator at some fixed location in the left exterior (i.e. set $X_1 \in R, X_2, \in L$ and $U_s = U(s)$)
\begin{equation}
	C(s) =  \left[\Phi(X_1;s), \phi_L(X_2) \right] = \left[U_{s}^\da \phi_R(X_1) U_{s}, \phi_L(X_2) \ri]
	  \ . 
\label{cdef}
\end{equation}
Now multiply both sides of the above equation by $U_v^\da$ and $U_{v}$. Since we are working at the level of free field, $C(s)$  
is a c-number and thus is unchanged by this conjugation. Using unitarity and the group property of $U_s$, we find 
\ie
C (s) 
&= [U_{s+v}^\da \phi_R (X_1) U_{s+v} , U_{v}^\da \phi_L (X_2) U_{v}]  \ .
\fe

\begin{figure}[h]
\begin{centering}
	\includegraphics[width=2.1in]{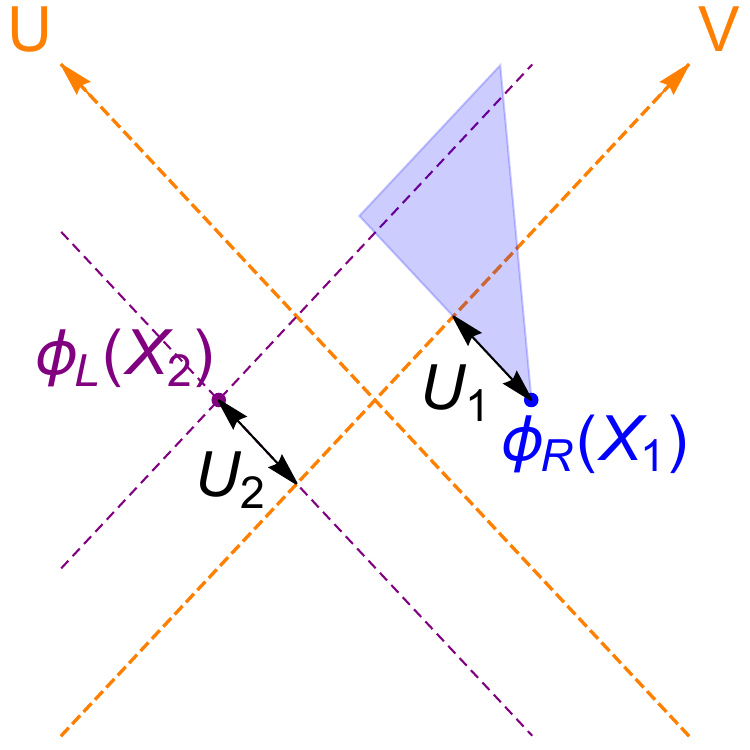} \qquad
	\includegraphics[width=2.1in]{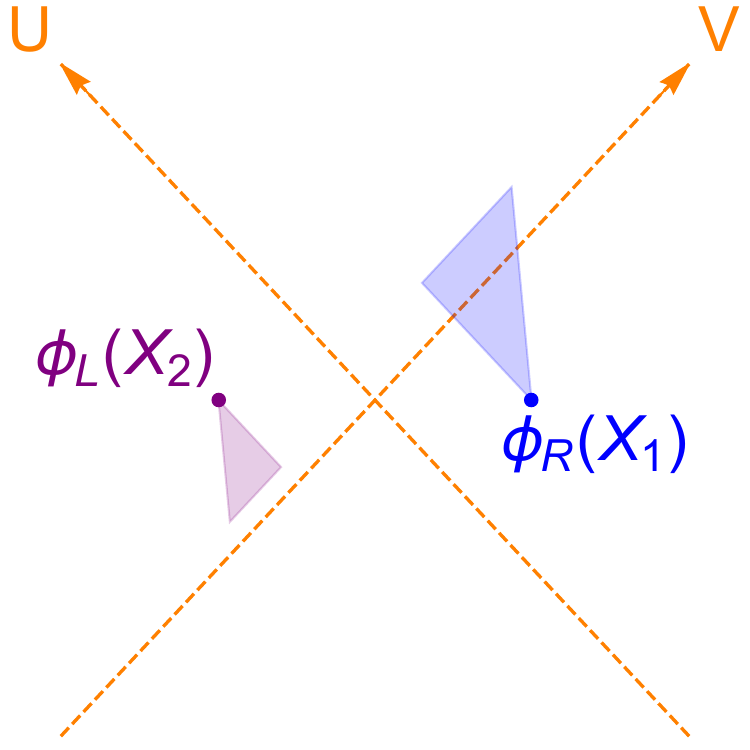}
\par\end{centering}
\caption{
Left: When a bulk field $\phi_R (X_1)$ with $X_1 \in R$ is transported by a null Kruskal coordinate distance $-U_1 + U_2$ (since $U_1 < 0$), it enters the lightcone of $\phi_L (X_2)$. The shaded region is a cartoon for the spread of $\Phi (X_1; s)$.  The orange dashed lines are event horizons, and the purple dashed lines give the light cones of $X_2$. Right: The commutator between the evolved right operator and fixed left bulk field is equal to the commutator of evolved left and right fields with the same difference of evolution parameters. We use this to evolve the left operator almost all the way to the past horizon. The commutator can only be non-zero then if the evolved right operator remains supported on left oscillators after now also applying the evolution that brought the left field to the horizon. The blue shaded region is a cartoon for the spread of $\Phi (X_1; s+v)$ and the purple shaded region is a cartoon for the spread of $\Phi (X_2; v)$. The boundaries and singularities suppressed in each figure. 
}
\label{fig:commLightcone}
\end{figure}

Recall that $\Phi(X;s)$ has support on left operators only for $s > s_0(X) = - U(X)$. From $J$ conjugation 
 $U_{v}^\da \phi_L (X_2) U_{v}$ should take $\phi_L$ closer to the past horizon for $v < 0$, but it will not have any 
 dependence on the right operators for $v > -s_0(X_2) = - U(X_2)$, see Fig.~\ref{fig:commLightcone}. Now take $v = - s_0 (X_2) + \ep$ where $\ep > 0$ is an infinitesimal number. With such a $v$, $U_{v}^\da \phi_L (X_2) U_{v}$ still lies in the $L$ region. To have a nonzero commutator we need $\Phi(X_1;s+v)$ to have support on the left, which requires 
\be 
s+ v > s_0 (X_1) \quad \to \quad s > s_0 (X_1) + s_0 (X_2) = - U(X_1) + U(X_2) \ ,
\ee
Since $\Phi (X_1;s)$ must enter the lightcone of $\phi_L(X_2)$ in order for the commutator to become non-zero, the above equation implies that the support of $\Phi (X_1;s)$ must lie in the region $U \leq U_1 +s$, see Fig.~\ref{fig:commLightcone}.

\subsection{Transformation of a boundary operator} \label{sec:bdtr}

We now consider the evolution under $U(s)$  of a boundary operator $\sO_R (x)$, 
\bega 
\sO_R (x; s) = U(s)^\da \sO_R (x) U(s) =  \sum_{k'}   u_{k'}^{(R\b)} (x; s) a_{k'}^{(\b)} , \\
\sO_R (x) = \sum_k  u_k^{(R)} (x) a_k^{(R)}, \quad u^{(R)}_k(x) = N_k e^{-i \om \eta + i q \chi} , \quad   u_{k'}^{(R\b)} (x; s)  = \sum_k u_k^{(R)} (x) \Lam_{k k'}^{R \b}  (s)  \ .
\end{gather} 
$ u_{k'}^{(R\b)} (x; s)$ can be obtained from~\eqref{rFromJ}--\eqref{lFromJ} by taking $w \to 0$ and 
and stripping off the factor of $w^{\Delta}$. $J_{1,2,3}$ then simplify to 
\bega\label{jjj1}
J_1 =(2c)^{-i \om} \int \frac{d\omega'}{2\pi} \Ga\le(\bar q_+ - {i \om' \ov 2} \ri) \Ga\le(\bar q_- - {i \om' \ov 2} \ri)  \Gamma(i \omega' + \ep) (2c)^{-i\omega'}
\\
\label{jjj2}
J_2 
=(2c)^{-i \om} \int \frac{d\omega'}{2\pi} \sinh \pi (\om' +\om) \Ga\le(\bar q_+ - {i \om' \ov 2} \ri) \Ga\le(\bar q_- - {i \om' \ov 2} \ri)  \Gamma(i \omega' + \ep) (2c)^{-i\omega'}
\\
\label{jjj3}
J_3 
=(2c)^{i \om} \int \frac{d\omega'}{2\pi} \sinh \pi \om' \Ga\le(q_- - {i \om' \ov 2} \ri) \Ga\le(q_+ - {i \om' \ov 2} \ri)  \Gamma(i \omega' + \ep) (2c)^{-i\omega'} \\
a(w=0) = 2 c, \quad c \equiv  |s| e^{\eta} \ .
\end{gather} 
From~\eqref{opw1}--\eqref{opw2} we now have $s_0 = s_2= e^{-\eta}$ as on the boundary $UV =-1$. For $c < 1$ ($c > 1$) we can close the contours of the above integrals in the  upper (lower) half plane.  

Consider first $J_1$ which is relevant for $s < 0$. For any value of $c$, we find that $J_1$ can be expressed as another hypergeometric function, and (see Appendix~\ref{app:der1} for a derivation)
\bega \label{enn0}
 u_k^{(RR)} (x; s) = N_k e^{-i \om \eta + iq\chi}  J_k(s) = u_k^{(R)} (x)J_k(s)  \\
J_k(s) = {\Ga (\bar q_++\ha) \Ga (\bar q_- + \ha) \ov  \sqrt{\pi} \Ga (\De - i \om +\ha)} F \le(2 \bar q_+,2 \bar q_-; \De-i \om +\ha; {1 + s e^{\eta} \ov 2} \ri)  \ .
 \label{enn}
\end{gather} 
Clearly the transformation is not point-wise. For $s > 0$, the evolution is described by $J_2, J_3$. 
For $c < 1$ (i.e. $s < s_0$) we have $J_3 =0$ and $J_2$ is such that we again recover~\eqref{enn0}--\eqref{enn}. 

The argument of the hypergeometric function in~\eqref{enn} becomes $1$ for $s = s_0 = e^{-\eta}$ ($c=1$). 
The behavior of hypergeometric function $F (a,b;c; z)$ at $z=1$ depends on ${\rm Re} (c-a-b)$: it is divergent
for ${\rm Re} (c-a-b) < 0$.\footnote{This can be seen by the transformation $ F(a,b;c; z) = (1-z)^{c-a-b} F(c-a, c-b; c; z) $.}
 Now  ${\rm Re} (\De-i \om +\ha - 2 \bar q_+ - 2 \bar q_-) =- \De + \ha  $, thus~\eqref{enn0} becomes singular for any operator with $\De > \ha$. In other words, $\sO_R(x;s)$ becomes singular for 
$s = s_0$. Recall that $s_0$  is precisely the Kruskal $U$ distance between initial point $\eta$ and $\eta = +\infty$. 

To understand the action~\eqref{enn0}--\eqref{enn} of $U(s)$ on a boundary operator $\sO_R (x)$ a bit further, now consider its support in the position space. For this purpose it is convenient to introduce an evolution function $G(x, x'; s)$ defined by 
\be \label{vsh1}
\sO_R (x; s) =  \int d^2 x' \, G (x, x'; s) \sO_R (x') 
\ee
where 
\be \label{GsInt}
G (x,x'; s) =  \sum_k J_k (s) e^{- i \om (\eta-\eta') + i q (\chi - \chi')}  \ .
\ee
To understand the support in $\eta$, consider the $\om$-integral in~\eqref{GsInt}, 
\be \label{Gkl}
G_q (\eta, \eta'; s) \equiv \int {d \om \ov 2 \pi} \, e^{- i \om (\eta-\eta')} J_k (s) \ .
\ee 
Notice that $J_k (s)$ has no pole in the upper half $\om$-plane, and  has the 
 asymptotic behavior 
 \be \label{ehnk}
 J_k(s) 
= \le(1- s e^{\eta}  \ri)^{\ha - \De + i \om}  \le(1 + O(|\om|^{-1}) \ri) , \quad |\om| \to \infty \ .
 \ee
 Given that $J_k (s) \propto O(1)$ for real $\om \to \pm \infty$,~\eqref{Gkl} has to be treated with a bit of care. 
 By adding and subtracting $ \le(1- s e^{\eta}  \ri)^{\ha - \De + i \om}  $ in the integrand we can rewrite it as 
 \be\label{Gkl1} 
 G_q (\eta, \eta'; s) \equiv \le(1- s e^{\eta}  \ri)^{\ha - \De} \de (\lam) + \tilde G_q , \quad \tilde G_q = 
 \int {d \om \ov 2 \pi} \, e^{- i \om \lam} \tilde J_k (s)
 \ee
 where 
 \be 
 \lam \equiv \eta-\eta' -  \log \le(1- s e^{\eta}  \ri) , \quad  \tilde J_k (s) \equiv J_k(s) \le(1- s e^{\eta}  \ri)^{-i \om}   - \le(1- s e^{\eta}  \ri)^{\ha - \De}  \ . 
  \ee
Now $\tilde J_k (s) \to 0$ along the real axis as $\om \to \pm \infty$ and it only has poles in the lower half $\om$-plane. 
 
 We can close the contour for $\om$-integration in $\tilde G_q$ of~\eqref{Gkl1} in the upper half $\om$-plane if 
\be 
\eta - \eta' - \log (1-  s e^{\eta}) < 0 
\ee
for which the integral gives zero. We conclude that $G_q(\eta, \eta';s)$ only has support for 
\be \label{eyn}
\eta' \leq  \eta - \log (1-  s e^{\eta})  \quad \to \quad U' \leq   U +s, \quad U' = -e^{-\eta'}, \quad U = - e^{-\eta} 
\ee
where $U', U$ are respectively the boundary Kruskal coordinates for $\eta'$ and $\eta$. 
We note that equation~\eqref{eyn} agrees precisely with~\eqref{bdrytshift} with the identification of $\eta = 2 \pi t$, providing a nontrivial further consistency check of our identification of the bulk region $\tilde \sX_{0}$ as the bulk dual of $\sN = \sX_0$.

As $s \to s_0 =e^{-\eta}$, the support of $G_q$ and thus $G (x, x';s)$ covers the full $\eta'$ axis and 
$G (x, x';s)$ is singular at $s =s_0$ as a result of the singular behavior of $J_k (s)$. 
 This indicates that we cannot extend the action of $U(s)$ beyond $s_0$.\footnote{Integrals for $J_2$ and $J_3$ appear to be well defined 
for $s > s_0$ (i.e. $c > 1$).  $J_3$ is now nonzero, i.e.  $\sO_R (x;s)$  now involves also $a_k^{(L)}$. 
 But due to the singular behavior as $s \to s_0$ from below,  {$\sO_R(x;s)$} for $s > s_0$ may not be meaningful.} 

Plugging~\eqref{Gkl1} into~\eqref{vsh1} we find that 
\bega \label{henl}
\sO_R (x;s) = \le(1- s e^{\eta}  \ri)^{\ha - \De}  \sO_R (x_s) + \tilde \sO_R (x;s) \\
\tilde \sO_R (x;s)  =\int d^2 x' \, \tilde G (x,x';s)  \sO_R(x') , \quad  \tilde G (x,x';s) =  \sum_q e^{i q (\chi-\chi')} \tilde G_q (\eta, \eta';s)
 \label{henl1}
\end{gather} 
where the first term is a point-like transformation with 
\be 
x_s = (\eta_s, \chi), \quad \eta_s = \eta - \log (1-s e^{\eta}) \;\; \text{or} \; \; U_s = U + s\ .  
\ee
From~\eqref{Gkl1}, $\tilde G_q (\eta, \eta';s) $ can be written in a form 
\be \label{ojw}
\tilde G_q (\eta, \eta';s) =- i \sum_\pm \sum_{n=0}^\infty e^{-\om_n^{(\pm)} (q) (\eta_s - \eta')} c_n^{(\pm)} (q) , \quad
\om_n^{(\pm)} (q) = (2n+1+\De) \pm i q 
\ee
where $-i \om_n^{(\pm)}$ are the poles of $\tilde J_k (s)$ in the lower half $\om$-plane and $c_n^{(\pm)}$ are the corresponding residues. Thus $\tilde G_q$ only has a small exponential tail away from $\eta_s$. 
We are not able to evaluate the $q$-sum in~\eqref{henl1} explicitly, but from~\eqref{ojw} we expect that 
\be \label{yenl}
G (x,x';s) \sim e^{- (\De+1) (\eta_s -\eta)} + \cdots 
\ee
where $\cdots$ denotes higher exponential suppressions in $\eta_s - \eta$. 
We thus find that the support of $G(x,x';s)$ is localized around $\eta_s$ with a small exponential tail away from it.

\subsection{Summary}\label{sec:wfe}

For a general bulk point $X_0 = (\eta_0, w_0, \chi_0) = (U_0, V_0, \chi_0)$  the transformation~\eqref{rFromJ}--\eqref{Adef}. 
is not point-wise and rather complicated. As outlined there we can evaluate the integrals $J_{1,2,3}$ for any $s$ using residues which results in an 
infinite sums of hypergeometric functions (some of which can be summed to Appell functions). The analyses of these infinite sums (or Appell functions) appear intricate and will be not be treated here. From last few subsections we have found that: 

\ben 

\item  For $X_0$ close to the past horizon $V_0 \to 0$, we have $\Phi (X_0; s) = \phi (X_s)$ where 
$X_s = (U_0+s, V_0, \chi_0)$ is obtained by $X_0$ by a null shift. We can view this as an indication of an emergent horizon symmetry. 

\item While for a general point the action of $U(s)$ is nonlocal, the support of $\Phi (X_0;s)$ respects the sharp causal structure implied by 
the event horizon: (i)  There exists a critical value $s_0 = - U_0$ after which $\Phi (X_0;s)$ develops dependence on $a_k^{(L)}$, which signals crossing the horizon; (ii) 
It starts having nontrivial commutators with $\phi_L (X_1)$ for $s > - U_0 + U(X_1)$. Both imply that the support of $\Phi (X_0;s)$ lies in the region $U \leq U_0 + s$, consistent with the proposal of the bulk dual of Fig.~\ref{fig:propEW}. 

\item For a boundary operator $\sO_R (x_0)$ (i.e. $w_0 \to 0$ limit of a bulk field) with $x_0 = (\eta_0, \chi_0)$, the evolved operator remains on the boundary, and we can show explicitly that the support of $\sO_R (x_0; s)$ lies in the region $U \leq U_0 +s$ where now $U = -e^{-\eta}$ and $U_0 = - e^{-\eta_0}$. In particular, $\sO_R (x_0; s)$ contains a local piece proportional to $\sO_R (x_s)$ with $x_s = (U_0+s, \chi_0)$ and a non-local piece which is still mostly supported near the time slice $\eta_s = \eta_0 - \log (1-s e^{\eta_0})$.  
The action of $U(s)$ becomes singular at $s=s_0 = - U_0$. 

\een


In next section we will show that the transformation of a bulk field becomes much simpler in the large $\De$ limit, and in fact becomes a point-wise transformation.


\section{A point-wise transformation in the large mass limit} \label{sec:largeMass}

In this section we consider the evolution of bulk fields under $U(s)$ in the large mass limit (or large dimension $\De$ limit). Interestingly we find that in this limit the evolution becomes point-wise when we average over the spatial manifold of the boundary theory. 


\subsection{General setup and summary of results}

Evolution of a bulk field, initially at a point $X_0 = (\eta_0, w_0, \chi_0) = (U_0, V_0, \chi_0) \in R$, is given by~\eqref{tena2}--\eqref{tena1} which we copy here for convenience 
 \bega \label{tena4}
  \Phi(X_0; s) = U^{\da} (s) \phi_R (X_0) U(s) = \sum_{k'}  v_{k'}^{(R\b)} (X_0; s) a_{k'}^{(\b)},  \\
   v_{k'}^{(RR)} (X_0; s)  = \sum_k v_k^{(R)} (X_0) \Lam_{k k'}^{R R}  (s)   , \quad
 v_{k'}^{(RL)} (X_0; s)  = \sum_k v_k^{(R)} (X_0) \Lam_{k k'}^{R L}  (s)   \ .
  \label{tena3}
\end{gather} 
Recall that the mass $m$ of $\phi$ is related to the dimension $\De$ of the corresponding boundary operator $\sO$ by 
\begin{equation}
	\Delta = \frac{d}{2} + \nu, \qquad \nu = \sqrt{\frac{d^2}{4} + m^2 }  \ .
\end{equation}
We will consider the large $\nu$ limit and expand various quantities in~\eqref{tena4}--\eqref{tena3} in $1/\nu$.\footnote{$\nu$ is always $O(N^0)$. Equivalently we can expand in ${1 \ov m}$ or $1/\De$. At leading order, i.e. $O(\nu)$ all these expansions agree. But for higher orders, including the calculation of $O(\nu^0)$ prefactors, expanding in $1/\nu$ is the most natural and convenient.}  To define the limit, we will also scale frequency and spatial momenta as~\cite{Festuccia:2005pi} 
\begin{equation} \label{lnul} 
	\omega = \nu u, \quad q = \nu p , \quad {\mathrm k} = (u, p) \; \; {\rm fixed}, \quad \nu \to \infty  \ . 
\end{equation}
In the limit~\eqref{lnul} the various quantities in~\eqref{tena3} have the form 
\bega \label{kj1}
v_k^{(R)} (X_0)  = 
\begin{cases}
	\sum_{\gamma = \pm} A^{(\gamma)}_{\rk} (X_0) e^{ i \nu Z^{(\gamma)}_\rk (X_0)} \le(1 + O(\nu^{-1}) \ri) \ , &|u| > u_w \equiv {\sqrt{(1-w^2_0)(1+p^2 w_0^2)} \ov w_0} \cr
	A^{(f)}_{\rk} (X_0) e^{- \nu Z^{(f)}_\rk (X_0)} \le(1 + O(\nu^{-1}) \ri) \ , &|u| < u_w
\end{cases} \\
\Lam_{k k'}^{R \al} (s)   = \de_{q q'} B_{\rk \rk'}^{R\al} (s) e^{i \nu W^{R\al}_{\rk \rk'} (s)} \le(1 + O(\nu^{-1}) \ri) ,
\label{kj2}
\end{gather} 
where the first (second) line of~\eqref{kj1} is the wave function in the classically allowed (forbidden) region. Explicit expressions for these quantities are given in~\eqref{WKBPhaseFQ}--\eqref{LambdaPrefactorsFQ}.
We then find, 
\be \label{eax}
  v_{k'}^{(RR)} (X_0; s)  = \nu \le[\int_{|u| > u_w} {d u  \ov 2 \pi} \sum_{\gamma = \pm} A^{(\gamma)}_{\rk}  B_{\rk \rk'}^{RR} (s) e^{i \nu G^{(\gamma)}_R} + \int_{-u_w}^{u_w} {d u  \ov 2 \pi} A^{(f)}_{\rk} B_{\rk \rk'}^{RR} (s) e^{-\nu G^{(f)}_R} \ri],
\ee
where 
\be 
G^{(\gamma)}_R = Z^{(\gamma)}_\rk (X_0) + W^{RR}_{\rk \rk'} (s) ,  \quad
G^{(f)}_R = Z^{(\gamma)}_\rk (X_0) - i  W^{RR}_{\rk \rk'} (s)  \ .
\ee
A similar expression applies for $ v_{k'}^{(RL)} (X_0; s)$. Equation~\eqref{eax} can be evaluated using the saddle point (steepest descent) approximation.

Since we are mainly interested in how $\Phi (X_0;s)$ evolves with $s$ in the $(w, \eta)$ (or $(U,V)$)  plane, it is convenient to average it over the boundary spatial direction $\chi$, i.e. restricting to $q=0$ in all equations. 
In this case we find that the transformation is  point-wise 
\be \label{pwXfmLargeNu}
\Phi (X_0;s) =\lam_X  \phi (X_s), \quad \lam_X = \sqrt{1- s e^{\eta_0} \sqrt{1-w_0^2} }
\ee
with $X_s$ given in terms of Kruskal coordinates as
\be \label{heb}
U_s = U_0+ s, \qquad V_s 
= {V_0 \ov 1 - s V_0}  \ . 
\ee

Here are some remarks on the transformation~\eqref{pwXfmLargeNu}--\eqref{heb}: 

\ben 

\item At the horizon, $V_0=0,$ we have $V_s=0$ and $U_s = U_0+s$. 

\item At the boundary we have $V_0 = -{1 \ov U_0}$ giving $V_s = -{1 \ov U_0+s} = - {1 \ov U_s} $, so a point initially on the boundary remains on the boundary. 

\item Given that a boundary operator can be found by $\sO_R (x) = \lim_{w \to 0} w^{-\De} \phi_R (X)$, we have from~\eqref{pwXfmLargeNu}--\eqref{zprim} 
\be 
\sO_R (x_0;s) = (1- s e^{\eta_0})^{\ha - \De} \sO_R (x_s) 
\ee
precisely giving the first term of~\eqref{henl} including the prefactor.  Thus the second term of~\eqref{henl} must be suppressed in the large $\nu$ limit, which is consistent with the expectation~\eqref{yenl}. 

\item For a generic initial point with $-1 < U_0 V_0 < 0$, the horizon is reached for $s= s_0 = - U_0 = \sqrt{1-w_0 \ov 1+w_0} ~e^{-\eta_0}$. 

\item For $s < 0$, the boundary is reached in the limit $s \rightarrow -\infty$. i.e.
\be 
U_s V_s = {U_0 V_0 +s V_0 \ov 1-s V_0} \to -1 , \quad s \to -\infty \ . 
\ee

\item Notice that the prefactor $\lam_X$ in~\eqref{pwXfmLargeNu} becomes zero for 
\be
s= s_1 \equiv {1 \ov \sqrt{1-w_0^2}} ~e^{-\eta_0}  
\ee
at which value we have $U_{s_1} V_{s_1} =1$, i.e. the location of the black hole singularity. 
For $s > s_1$, $\phi (X_s)$ is no longer defined, but the left hand side of~\eqref{pwXfmLargeNu} appears still to be well defined. 
Note that ${s_1 \ov s_2} = {1 \ov 1+w_0}  \leq 1$, where $s_2$  was introduced in~\eqref{opw2}.

\item At $s = s_2 = {1 \ov V_0}$ we have $V_s \to \infty$. 


\item  Equation~\eqref{heb} does not appear to correspond to any geodesic motion.

\item Equation~\eqref{heb} also applies if the initial point $X_0$ lies in the $L$ region. 

\begin{figure}[h]
\begin{centering}
	\includegraphics[width=2.5in]{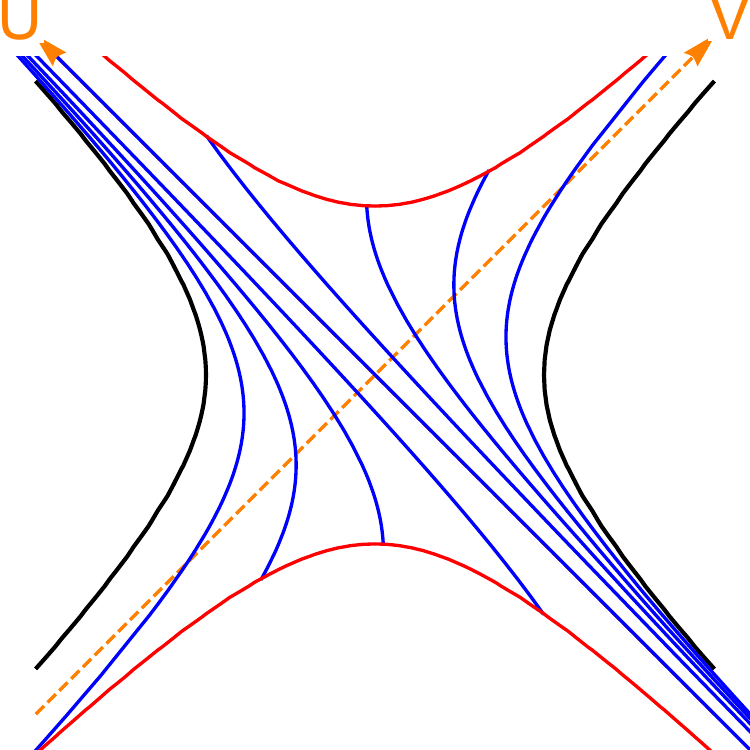} \qquad \includegraphics[width=2.5in]{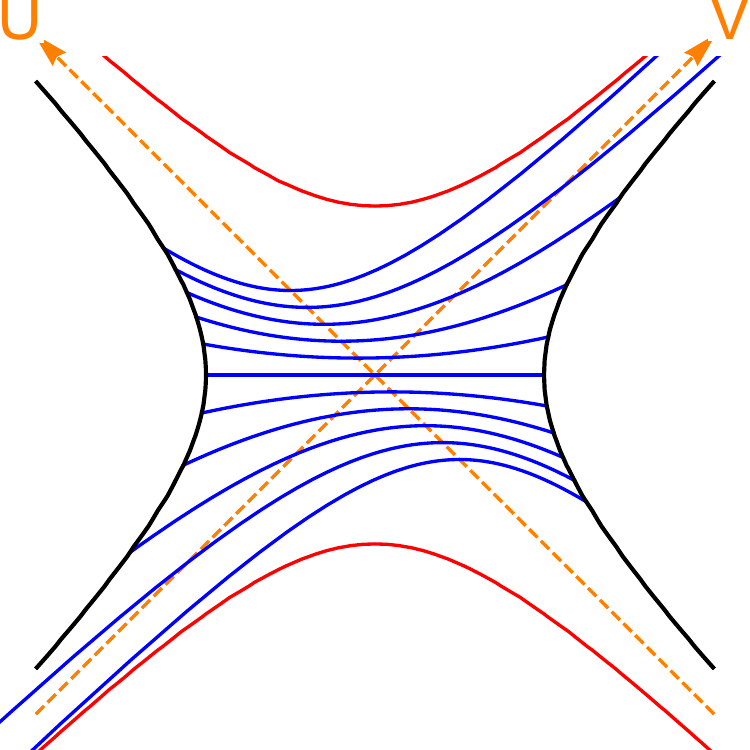}
\par\end{centering}
\caption{The left plot gives trajectories of~\eqref{heb}. The right plot gives constant $s$ surfaces evolved from the $\eta =0$ slice.  
The orange dashed lines are the event horizons, black solid lines are the boundaries, while the red solid lines are the singularities.   }
\label{fig:uEvolvedObserver}
\end{figure}

\een
The trajectories following from~\eqref{heb} are shown in Fig.~\ref{fig:uEvolvedObserver}.

\begin{figure}[h]
\begin{centering}
	\includegraphics[width=2.5in]{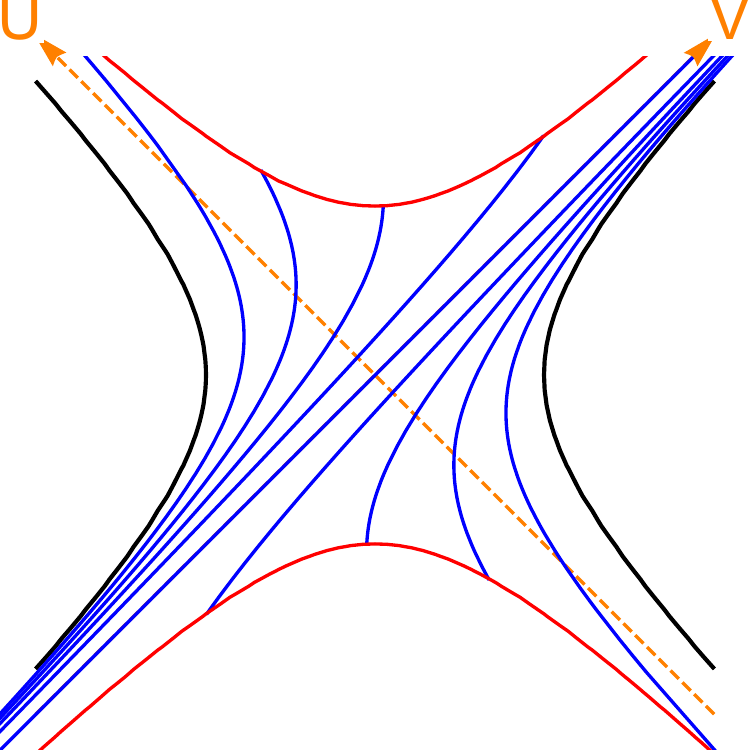} \qquad \includegraphics[width=2.5in]{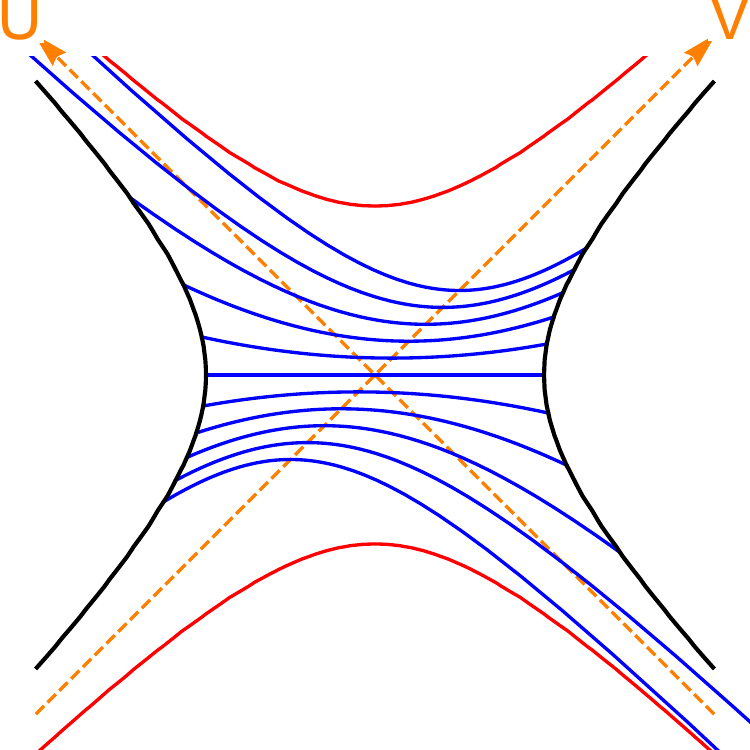}
\par\end{centering}
\caption{The counterparts of Fig.~\ref{fig:uEvolvedObserver} when using $\sN$ as in the right plot of Fig.~\ref{fig:tShiftBdryBH}.
}
\label{fig:vEvolvedObserver}
\end{figure}

By choosing $\sN$ to be the algebra associated with region in the right plot of Fig.~\ref{fig:tShiftBdryBH}, we can similarly construct 
unitary evolutions as above but with {the roles of Kruskal $U$ and $V$ swapped}. See Fig.~\ref{fig:vEvolvedObserver} for the corresponding flow trajectories.

We will now describe the calculation of~\eqref{eax} in detail.

\subsection{Saddle-point equations}

It is useful to first notice that the Hermitian conjugation property of our transformation, $\Lambda^{R\alpha}_{(-k)(-k')}(s) = \le(\Lambda^{R\alpha}_{kk'}(s)\ri)^*$, and $v^{(R)}_{-k}(X_0) = \le(v^{(R)}_{k}(X_0)\ri)^*$ imply 
\be \label{uPrimePos}
	 v^{(R\alpha)}_{-k'}(X_0; s) = \sum_k v^{(R)}_k(X_0) \Lambda^{R\alpha}_{k(-k')}(s) = \sum_k v^{(R)}_{-k}(X_0) \Lambda^{R\alpha}_{(-k)(-k')}(s) = \le(  v^{(R\alpha)}_{k'}(X_0; s) \ri)^* \ .
\ee
Thus \eqref{uPrimePos} implies that the results for $u'<0$ can be immediately obtained from those with $u'>0$, so we restrict to $u' > 0$ for the calculations of this section.
From the expressions of Appendix~\ref{app:WKB} we find $G^{(\pm)}_R $ in~\eqref{eax} can be written as 
\be \label{GR}
\begin{aligned}
	G^{(\pm)}_R &= -u\log |c| +{|u| \ov 2}\log(1-w_0^2) - (u' - u)\log(i(u - u')) -{i\pi \ov 2}\ep(s)|u| -i\log(1-iu) \\
	&- |u|\log\le(|u| \pm w_0 \sqrt{u^2 - u_w^2}\ri) + i \log\le(1 \mp i w_0 \sqrt{u^2 - u_w^2}\ri) -\theta(-u)u\log(1+u^2) \\
	&+ u'\log|s| -{i \ov 2}\log\le({1+iu' \ov 1-iu'}\ri) + {u' \ov 2}\log(1+u'{}^2) + {i\pi \ov 2}\ep(s)|u'| -i\log w_0 \ .
\end{aligned}
\ee


It will turn out that $G^{(f)}_R$ does not lead to any saddle point {and the contribution from the second term in~\eqref{eax} 
is always suppressed compared with that from the first term.} We will thus not give the explicit expression of $G^{(f)}_R$ here, leaving it to Appendix~\ref{app:WKB}. For our discussion below it is convenient to introduce 
\be
\label{bcde}
\rb \equiv {s e^{\eta_0} \ov \sqrt{1-w^2_0}}, \qquad c \equiv se^{\eta_0} \ .
\ee


The saddle point equation of~\eqref{GR} can be written as 
\be \label{dGR}
\begin{aligned}
 -\log|\rb| +\log(i(u-u')) - \log\le(|u|\pm\ep(u)w_0 \sqrt{u^2-u_w^2} \ri) - {i\pi \ov 2}\ep(s)\ep(u) = 0 \ .
 \end{aligned}
\ee
which leads to 
\be \label{sqrtsDGR}
	(1-\rb)u - u' = \pm \rb w_0 \sqrt{u^2-u_w^2} , 
\ee 
whose solutions are given by
\be \label{ehn}
u^{(\pm)}_c = { (1- \rb) u' \pm \rb w_0 \sqrt{u'{}^2 - u_{w(s)}^2 } \ov  1- 2 \rb + c^2} , \quad u_{w(s)}^2 =  u_w^2 (1- 2 \rb + c^2)  \ .
\ee
The solution is real for $u' > u_{w(s)}$. We can also check that both roots~\eqref{ehn} satisfy the requirement $u'> u^{(\pm)}_c > u_w$ for $s < 0$ and $u^{(\pm)}_c > u,~u^{(\pm)}_c > u_w$ for $s > 0$.
For $s<s_0$ the root $u^{(+)}_c$ above satisfies $\ep((1- b) u^{(+)}_c - u') = \ep(s)$, so it is only a proper solution for~\eqref{sqrtsDGR} with the plus sign on the right side. Thus this is a saddle solution for $G^{(+)}_c$.  Similarly, $u^{(-)}_c$  is  a saddle for $G^{(-)}_c$. For these real roots we have the following behaviour as functions of $s$. We have $u_c^{(\pm)} > 0$ for all $s < s_0$. As $s \to s_0^-$, $u_c^{(+)} \to \infty$ while $u_c^{(-)}$ is finite and we have $u^{(+)}_c < 0,~u^{(-)}_c > 0$ for $s_0 < s < s_2$. Finally, as $s \to s_2^-$, $u_c^{(-)} \to \infty$ while $u_c^{(+)}$ is finite and for $s > s_2$ we have $u^{(\pm)}_c < 0$.

For $s < s_0$ and $s > s_2$, $u_{w(s)}^2 > 0$, so the solutions~\eqref{ehn} are not real for all real $u'$. In particular, for $u'{}^2 < u_{w(s)}^2$ the solutions are complex.\footnote{Note that $u_{w(s)}$ is exactly the critical frequency separating classically allowed and classically forbidden frequencies $u'$ when $w = w_s$ as in~\eqref{zprim}} For such small values of $u'$, the steepest descent contour should only pass through $u^{(-)}_c$, so there is only one contribution to the saddle point evaluation of the integral for $u'{}^2 < u_{w(s)}^2$. Note that for $s_0 < s < s_2$, $u_{w(s)}^2 < 0$, so both roots are always real, independent of the value of $u'$, in this region of $s$. 

We will now argue that we need only consider the roots $u^{(\pm)}_c$ when they are positive. Even when $u^{(\pm)}_c < 0$ is a genuine saddle point, its contribution will be subleading. The magnitude of the result of saddle point integration in~\eqref{eax}
 is controlled by $\Im G^{(\pm)}_R |_{\rk_c^{(\pm)}}$. For $s < 0$ we always have $u_c^{(\pm)} < u'$, so 
\be \label{sadNegS}
	\Im G^{(\pm)}_R |_{\rk_c^{(\pm)}} = {\pi \ov 2} \le(|u_c^{(\pm)}| - |u'| + |u_c^{(\pm)} - u'|\ri) = 0 , \quad s < 0 \ ,
\ee
while for $s > 0$ we either have $u_c^{(\pm)} > u' > 0$ or $u_c^{(\pm)} < 0$ giving
\be \label{excNegSad}
	\Im G^{(\pm)}_R |_{\rk_c^{(\pm)}} = {\pi \ov 2} \le(|u'| - |u_c^{(\pm)}| + |u_c^{(\pm)} - u'|\ri) = 
	\begin{cases}
		0 , &u_c^{(\pm)} > u' > 0 \\
		\pi u' , &u_c^{(\pm)} < 0 
	\end{cases}
	 \ .
\ee
We then see that the contributions of $u_c^{(+)}$ for $s > s_0$ and of $u_c^{(-)}$ for $s > s_2$ are suppressed by $\exp(-\pi \nu u')$, which are subdominant. 
 Thus, for $s < s_0$ we have leading order contributions from both $u_c^{(+)}$ and $u_c^{(-)}$, for $s_0 < s < s_2$ we have a leading order contribution from $u_c^{(-)}$ and for $s > s_2$ both contributions are exponentially small.

Now we turn to $ v^{(RL)}_k(X_0; s)$, where there is only a non-trivial calculation for $s > 0$.
\be \label{GL}
\begin{aligned}
	G^{(\pm)}_L &= -u\log |c| +{|u| \ov 2}\log(1-w_0^2) + (u' + u)\log(i(u + u')) -{i\pi \ov 2}(2|u+u'|-|u|-|u'|) -i\log(1-iu) \\
	&- |u|\log\le(|u| \pm w_0 \sqrt{u^2 - u_w^2}\ri) + i \log\le(1 \mp i w_0 \sqrt{u^2 - u_w^2}\ri) -\theta(-u)u\log(1+u^2) \\
	&- u'\log|s| +{i \ov 2}\log\le({1+iu' \ov 1-iu'}\ri) - {u' \ov 2}\log(1+u'{}^2) -i\log w_0 \ .
\end{aligned}
\ee
It can be shown that $G^{(\pm)}_L$ has no real saddle point with $\min\{-u_w, -u'\} < u < u_w$, and the saddle point equation 
outside this region can be written as 
\be \label{dGL}
\begin{aligned}
	-\log|\rb| +\log(i(u+u')) - \log\le(|u|\pm\ep(u)w_0 \sqrt{u^2-u_w^2} \ri) - {i\pi \ov 2}\ep(u) = 0 \ .
\end{aligned}
\ee
which leads to 
\be \label{sqrtsDGL}
	(1-\rb)u + u' = \pm \rb w_0 \sqrt{u^2-u_w^2} 
\ee 
with solutions 
\be \label{ehnL}
u^{(\pm)}_d = { -(1- \rb) u' \mp \rb w_0 \sqrt{u'{}^2 - u_{w(s)}^2 } \ov  1- 2 \rb + c^2} = -u^{(\pm)}_c , \quad u_{w(s)}^2 =  u_w^2 (1- 2 \rb + c^2)  \ . 
\ee
Recall that here  $\rb > 0$.
Again we have real saddle points for $u'{}^2 > u_{w(s)}^2$, whose behavior we now discuss. For  $s < s_0$, we have $u_d^{(\pm)} < 0$, while for $s_0 < s < s_2$ we have $u_d^{(+)} > 0$ and $u_d^{(-)} < 0$. Finally, for $s > s_2$ we have $u_d^{(\pm)} > 0$. From~\eqref{eaxL} we see that the magnitude of the contributions of the saddle points are controlled by $\Im G^{(\pm)}_L |_{\rk_d^{(\pm)}}$, so we will again find that the contributions from saddle points $u^{(\pm)}_d < 0$ are exponentially small. We have
\be \label{excNegSadL}
	\Im G^{(\pm)}_L |_{\rk_d^{(\pm)}} = 
	\begin{cases}
		0 , &u_d^{(\pm)} > 0 \\
		\pi u' , &u_d^{(\pm)} < 0 
	\end{cases}
	 \ ,
\ee
Thus the contributions of saddle points with $u_d^{(\pm)} < 0$ are suppressed by $\exp(-\pi \nu u)$. Thus, for $s < s_0$ both contributions are exponentially small, for $s_0 < s < s_2$ we have an $O(1)$ contribution from $u_d^{(+)}$ and for $s > s_2$ there are $ O(1)$ contributions from both $u_d^{(+)}$ and $u_d^{(-)}$.

\subsection{Transformed wave functions}

Evaluating~\eqref{eax} at the saddle point we find 
\be \label{spRightRes}
   \\
v_{k'}^{(RR)} (X_0; s)  = \sum_{\gamma = \pm} \sqrt{i \nu \ov 2 \pi}  A^{(\gamma)}_{\rk_c^{(\gamma)}} (X_0) B_{\rk_c^{(\gamma)} \rk'}^{RR} (s) K_R^{(\gamma)}{}^{-\ha} e^{i \nu G^{(\gamma)}_R |_{\rk_c^{(\gamma)}}}   \le(1 + O(\nu^{-1}) \ri) 
\ee
where
\be
\qquad K^{(\gamma)}_R = \p_u^2 G^{(\gamma)}_R |_{u_c^{(\gamma)}}  \ ,
\ee
and $\rk_c^{(\gamma)} = (u_c^{(\gamma)}, 0)$ with $u_c^{(\gamma)}$ the saddle point for $G^{(\gamma)}_R$
and note that there is no saddle point for $|u| < u_w$. Similarly we have 
\bega \label{eaxL}
v_{k'}^{(RL)} (X_0; s) = \sum_{\gamma = \pm} \sqrt{i \nu \ov 2 \pi}  A^{(\gamma)}_{\rk_d^{(\gamma)}} (X_0) B_{\rk_d^{(\gamma)} \rk'}^{RL} (s) K_L^{(\gamma)}{}^{-\ha} e^{i \nu G^{(\gamma)}_L |_{\rk_d^{(\gamma)}}}   \le(1 + O(\nu^{-1}) \ri) , \\
G^{(\gamma)}_L = Z^{(\gamma)}_\rk (X_0) + W^{RL}_{\rk \rk'} (s) , \qquad K^{(\gamma)}_L = \p_u^2 G^{(\gamma)}_L |_{u_d^{(\gamma)}}  \ ,
\end{gather}
where $\rk_d^{(\gamma)} = (u_d^{(\gamma)}, 0)$ with $u_d^{(\gamma)}$ the saddle point for $G^{(\gamma)}_L$. Explicit expressions for the quantities appearing in~\eqref{spRightRes} and~\eqref{eaxL} at general values of $k$ and $k'$ are given in~\eqref{LambdaPrefactorsFQ}, \eqref{WKBPhase}, \eqref{WKBPrefactor}, \eqref{LambdaPhases}.

We now show that our transformation in the large mass limit, as described by \eqref{spRightRes}--\eqref{eaxL} with saddle points~\eqref{ehn} and~\eqref{ehn}, is exactly the point-wise transformation \eqref{pwXfmLargeNu}. 

\subsubsection{Outside the horizon}

Here we consider $s < s_0$. We first restrict to the case of $u' > u_{w(s)}$. 

We begin with the calculation of $v^{(RR)}_k(X_0; s)$. Explicitly evaluating the quantities in \eqref{spRightRes} at the respective saddle points for $G^{(+)}_R$ and $G^{(-)}_R$ is quite complicated. Repeatedly using that $u^{(\pm)}_c$ solves the saddle point equation in the form~\eqref{sqrtsDGR} and recalling that we only need to consider $u^{(\pm)}>0$, $G^{(\pm)}_R$ can be brought to the form
\be \label{GRonSP}
\begin{aligned}
	G^{(\pm)}_R|_{\rk^{(\pm)}_c} &= -u'\log(\ep(s)(u^{(\pm)}_c - u')) + i\log\le( 1 \mp i w_0 \sqrt{u^{(\pm)}_c{}^2 - u_w^2} \ri) -i\log(1-i u^{(\pm)}_c)\\
	&+ u'\log |s| -{i \ov 2}\log\le({1+iu' \ov 1-iu'}\ri) + {u' \ov 2}\log(1+u'{}^2) -i\log w_0 \ .	
\end{aligned}
\ee
Comparing~\eqref{GRonSP} with~\eqref{WKBPhase} we see that $G^{(\pm)}_R|_{\rk^{(\pm)}_c}$ will equal to $Z^{(\pm)}_{\rk'}(X')$ for $X'=(\eta',w',\chi')$ only if
\bega\label{csa1}
\log |\rb'|  +  \log \le(u' \pm w' \sqrt{u'^2 - u_{w'}^2}\ri) =     
   \log (\ep(s)(u^{(\pm)}_c - u')) ,  \\
   \log {w_0 \ov w'} 
  =    
 \log {1 - i u' \ov  1- i u^{(\pm)}_c}  
+ \log {1 \mp i w_0 \sqrt{u^{(\pm)}_c{}^2-u_w^2}  \ov 1 \mp  i w' \sqrt{u'^2  - u_{w'}^2 } } \ .
\label{csa2} 
\end{gather} 
The above equations mean that $u'$ and $u^{(\pm)}_c$ should play symmetrical roles, with $u^{(\pm)}_c$ being the saddle point for the
transformation with parameter $s$ from $(\eta_0,w_0)$ at frequency $u'$, and $u'$ being the saddle point for the transformation by $-s$ 
from the point $(\eta', w')$ at frequency $u^{(\pm)}_c$. 
That such $(\eta',w')$ exist (as they cannot depend on $u'$) is highly nontrivial. 

Now it can be checked that for $\eta' = \eta_s, w' =w_s$, with
\bega \label{zprim}
e^{2\eta_s} 
=  {e^{2\eta_0} \ov 1- {2\rb}+ c^2 }, 
\quad w_s  
= {w_0 \ov 1- \rb (1-w^2_0)  } , 
\end{gather} 
which in terms of Kruskal coordinates gives~\eqref{heb}, 
equations~\eqref{csa1}--\eqref{csa2} are satisfied. 
To see this we note that 
\bega 
1-w'^2 = (1-w_0^2) {1- 2\rb + c^2 \ov (1 - \rb (1-w_0^2))^2} \quad \to \quad e^{2 \eta'} {1-w'^2 \ov w'^2} = e^{2\eta} {1-w_0^2 \ov w_0^2} \cr
\quad \to \quad e^{2 \eta'}  u_{w'}^2 = e^{2\eta} u_w^2 , \quad \rb' = \rb {1 - \rb (1-w_0^2) \ov 1- 2\rb + c^2}
\end{gather} 
and the inverse transformation is given by 
 \be\label{newp}
 w_0  
  = {w' \ov 1 + \rb' (1-w'^2)}, \quad  e^{\eta} = {e^{\eta'} \ov \sqrt{1+ 2\rb' + c'^2}}  ,\quad
 \rb' = {s e^{\eta'} \ov \sqrt{1-w'^2}} = {c' \ov \sqrt{1-w'^2}} \ .
\ee
Also note  
\be 
1 + \rb' (1-w'^2) ={1 \ov 1- \rb (1-w_0^2)}, \quad 1+ 2\rb' + c'^2 = {1 \ov 1 - 2\rb + c^2}  \ .
\ee

From the above relations and~\eqref{ehn} we then have 
\bega 
u^{(\pm)}_c - u'  = \rb {1-  \rb (1-w_0^2) \ov 1- 2\rb +c^2} \le(u' \pm w' \sqrt{u'^2 - u_{w'}^2} \ri) = \rb' \le(u' \pm w' \sqrt{u'^2 - u_{w'}^2} \ri)
\end{gather} 
which gives~\eqref{csa1}. We note by passing 
\be 
u^{(\pm)}_c - u'  = \rb' \le(u' \pm w' \sqrt{u'^2 - u_{w'}^2} \ri) = \rb \le(u^{(\pm)}_c \pm w_0 \sqrt{u^{(\pm)}_c{}^2 - u_w^2} \ri) \ . 
\ee
Similarly we have 
\bega 
{1- i u' \ov 1- i u^{(\pm)}_c}  \le(1 \mp  i w_0 \sqrt{u^{(\pm)}_c{}^2  - u_w^2 }\ri) = (1-\rb(1-w_0^2))\le(1 \mp i w' \sqrt{u'^2-u_{w'}^2}\ri) \cr
= {w_0 \ov w'} \le(1 \mp i w' \sqrt{u'^2-u_{w'}^2}\ri) \ ,
\end{gather} 
which gives~\eqref{csa2}. We have then shown 
\be \label{outRGatSP}
\begin{aligned}
	G^{(\pm)}_R |_{\rk_c^{(\pm)}} &= Z^{(\pm)}_{\rk'}(X_s) \ .
\end{aligned}
\ee

Now let us look at the prefactor of~\eqref{spRightRes}. 
 Computing the second derivative at the saddle point we have
\be \label{outRKatSP}
\begin{aligned}
	K^{(\pm)}_R  &= {1 \ov u^{(\pm)}_c - u'} - {1 \pm {u^{(\pm)}_c w_0 \ov \sqrt{u^{(\pm)}_c{}^2 - u_w^2}} \ov u^{(\pm)}_c \pm \sqrt{u^{(\pm)}_c{}^2 - u_w^2}} = {\pm \rb w_0 \sqrt{u'{}^2-u_{w(s)}^2} \ov (u^{(\pm)}_c - u')\le(\le(1-\rb\ri)u^{(\pm)}_c - u'\ri)  } \ ,
\end{aligned}
\ee
where we used~\eqref{sqrtsDGR} twice to replace all terms with square roots involving $u^{(\pm)}_c$ and then the explicit form of the root~\eqref{ehn}. From~\eqref{LambdaPrefactorsFQ} we have
\be \label{outRBatSP}
\begin{aligned}
	B^{RR}_{\rk_c^{(\pm)} \rk} &= \sqrt{2\pi \ov i\nu (u_c^{(\pm)} - u') } \ .
\end{aligned}
\ee
From~\eqref{WKBPrefactor} we have
\be \label{outRAatSP}
\begin{aligned}
	A^{(\pm)}_{\rk_c^{(\pm)}}(X_0) &= {w_0^{\ha} ~e^{\pm {i\pi \ov 4}} \ov \sqrt{2\nu}} \le({1 \ov u_c^{(\pm)}{}^2 - u_w^2}\ri)^{{1 \ov 4}} = {w_0^{\ha} ~e^{\pm {i\pi \ov 4}} \ov \sqrt{2\nu}} \sqrt{{ \pm \rb w_0  \ov \le(1-\rb \ri) u^{(\pm)}_c - u'}} \ ,
\end{aligned}
\ee
where we use~\eqref{sqrtsDGR}.
Putting all the prefactor contributions together we obtain
\be \label{outRPfAtSP}
	A^{(\pm)}_{\rk_c^{(\pm)} }(X_0) B^{RR}_{\rk_c^{(\pm)} \rk} K^{(\pm)}_R{}^{-\ha} = \sqrt{2\pi \ov i\nu} {w_0^{\ha} ~e^{\pm {i\pi \ov 4}} \ov \sqrt{2\nu}} \le({1 \ov u'{}^2 - u_{w(s)}^2}\ri)^{1 \ov 4} = \sqrt{2\pi \ov i\nu} \sqrt{{w_0 \ov w_s}} A^{(\pm)}_{\rk'}(X_s) ,
\ee
so the final result of the saddle point calculation is 
\be \label{vHatROut}
	 v^{(RR)}_{k'}(X_0;s) = \sqrt{1 - se^{\eta_0}\sqrt{1-w_0^2}} ~\sum_{\gamma = \pm} A^{(\gamma)}_{\rk'}(X_s) e^{i\nu Z^{(\gamma)}_{\rk'}(X_s)} \ ,
\ee
giving~\eqref{pwXfmLargeNu}.

The story for $u' < u_{w(s)}$ is rather similar, except that as mentioned earlier the saddle point is now complex. Through rather parallel calculations we again find~\eqref{pwXfmLargeNu}.

For $ v^{(RL)}_{k'}(X_0;s)$ as noted earlier $G^{(\pm)}_L$ only has saddle points with $u^{(\pm)}_d < 0$ for $0 < s < s_0$ which 
give rise to contributions which are exponentially suppressed. Thus, at the order we are working with, for $s < s_0$, the saddle point calculation gives 
\be \label{vHatLOut}
	 v^{(RL)}_{k'}(X_0;s) = 0 \ ,
\ee
 as required by \eqref{pwXfmLargeNu} for an operator in the right exterior region.

\subsubsection{Inside the horizon}

Now consider $s > s_0$. Since $\phi (X_s)$ ceases to make sense (and so does equation~\eqref{pwXfmLargeNu}) beyond the singularity, which is at $s =s_1$, we will first consider $s < s_1$.

We begin with the calculation of $ v^{(RR)}_k(X_0; s)$. The saddle point $u^{(+)}_c$ is now negative and gives a subdominant  contribution which can be dropped. The only contribution is then from $u^{(-)}_c$. Explicitly evaluating the quantities in \eqref{eax} at the saddle point for $G^{(-)}_R$ and comparing with~\eqref{WKBPhaseFutInt} and~\eqref{WKBPrefactorFutInt} we find
\be \label{inRGatSP}
\begin{aligned}
	G^{(-)}_R |_{\rk_c^{(-)}} &= Z^{(\sF)}_{\rk'}(X_s) 
\end{aligned}
\ee
and
\be \label{inRPfAtSP}
	A^{(-)}_{\rk_c^{(-)} }(X_0) B^{RR}_{\rk_c^{(-)} \rk} K^{(-)}_R{}^{-\ha} = \sqrt{2\pi \ov i\nu} {w_0^{\ha} ~e^{- {i\pi \ov 4}} \ov \sqrt{2\nu}} \le({1 \ov u'{}^2 - u_{w(s)}^2}\ri)^{1 \ov 4} = \sqrt{2\pi \ov i\nu} \sqrt{{w_0 \ov w_s}} A^{(\sF)}_{\rk'}(X_s)  
\ee
where $X_s$ is given by 
\bega \label{zprimFut}
e^{2\eta_s} =  - {e^{2\eta_0} \ov 1- {2\rb}+ c^2 }, 
\quad w_s  = {w_0 \ov 1- \rb (1-w^2_0)  } \ .
\end{gather} 
Equation~\eqref{zprimFut} gives~\eqref{heb} with $X_s \in F$. 

The final result of the saddle point calculation is then
\be \label{vHatRIn}
	 v^{(RR)}_{k'}(X_0;s) = \sqrt{1 - se^{\eta_0}\sqrt{1-w_0^2}}  ~A^{(\sF)}_{\rk'}(X_s) e^{i\nu Z^{(\sF)}_{\rk'}(X_s)} \ ,
\ee
giving \eqref{pwXfmLargeNu}, now with $X_s \in F$. 

For $ v^{(RL)}_{k'}(X_0;s)$ we now have the saddle point $u^{(+)}_d > 0$ of $G^{(+)}_L$ giving a leading order contribution. Explicitly evaluating the quantities in \eqref{eaxL} at the saddle point and comparing with~\eqref{WKBPhase} we find
\be \label{inLGatSP}
\begin{aligned}
	G^{(+)}_L |_{\rk_d^{(+)}} &= Z^{(\sF)}_{\rk'}(X_s) + 2 u \eta_s
\end{aligned}
\ee
and
\be \label{inLPfAtSP}
	A^{(+)}_{\rk_d^{(+)} }(X_0) B^{RL}_{\rk_d^{(+)} \rk} K^{(+)}_L{}^{-\ha} = \sqrt{2\pi \ov i\nu} {w_0^{\ha} ~e^{- {i\pi \ov 4}} \ov \sqrt{2\nu}} \le({1 \ov u'{}^2 - u_{w(s)}^2}\ri)^{1 \ov 4} = \sqrt{2\pi \ov i\nu} \sqrt{{w_0 \ov w_s}} A^{(\sF)}_{\rk'}(X_s) \ .
\ee

Thus, for $s_0 < s < s_1$, the saddle point calculation gives
\be \label{vHatLIn}
	 v^{(RL)}_{k'}(X_0;s) = \sqrt{1 - se^{\eta_0}\sqrt{1-w_0^2}}  ~A^{(\sF)}_{\rk'}(X_s) e^{2i\nu u \eta_s + i\nu Z^{(\sF)}_{\rk'}(X_s)} \ .
\ee
Note that the left and right mode functions, $v^{(L)}_k(X)$ and $v^{(R)}_k(X)$, are identical in the $F$ region except that the former involves $e^{i\om\eta-iq\chi}$ and the latter $e^{-i\om\eta+iq\chi}$, which are exactly reproduced by~\eqref{vHatRIn} and~\eqref{vHatLIn}. For more details, see~\eqref{vRvLBTZinF} and~\eqref{largeNuWFApp}.

For  $s_1 < s < s_2$, the expressions for the exponents and prefactors are the same except that we can longer compare them with $\phi (X_s)$.


\subsection{Comments on initial operators localized in $\chi$}
 
We now quickly comment on the computation of $\Phi(X_0;s)$ which is no longer averaged in the $\chi$-direction. 
Unlike the $q=0$ case, we cannot explicitly verify that the transformation is point-wise in the large mass limit as the transformation can now involve a change of the $\chi$ coordinate as well. Moreover, the saddle points are now solutions of a quartic equation and thus are very complicated functions of the parameters of the evolution. We will leave such analysis to the future. 


\section{Conclusions and discussions} \label{sec:concDisc}

In this paper we discussed in detail how to construct emergent bulk ``infalling'' times in the boundary theory. Their construction is a consequence of emergent type III$_1$ algebras and an associated half-sided modular inclusion/translation structure. 
We discussed explicitly two choices of such times which at the horizon correspond to uniform (in the transverse spatial directions) null $U$ or $V$ translations. There are an infinite number of others. 
For example, we can choose the subalgebra $\sN$ to be either of those depicted in Fig.~\ref{fig:arbT}, which should give rise to bulk infalling evolutions which are non-uniform in the transverse spatial directions.
 Alternatively, instead of taking the cyclic and separating vector to be the GNS vacuum $\ket{\Om_0}$, we can choose other vectors. The simplest possibilities are obtained by acting unitaries from $\sY_R$ and $\sY_L$ on $\ket{\Om_0}$, i.e. 
$ V_L W_R \ket{\Om_0}, \quad V_L \in \sY_L, \quad W_R \in \sY_R$
which results in a $\tilde U(s) = V_L W_R U (s) W_R^\da V_L^\da$ with $U(s)$ the evolution operator corresponding to $\ket{\Om_0}$. 

Our discussion can also be generalized to other entangled states of CFT$_R$ and CFT$_L$. A simple variant is to act on $\ktfd$ by a left unitary $U_L$ which does not change the reduced density matrix $\rho_\b$ of the CFT$_R$, i.e. 
$\ket{\Psi} = U_L \ktfd $. 
The story depends on whether $\ket{\Psi}$ lies in the {the image of the} GNS Hilbert space built from $\ktfd$. If $\ket{\Psi}$ lies in {the image of the} the GNS Hilbert space, 
the bulk geometry is still described by the eternal black hole, now with some small excitations on the left due to insertion of $U_L$. The construction of $U(s)$ is the same as that for $\ktfd$.  When $\ket{\Psi}$ does not lie in the GNS Hilbert space, which happens when $U_L$ changes the energy of the system by an amount which scales with $N$, {the story is different}. We need to work with the  GNS space $\sH_{\Psi}^{\rm GNS}$ associated with $\ket{\Psi}$, which does not overlap with that associated with $\ktfd$, and the corresponding representations $\sY_{L,R}$ of single-trace operator algebras are also different from those of associated with $\ktfd$.\footnote{The appearance of a different representation in this case is also required by the duality since the bulk geometry is also modified.} In this case there is no simple relation between $U(s)$ for $\ket{\Psi}$ with those for $\ktfd$ as they act on different GNS Hilbert spaces. 

There are many future questions to explore. We already mentioned some in Sec.~\ref{sec:impl}. Here we highlight a few more: 

\ben 

\item From a generic bulk point $X \in R$, the flow~\eqref{heb} reaches the future singularity for a finite value of $s$. 
We have not seen a sharp signature of the singularity either from~\eqref{tena1}--\eqref{Adef} or the leading expressions in the large mass limit except that the prefactor $\lam_X$ in~\eqref{pwXfmLargeNu} goes to zero at the singularity. 
It is possible that the signature of the singularity is weakened by the nonlocal nature of the $U(s)$ evolution and is more subtle to detect. 
The singularity should signal the breakdown of the $U(s)$ evolution, which is the way gravity tells us of its emergent nature. 
 It is clearly of great interest to understand the emergence of the singularity better and its possible resolution using our approach. 

\item Our discussions have been restricted to the leading order in the $1/N$ expansion: in the bulk we have a free field in a curved spacetime while on the boundary we have a generalized free field theory. We expect the general structure we uncovered should persist to any finite order in the $1/N$ expansion. Including higher order corrections corresponds to including gravitational physics in the bulk, which could lead to a much richer structure, in particular when including $1/N$ corrections to all orders~\cite{WittenNew}. 

\item It is  of great interest to understand better how the type III$_1$ structure emerges in the large $N$ limit. Systems like the SYK model or matrix quantum mechanics, should provide laboratories. 
In fact, a better understanding of the continuum limit of local operator algebras of a quantum field theory should be very instructive. 

\item The discussion here should also be generalizable to single-sided black holes including evaporating ones. We expect such constructions can shed new light on the information loss problem. 

\item We also expect that the manner in which an in-falling time emerges from the boundary theory here should teach us valuable lessons about holography for asymptotically flat and cosmological spacetimes. This should be especially helpful for understanding time in cosmological spacetimes including de Sitter.

\een

\vspace{0.2in}   \centerline{\bf{Acknowledgements}} \vspace{0.2in}
We would like to thank Wentao Cui, Netta Engelhardt,  Daniel Harlow,  Krishna Jalan, Juan Maldacena,  Leonard Susskind, and Edward Witten  for discussions, and Horacio Casini for communications. We also would like to thank Edward Witten for sharing a draft of~\cite{WittenNew}. 
This work is supported by the Office of High Energy Physics of U.S. Department of Energy under grant Contract Number  DE-SC0012567 and DE-SC0020360 (MIT contract \# 578218).
SL acknowledges the support of the Natural Sciences and Engineering Research Council of Canada (NSERC).

\appendix

\section{Details of the GNS construction} \label{app:GNS} 

In this appendix we discuss some details of the GNS construction of~\ref{sec:TFD}. 

For each operator $a \in \hat \sA_{\rm TFD}$ we associate a state $\ket{a}$, with the inner products among them given by~\eqref{inp1}.  The set of operators $y \in \hat \sA_{\rm TFD}$ such that $\vev{y|y} = 0$ is denoted by $\sJ$.
 $\sJ$ is a left ideal, as $a y \in \sJ$ for $\forall a \in \hat \sA_{\rm TFD},~ y \in \sJ$, and is  called the Gelfand ideal. The GNS Hilbert space $\sH_{\rm TFD}^{\rm (GNS)}$  is the completion of $\hat \sA_{\rm TFD} / \sJ$. 

Equation~\eqref{ehen1} implies that for each equivalence class in $\hat \sA_{\rm TFD} / \sJ$ we may choose a representative in the subalgebra $\sA_{R, \rm TFD}$, i.e. for any $a \in \hat \sA_{\rm TFD}$ there exists $A_R \in \sA_{R, \rm TFD}$ such that 
\be\label{ewn}
[a] = [A_R] \ .
\ee 
To see  this, consider an $a \in \hat \sA_{\rm TFD}$ of the form 
\be \label{ewn1} 
a = B_R C_L , \quad C_L =\prod_{i=1}^{n} \sO_L\le(\rt_i,x_i\ri)  \ ,
\ee
From~\eqref{ehen1} 
\be
	y_i = \sO_L\le(\rt_i,x_i\ri) - \sO_R\le(\rt_i + {i\b \ov 2}, x_i\ri)  \in \sJ , \quad i = 1, \cdots , n  \ .
\ee
We can then write 
\be 
 \sO_L\le(\rt_i,x_i\ri)  = \sO_R\le(\rt_i + {i\b \ov 2}, x_i\ri) + y_i 
 \ee
and $C_L$ can be written as 
\bega 
C_L =  \prod_{i=1}^{n-1} \sO_L\le(\rt_i,x_i\ri) \le(\sO_R\le(\rt_n + {i\b \ov 2}, x_n\ri) + y_n\ri)\cr
= \sO_R\le(\rt_n + {i\b \ov 2}, x_n \ri)   \prod_{i=1}^{n-1} \sO_L\le(\rt_i,x_i\ri)  + c  y_n 
\end{gather} 
for some $c \in \hat \sA_{\rm TFD}$. Note  $c y_n \in \sJ $ as $\sJ$ is a left ideal. Continuing this process repeatedly we reach at the end 
\be 
C_L = \sO_R\le(\rt_n + {i\b \ov 2}, x_n\ri)   \sO_R\le(\rt_{n-1} + {i\b \ov 2}, x_{n-1} \ri)  \cdots  \sO_R\le(\rt_1 + {i\b \ov 2}, x_1\ri) 
+ \tilde y , \quad \tilde y \in \sJ 
\ee 
which gives~\eqref{ewn}. The discussion immediately generalizes to sums of operators of the form~\eqref{ewn1}. 
Note that the representative $A_R$ in~\eqref{ewn} is unique, as if there is another $A_R'$ also satisfying $[a] = [A_R']$, we then 
have $A_R - A_R' \in \sJ$ but this cannot be the case since  $\ket{\Psi_\b}$ is separating for $\sA_{R, \rm TFD}$. 

We thus conclude that $\sH_{\tfd}^{\rm (GNS)}$ can be generated by $\sA_{R, \rm TFD}$ alone.

\section{Verification of $U(1)$ properties}  \label{app:u1}

In this appendix we show that the group property of $ \Lam^{\al \b}_{k k'}  (s)$ is satisfied, i.e. 
\be  \label{heo210}
\Lam_{kk'}^{\al \b} (s_1)  \Lam_{k' k''}^{\b \ga} (s_2) = \Lam_{k k''}^{\al \ga} (s_1 + s_2) \ .
\ee
 Recall $ \Lam^{\al \b}_{k k'}  (s)$ are given in terms of $ C_{k k'}  (s)$ by~\eqref{lam1}--\eqref{lam2}. 

\subsection{$s_1$ and $s_2$ of the same sign} 

We first show that~\eqref{heo210} follows from~\eqref{dgb} when $s_1$ and $s_2$ are of the same sign.

For $s_1$ and $s_2$ both negative,~\eqref{heo21} trivially follows from~\eqref{dgb} for $\alpha = R$ and either choice of $\beta$.  For $\alpha = L$ we need 
\bega \label{cc1}
\Lam^{LL}_{k k'} (s_1) \Lam_{k' k''}^{LL} (s_2) +  \Lam^{LR}_{k k'} (s_1) \Lam_{k' k''}^{RL} (s_2)= 
\Lam_{k k''}^{LL} (s_1 + s_2) ,  \\
\label{cc2} 
 \Lam^{LR}_{k k'} (s_1) \Lam_{k' k''}^{RR} (s_2) + \Lam^{LL}_{k k'} (s_1) \Lam_{k' k''}^{LR} (s_2)  = 
\Lam_{k k''}^{LR} (s_1 + s_2)  \ .
\end{gather} 
For~\eqref{cc1} we have 
\begin{equation}
\begin{aligned}
 \frac{\sinh\pi\omega}{\sinh\pi\omega'}C_{-k-k'}(s_1) \frac{\sinh\pi\omega'}{\sinh\pi\omega''} C_{-k'-k''}(s_2)= \frac{\sinh\pi\omega}{\sinh\pi\omega''} C_{-k-k''} (s_1 + s_2) 
\end{aligned}
\label{eq:u1}
\end{equation}
which is automatically satisfied. For~\eqref{cc2}, the left hand side has the form 
\begin{equation}
\begin{aligned}
	& \frac{\sinh\pi(\omega+\omega')}{\sinh\pi\omega'} C_{-k k'}(s_1) C_{k'k''}(s_2) + \frac{\sinh\pi\omega}{\sinh\pi\omega'} C_{-k-k'}(s_1) \frac{\sinh\pi(\omega'+\omega'')}{\sinh\pi\omega''} C_{-k' k''}(s_2)\\
	&= \left(\frac{\sinh\pi(\omega'+\omega)}{\sinh\pi\omega'} - \frac{\sinh\pi\omega}{\sinh\pi\omega'} \frac{\sinh\pi(\omega'' - \omega')}{\sinh\pi\omega''} \right)C_{-k k'}(s_1) C_{k'k''}(s_2) \\
	&= \frac{\sinh\pi(\omega+\omega'')}{\sinh\pi\omega''} C_{-kk''}(s_1 + s_2) =  \Lambda^{LR}_{k k''}(s_1 + s_2) \ .
\end{aligned}
\label{eq:u1prop-LRForNegNeg}
\end{equation}
For $s_1, s_2 > 0$ we have the same story due to the symmetry between $L$ and $R$ in changing $s < 0$ to $s > 0$.

\subsection{Opposite signs} 

For $s_1$ and $s_2$ of opposite sign the situation is more complicated, as we know that it must be since there must be some kind of transition when $s_1 + s_2$ changes sign. For $\al =R$, we need
\bega \label{cc3}
 \Lam^{RR}_{k k'} (s_1) \Lam_{k' k''}^{RR} (s_2)  + \Lam^{RL}_{k k'} (s_1) \Lam_{k' k''}^{LR} (s_2)   = 
\Lam_{k k''}^{RR} (s_1 + s_2) ,  \\
\label{cc4} 
 \Lam^{RR}_{k k'} (s_1) \Lam_{k' k''}^{RL} (s_2) + \Lam^{RL}_{k k'} (s_1) \Lam_{k' k''}^{LL} (s_2)  = 
\Lam_{k k''}^{RL} (s_1 + s_2)  \ .
\end{gather} 
For $s_1 < 0 , s_2 > 0$,~\eqref{cc3} can be written more explicitly as 
\bega
\label{hsj} 
C_{k k'} (s_1)  {\sinh \pi \om' \ov \sinh \pi \om''}   C_{k' k''} (-s_2)  = \bca C_{k k''} (s_1 + s_2)  & s_1 + s_2 < 0 \cr
 {\sinh \pi \om \ov \sinh \pi \om''}  C_{k k''} (-s_1 - s_2) & s_1 + s_2 > 0  \eca, 
  \end{gather}
while for $s_1 > 0 , s_2 < 0$
\bega \label{hsj2} 
 {\sinh \pi \om \ov \sinh \pi \om'}   C_{k k'} (-s_1)  C_{k' k''} (s_2)  + 
 {\sinh \pi (\om + \om') \ov \sinh \pi \om'} C_{k -k'} (-s_1)  {\sinh \pi (\om' + \om'') \ov \sinh \pi \om''} C_{- k' k''} (s_2)  \cr
 =  C_{k k'} (-s_1)  { \sinh \pi (\om + \om'' -\om') \ov \sinh \pi \om'' } C_{k' k''} (s_2) \cr
 = \bca C_{k k''} (s_1 + s_2)  & s_1 + s_2 < 0 \cr
 {\sinh \pi \om \ov \sinh \pi \om''}  C_{k k''} (-s_1 - s_2) & s_1 + s_2 > 0  \eca \ . 
  \end{gather}
For $s_1 < 0 , s_2 > 0$~\eqref{cc4} can be written more explicitly as
\bega \label{hsj0}
C_{k k'} (s_1)  \sinh \pi (\om' + \om'') C_{k' -k''} (-s_2)   = \bca 0  & s_1 + s_2 < 0 \cr
  \sinh \pi (\om + \om'')  C_{k k''} (-s_1 - s_2) & s_1 + s_2 > 0  \eca, 
  \end{gather}
while for $s_1 > 0 , s_2 < 0$
\bega
\label{hsj1} 
 {\sinh \pi (\om + \om') \ov \sinh \pi \om'} C_{k -k'} (-s_1)  {\sinh \pi \om'  \ov \sinh \pi \om''} C_{- k' -k''} (s_2)  
 = C_{k k'} (s_1')  {\sinh \pi (\om - \om')    \ov \sinh \pi \om''} C_{k' -k''} (-s_2')  \cr
 =  \bca 0  & s_1' + s_2' > 0 \cr
  {\sinh \pi (\om + \om'') \ov \sinh \pi \om''} C_{k k''} (s_1' + s_2') & s_1' + s_2' < 0  \eca , \quad s_1' = - s_1 < 0, \; \; s_2' = - s_2 >0 \ .
  \end{gather}

From the ``transpose'' property~\eqref{ctrans} equations~\eqref{hsj1} and~\eqref{hsj0} are equivalent. 
Equation~\eqref{hsj2} becomes 
\be \label{oo1}
C_{k k'} (s_2)  \sinh \pi (\om + \om'' -\om')   C_{k' k''} (-s_1) 
 = \bca \sinh \pi \om  C_{k k''} (s_1 + s_2)  & s_1 + s_2 < 0 \cr
 \sinh \pi \om''   C_{k k''} (-s_1 - s_2) & s_1 + s_2 > 0  \eca  
\ee
while~\eqref{hsj} has the form 
\bega \label{oo2} 
C_{k k'} (s_1)  \sinh \pi \om'  C_{k' k''} (-s_2)  = \bca \sinh \pi \om'' C_{k k''} (s_1 + s_2)  & s_1 + s_2 < 0 \cr
 \sinh \pi \om   C_{k k''} (-s_1 - s_2) & s_1 + s_2 > 0  \eca \ .
  \end{gather} 
The independent equations are then~\eqref{oo1}--\eqref{oo2} and~\eqref{hsj0}, respectively. 
These relations readily follow from~\eqref{genSt} and the following identity 
\be
\int {d \om' \ov 2 \pi} I_{\om \om'} (s_1) \sinh \pi (\om' +a)  I_{\om' \om''} (s_2) =
I_{\om \om''} (|s_1  - s_2|) \sinh \pi (\tilde \om +a) , \quad s_{1,2} > 0 
\ee
where $\tilde \om = \om, \om''$ for $s_2 > s_1$ and $s_2 < s_1$. 
To see the identity, note 
\bega 
\int {d \om' \ov 2 \pi} I_{\om \om'} (s_1) \sinh \pi (\om' +a)  I_{\om' \om''} (s_2) \cr
= \int {d \om' \ov 2 \pi} s_1^{- i (\om -\om')} \Ga (i (\om-\om')+\ep) s_2^{- i (\om' -\om'')} \Ga (i (\om'-\om'')+\ep)) \sinh \pi (\om' +a)
\cr
=\sum_{n=0}^\infty {1 \ov n!}  \bca   \le({s_1 \ov s_2}\ri)^n s_2^{-i (\om -\om'')}  \Ga (i (\om -\om'') + n) \sinh \pi (\om +a) & s_2 > s_1 \cr
 \le({s_2 \ov s_1}\ri)^n s_1^{-i (\om -\om'')}  \Ga (i (\om -\om'') + n) \sinh \pi (\om'' +a)
  &  s_2 < s_1
\eca \cr
= \Ga (i (\om-\om'') +\ep) (s_a - s_b)^{-i (\om-\om'')} \sinh \pi (\tilde \om +a) 
\end{gather}  
where for $s_2 < s_1$  ($s_2> s_1$) we can close the contour in the upper (lower) half complex $\om'$-plane. In the above we have $s_a = {\rm max} (s_1, s_2), s_b = {\rm min} (s_1, s_2)$ and $\tilde \om$ as defined earlier.

\section{Review of AdS Rindler and BTZ black hole} \label{app:adsR}

In this section we elaborate more on the geometries of AdS-Rindler in $(2+1)$-dimensions and the BTZ black hole
reviewed in Sec.~\ref{sec:adsgeo}. We will set the AdS radius to be unity throughout. 


\subsection{AdS Rindler in $(2+1)$-dimension}

The Poincar\'e patch of AdS$_3$
\be 
ds^2 =  {1 \ov z^2} (- (dx^0)^2 + (dx^1)^2 + dz^2 )  =  {1 \ov z^2} (- dx^+ dx^- + dz^2 ), \quad x^\pm = x^0 \pm x^1
\ee
can be separated into four different AdS Rindler regions, labeled by $\sR, \sL, \sF, \sP$ corresponding respectively to regions with $(x^+ , x^-)$ having signs $(+,-)$, $(-,+)$, $(+,+)$, $(-,-)$. They have respectively $R, L,F,P$ Rindler regions of Minkowski spacetime $\RR^{1,1}$ as their boundaries ($z \to 0$). See Fig.~\ref{fig:rind}. In the BTZ coordinates $(\eta,w,\chi)$, which for $\sR$ region has the form 
\bega   \label{Rind21}
        z = w  e^{\chi} , \quad x^+  = e^{\xi^+} \sqrt{1-w^2} , \quad x^-  = - e^{- \xi^-} \sqrt{1-w^2}, \quad 
         \xi^\pm = \eta \pm \chi   ,  \\
        e^{2 \eta} = - {x^+ \ov x^-},   \quad  e^{2 \chi} = z^2 - x^+ x^-, \quad w^2 = {z^2 \ov z^2 - x^+ x^-}, \quad 1 - w^2 = {- x^+ x^-\ov z^2 - x^+ x^-} 
          \label{Rind22}
    \end{gather} 
the metric has the form 
\be \label{Rindc2}
    ds^2 =  \frac{1}{w^2}\left[-\left(1 - w^2  \right)d\eta^2 + \left(1 - w^2  \right)^{-1}dw^2 + d\chi^2 \right]  \ .
      \ee
The AdS Rindler horizon is at $w =1$ and the boundary is at $w=0$. 
The metric~\eqref{Rindc2} can be used to cover the other AdS Rindler regions with the transformation~\eqref{Rind21} suitably modified for each region. For example, for the $\sF$ region we can introduce 
\bega   \label{Rind31}
        z = w  e^{\chi} , \quad x^+  = e^{\xi^+} \sqrt{w^2-1} , \quad x^-  = e^{- \xi^-} \sqrt{w^2-1}
          ,  \\
        e^{2 \eta} =  {x^+ \ov x^-},   \quad  e^{2 \chi} = z^2 - x^+ x^-, \quad w^2 = {z^2 \ov z^2 - x^+ x^-} \ . 
          \label{Rind32}
    \end{gather} 
Notice that the last three equations of~\eqref{Rind32} remain the same as those in~\eqref{Rind22} except that now $w > 1$. 
Equations~\eqref{Rind31}--\eqref{Rind32}, however, only cover the part of the $\sF$ region with $z^2 - x^+ x^- > 0$ (to which we will refer as $\sF_1$ region), with $w = \infty$ corresponding to $z^2 = x^+ x^-$. 
For $z^2 - x^+ x^- < 0$ (to which we will refer as $\sF_2$ region)  the second equation of~\eqref{Rind32} no longer makes sense. 
For the $\sF_2$ region we can analogously introduce 
\bega   \label{Rind31F2}
        z = w  e^{\chi} , \quad x^+  = e^{\xi^+} \sqrt{1+w^2} , \quad x^-  = e^{- \xi^-} \sqrt{1+w^2}
          ,  \\
        e^{2 \eta} =  {x^+ \ov x^-},   \quad  e^{2 \chi} = x^+ x^- - z^2, \quad w^2 = {z^2 \ov x^+ x^- -z^2 } \ . 
          \label{Rind32F2}
    \end{gather} 
but the corresponding metric now has the form 
\be \label{rindMetF2P2}
    ds^2 =  \frac{1}{w^2}\left[\left(1 + w^2  \right)d\eta^2 + \left(1 + w^2  \right)^{-1}dw^2 - d\chi^2 \right]  \ .
\ee
Similarly for the $\sP$ region. Note that $\sF_1, \sP_1$ regions do not contain any points near the asymptotic boundary, while the boundaries of $\sF_2$ and $\sP_2$ regions are respectively the $F$ and $P$ regions of Minkowski spacetime $\RR^{1,1}$. 

Now consider a point $X = (x^-, x^+, z)$ in the $\sR$ region (with $x^- < 0, x^+> 0$) and a null shift  $X \to X_s = (x^+_s, x^-_s, z_s)$ with 
\be \label{nsg}
x^-_s = x^- + s, \quad x^+_s = x^+ , \quad z_s = z  \ .
\ee
For $x^-_s < 0$,  $X_s$ remains in the $\sR$ region. The corresponding transformation can be written in terms of the BTZ coordinates as 
\bega\label{nsh}
w_s = {w \ov \sqrt{1-a_s}} , \quad e^{\xi^-_s} = {e^{\xi^-} \sqrt{1-w^2} \ov \sqrt{1 - a_s}\sqrt{1 - a_s - w^2}} ,
\quad  e^{\xi^+_s} = {e^{\xi^+} \sqrt{1-w^2} \sqrt{1 - a_s} \ov \sqrt{1 - a_s - w^2}}   
\end{gather}
where 
\be
	a_s \equiv s e^{\xi^-} \sqrt{1-w^2} 
\ee
and $a_s < 1-w^2$ for this range of $s$. For $s \geq s_0 = e^{- \xi^-} \sqrt{1-w^2}$, the AdS Rindler horizon is crossed and we have 
 $x^-_s > 0$. For $z_s^2 - x^+_s x^-_s > 0$, we have $1 > a_s > 1-w^2$ and $X_s$ lies in the $\sF_1$ region. The corresponding transformation becomes 
\bega\label{nsh1}
w_s = {w \ov \sqrt{1-a_s}} , \quad e^{\xi^-_s} = {e^{\xi^-} \sqrt{1-w^2} \ov \sqrt{1 - a_s}\sqrt{ a_s -1 + w^2}} ,
\quad  e^{\xi^+_s} = {e^{\xi^+} \sqrt{1-w^2} \sqrt{1 - a_s} \ov \sqrt{a_s -1 + w^2}}    \ .
\end{gather}
Finally, for $x^-_s > 0,~ z_s^2 - x^+_s x^-_s < 0$, we have $a_s > 1$ and $X_s$ lies in the $\sF_2$ region. The corresponding transformation becomes 
\bega\label{nsh2}
w_s = {w \ov \sqrt{a_s-1}} , \quad e^{\xi^-_s} = {e^{\xi^-} \sqrt{1-w^2} \ov \sqrt{a_s-1}\sqrt{ a_s -1 + w^2}} ,
\quad  e^{\xi^+_s} = {e^{\xi^+} \sqrt{1-w^2} \sqrt{a_s-1} \ov \sqrt{a_s -1 + w^2}}    \ .
\end{gather}

\subsection{BTZ geometry}

The BTZ black hole can be obtained from the AdS Rindler metric~\eqref{Rindc2} by making $\chi$ compact~\cite{Banados:1992wn}. 
Now $w=\infty$ is a genuine singularity where the spacetime ends, and
 $w =1$ becomes an event horizon. The black hole has inverse temperature $\b =2 \pi$. 
 As usual the black hole spacetime can be extended to four regions by using the Kruskal coordinates (see Fig.~\ref{fig:casu}), which for $R$ and $F$ regions have the form 
\bega\label{krubtz11}
R: \quad  U  =
- \sqrt{1-w \ov 1+w} e^{- \eta} , \quad
 V  =
\sqrt{1-w \ov 1+w} e^{ \eta},  \\
F: \quad  U =
\sqrt{w -1\ov 1+w} e^{- \eta} , \quad
 V  =
\sqrt{w-1 \ov 1+w} e^{ \eta} \ .
\end{gather} 
 In terms of the Kruskal coordinates the metric has the form 
\be \label{krusth}
 ds^2  = - {4  \ov (1 +  UV )^2} d U d V + {(1- UV)^2 \ov (1+UV)^2}
  d  \chi^2 \ .
\ee
The event horizons lie at $U, V=0$, the boundary lies at $UV =-1$, and the singularity at $UV =1$. 

Consider a shift of a point $X = (U, V, \chi) \in R$ to $X_s = (U_s, V_s, \chi_s)$ with 
\be\label{BTZCoordShift}
	U_s = U + s, \quad V_s  = {V \ov 1 - sV}, \quad \chi_s = \chi \ .
\ee
For $X_s \in R$, this can be expressed in BTZ coordinates as
\bega \label{zprimApp}
e^{2\eta_s} 
=  {e^{2\eta} \ov 1- {2\rb}+ c^2 }, 
\quad w_s  
= {w \ov 1- \rb (1-w^2)  } \\
\rb \equiv {s e^{\eta} \ov \sqrt{1-w^2}}, \qquad c \equiv se^{\eta} \ .
\end{gather} 
$X_s$ crosses the horizon at $s = s_0 =  \sqrt{1-w_0 \ov 1+w_0} ~e^{-\eta_0}$ where $1- {2\rb}+ c^2 =0$. 
For $s > s_0$, i.e. $X_s \in F$,  the first equation in~\eqref{zprimApp} has an extra minus sign.

\section{Analytic continuations of bulk mode functions} \label{app:cont}

In this Appendix we give analytic continuations of (i)  mode functions in the $\sR$ and $\sL$ regions to the $\sF$ region 
for the AdS Rindler; (ii)  mode functions in the $R$ and $L$ regions to the $F$ region of the BTZ geometry.

\subsection{AdS-Rindler Mode functions}
The AdS-Rindler mode functions in the right/left AdS-Rindler regions are given by
\bega \label{bulkRind}
	v^{(R)}_k(X) = N_k ~e^{-i\omega\eta + iq\chi} w^{\De}\le(1 - w^2\ri)^{-\frac{i\om}{2}} F(\bar q_+ , \bar q_- ; \De, w^2) \cr
	\qquad = N_k ~(x^+)^{-i\om} \le(z^2 - x^+x^-\ri)^{-q_+}  F\le(\bar q_+ , \bar q_- ; \De, \frac{z^2}{z^2 - x^+x^-} \ri) \\
	v^{(L)}_k(X) = N_k ~e^{i\omega\eta - iq\chi} w^{\De}\le(1 - w^2\ri)^{\frac{i\om}{2}} F( q_+ , q_- ; \De, w^2) \cr
	\qquad = N_k ~(x^+)^{i\om} \le(z^2 - x^+x^-\ri)^{-\bar q_+}  F\le( q_+ ,  q_- ; \De, \frac{z^2}{z^2 - x^+x^-} \ri)  .
\end{gather}
where we have also expressed them in terms of Poincare coordinates. 

From the usual Unruh procedure, we can obtain mode functions $w^{(R, L)}_k$ corresponding to the Poincare vacuum $\ket{0}_{\rm bulk}$ by analytically continuing $v^{(R,L)}_k$ in the complex  planes:\footnote{$w^{(R, L)}_k$ with $\om > 0$ ($\om <0)$ are obtained from continuing in the lower (upper) complex $x^-,~x^+$ planes.}
\bega \label{vk00}
w^{(R)}_k  =  {1 \ov \sqrt{2 \sinh \pi |\om|}} \left( e^{ {\pi |\om| \ov 2} } v^{(R)}_k + e^{-{\pi |\om| \ov 2}}   v^{(L)}_{-k}  \ri), \\
w^{(L)}_k  =  {1 \ov \sqrt{2 \sinh \pi |\om|}} \left( e^{ {\pi |\om| \ov 2} }  v^{(L)}_k + e^{-{\pi |\om| \ov 2}}   v^{(R)}_{-k}  \ri)  
\label{vk01}
\end{gather}
and the inverse are given by 
\bega \label{vk1}
v^{(R)}_k  =  {1 \ov \sqrt{2 \sinh \pi |\om|}} \left( e^{ {\pi |\om| \ov 2} } w^{(R)}_k - e^{-{\pi |\om| \ov 2}}   w^{(L)}_{-k}  \ri)
, \\
v^{(L)}_k  =  {1 \ov \sqrt{2 \sinh \pi |\om|}} \left( e^{ {\pi |\om| \ov 2} }  w^{(L)}_k - e^{-{\pi |\om| \ov 2}}   w^{(R)}_{-k}  \ri)  \ .
\label{vk2}
\end{gather}

By construction $w^{(R, L)}_k$ are analytic in $x^\pm$ and are thus defined for all $x^\pm$. We can then use~\eqref{vk1}--\eqref{vk2} to ``continue'' $v^{(R,L)}_k$ to the $\sF$ and $\sP$ regions. Note that thus constructed $v^{(R,L)}_k$ are not analytic at the Rindler horizons. We then find that 
\be \label{contRindvR}
\begin{aligned}
	v^{(R)}_k(X) &= N_k w^{\Delta} e^{-i\om\eta + i q \chi}  
	\cdot \begin{cases}
		-i \frac{\sin\pi q_-}{\sinh\pi\om} (w^2 + 1)^{-{i\om \ov 2}} F\le(\bar q_+, \bar q_- ; \De ; -w^2\ri), & \sF_2 \\
		 \frac{\Gamma(\De)\Gamma(i\om)}{\Ga(q_+)\Ga(q_-)} (w^2 - 1)^{-{i\om \ov 2}} F\le( \bar q_+ , \bar q_- ; 1 - i\om ; 1-w^2  \ri) ,&  \sF_1 \\
		(1 - w^2)^{-{i\om \ov 2}} F\le(\bar q_+, \bar q_- ; \De ; w^2\ri) ,&  \sR \\
		0 ,&  \sL \\
		\frac{\Gamma(\De)\Gamma(-i\om)}{\Ga(\bar q_+)\Ga(\bar q_-)} (w^2 - 1)^{{i\om \ov 2}} F\le(  q_+ ,  q_- ; 1 + i\om ; 1-w^2  \ri) ,&  \sP_1 \\
		i \frac{\sin\pi \bar q_+}{\sinh\pi\om} (w^2 + 1)^{{i\om \ov 2}} F\le( q_+,  q_- ; \De ; -w^2\ri) ,& \sP_2 
	\end{cases}
\end{aligned}
\ee
and
\be\label{contRindvL}
	\begin{aligned}
	v^{(L)}_k(X) &= N_k w^{\Delta} e^{i\om\eta - iq\chi}
	\cdot \begin{cases}
		 -i \frac{\sin\pi q_+}{\sinh\pi\om} (w^2 + 1)^{-{i\om \ov 2}} F\le(\bar q_+,  \bar q_- ; \De ; -w^2\ri),& \sF_2 \\
		\frac{\Gamma(\De)\Gamma(i\om)}{\Ga(q_+)\Ga(q_-)} (w^2 - 1)^{-{i\om \ov 2}} F\le( \bar q_+ , \bar q_- ; 1 - i\om ; 1-w^2  \ri),&  \sF_1 \\
		0 ,&  \sR \\
		(1 - w^2)^{{i\om \ov 2}} F\le( q_+,  q_- ; \De ; w^2\ri),&  \sL \\
		\frac{\Gamma(\De)\Gamma(-i\om)}{\Ga(\bar q_+)\Ga(\bar q_-)} (w^2 - 1)^{{i\om \ov 2}} F\le(  q_+ ,  q_- ; 1 + i\om ; 1-w^2  \ri) ,&  \sP_1 \\
		i \frac{\sin\pi \bar q_-}{\sinh\pi\om} (w^2 + 1)^{{i\om \ov 2}} F\le( q_+,  q_- ; \De ; -w^2\ri) ,& \sP_2  \\
	\end{cases}
\end{aligned}
\ .
\ee

\subsection{BTZ mode functions}

The story is completely parallel for continuing mode functions in BTZ, except now we use the Kruskal coordinates,  in terms of which 
\bega \label{BTZExtModes}
	v^{(R)}_k(X) = N_k ~e^{iq\chi} (2V)^{-i\om} (1+UV)^{\De} (1-UV)^{-\De + i\om} F\le(\bar q_+ , \bar q_- ; \De, \le(\frac{1+UV}{1-UV}\ri)^2 \ri) ,\\
	v^{(L)}_k(X) = N_k ~e^{-iq\chi} (-2V)^{i\om} (1+UV)^{\De} (1-UV)^{-\De - i\om} F\le( q_+ ,  q_- ; \De, \le(\frac{1+UV}{1-UV}\ri)^2 \ri) \ .
\end{gather}
Again by first constructing $w^{(R,L)}_k$ in the $R$ and $L$ regions using~\eqref{vk00}--\eqref{vk01}, analytically continuing $w^{(R,L)}_k$ to other regions, we then use~\eqref{vk1}--\eqref{vk2} to find the corresponding $v^{(R,L)}_k$ in other regions. We find 
\be \label{contBTZvR}
\begin{aligned}
	v^{(R)}_k(X) &= N_k \le(\frac{1+UV}{1-UV}\ri)^{\Delta} e^{iq \chi} 
	\cdot \begin{cases}
		(2V)^{-i\omega} \left(1 - UV\right)^{i\om} \frac{\Gamma(\De)\Gamma(i\om)}{\Ga(q_+)\Ga(q_-)} F\le( \bar q_+ , \bar q_- ; 1 - i\om ; {-4UV \ov (1-UV)^2}  \ri) ,&  F \\
		(2V)^{-i\omega} (1-UV)^{i\om} F\le(\bar q_+ , \bar q_- ; \De, \le(\frac{1+UV}{1-UV}\ri)^2 \ri),&  R \\
		0 ,&  L \\
		(-2U)^{-i\omega} \left(1 - UV\right)^{-i\om} \frac{\Gamma(\De)\Gamma(-i\om)}{\Ga(\bar q_+)\Ga(\bar q_-)} F\le( q_+ ,  q_- ; 1 + i\om ; {-4UV \ov (1-UV)^2}  \ri),&  P \\
	\end{cases}
\end{aligned}
\ee
and
\be \label{contBTZvL}
	\begin{aligned}
	v^{(L)}_k(X) &= N_k \le(\frac{1+UV}{1-UV}\ri)^{\Delta} e^{-iq \chi} 
	\cdot \begin{cases}
		(2U)^{-i\omega} \left(1-UV\right)^{i\om} \frac{\Gamma(\De)\Gamma(i\om)}{\Ga(q_+)\Ga(q_-)} F\le( \bar q_+ , \bar q_- ; 1 - i\om ; {-4UV \ov (1-UV)^2}  \ri),&  F \\
		0 ,&  R \\
		(-2V)^{i\omega} (1-UV)^{- i\om} F\le( q_+ ,  q_- ; \De, \le(\frac{1+UV}{1-UV}\ri)^2 \ri),&  L \\
		(-2V)^{i\omega} \left(1-UV\right)^{- i\om} \frac{\Gamma(\De)\Gamma(-i\om)}{\Ga(\bar q_+)\Ga(\bar q_-)} F\le( q_+ ,  q_- ; 1 + i\om ; {-4UV \ov (1-UV)^2}  \ri) ,&  P \ . \\
	\end{cases}
\end{aligned}
\ee

Note that~\eqref{contBTZvR} and~\eqref{contBTZvL} may be also expressed in terms of BTZ coordinates. For example, in the $F$ region we have
\be \label{vRvLBTZinF}
\begin{aligned}
	v^{(R)}_k(X) &= N_k w^{\Delta} e^{-i\om\eta + i q \chi} \frac{\Gamma(\De)\Gamma(i\om)}{\Ga(q_+)\Ga(q_-)} (w^2 - 1)^{-{i\om \ov 2}} F\le( \bar q_+ , \bar q_- ; 1 - i\om ; 1-w^2  \ri) \\
	v^{(L)}_k(X) &= N_k w^{\Delta} e^{i\om\eta - iq\chi} \frac{\Gamma(\De)\Gamma(i\om)}{\Ga(q_+)\Ga(q_-)} (w^2 - 1)^{-{i\om \ov 2}} F\le( \bar q_+ , \bar q_- ; 1 - i\om ; 1-w^2  \ri) \ .
\end{aligned}
\ee

Near the horizon, i.e. taking $UV \to 0$ in~\eqref{contBTZvR} and~\eqref{contBTZvL}, we then have
\be \label{NearcontBTZvR}
\begin{aligned}
	v^{(R)}_k(X) &= {e^{iq \chi} \ov \sqrt{2|\om|} } 
	\cdot \begin{cases}
		e^{i\de_k} V^{-i\om} ,&  F \\
		e^{i\de_k} V^{-i\om} + e^{-i\de_k} (-U)^{i\om} ,&  R \\
		0 ,&  L \\
		e^{-i\de_k} (-U)^{i\om},&  P \\
	\end{cases}
\end{aligned}
\ee
and
\be \label{NearcontBTZvL}
	\begin{aligned}
	v^{(L)}_k(X) &= {e^{-iq \chi} \ov \sqrt{2|\om|} } 
	\cdot \begin{cases}
		e^{i\de_k} U^{-i\om},&  F \\
		0 ,&  R \\
		e^{i\de_k} U^{-i\om} + e^{-i\de_k} (-V)^{i\om} ,&  L \\
		e^{-i\de_k} (-V)^{i\om} ,&  P \ . \\
	\end{cases}
\end{aligned}
\ee
In each case the phase shift is given by
\be \label{phaseShiftApp}
	e^{i\de_k} = {\Ga(i\om) | \Ga(q_-) \Ga(q_+) | \ov |\Ga(i\om)| \Ga(q_-) \Ga(q_+)} e^{-i\om \log 2} \ .
\ee

\section{Mode expansions in the boundary} \label{app:bdlim}

Here we discuss the mode expansions for the generalized free field theories resulting from a two-dimensional CFT in the large $N$ limit for two cases:  (i) in vacuum restricted to a Rindler region; (ii) at finite temperature (dual to a BTZ black hole). 
A convenient way to obtain both is to take boundary limit of the corresponding bulk mode expansions.



For the boundary CFT$_{R, L}$ dual to a BTZ black hole, the boundary mode expansion for the dual operator $\sO_\al$ can be obtained by taking $w \to 0$ limit of~\eqref{ebs} and  stripping off the $w^\De$ factor, which gives 
\bega 
\sO_{\al}  (x) = \lim_{w \to 0} w^{-\De} \phi^{(\al)} (X) =  \int \frac{d^2 k}{(2\pi)^2} \,
u_k^{(\al)}(x) a_{k}^{(\al)} \\
u_k^{(R)} (x) =  N_{k}  e^{i k \cdot x } ,\quad u_k^{(L)} (x) =  N_{k}  e^{-i k \cdot x } \ .
\label{vask}
\end{gather}

In the AdS-Rindler case (with non-compact $\chi$), the boundary limit should now be defined by using the Poincar\'e radial coordinate $z$ 
and stripping off a factor of $z^\De$, which gives 
\bega \label{pcExtDict}
\sO_{\al} (x) = \lim_{z \to 0} z^{-\De} \phi^{(\al)} (X) = \int \frac{d^2 k}{(2\pi)^2} \, 
 u_k^{(\al)}(x) a_{k}^{(\al)}   \ .
\end{gather} 
Now due to the difference between $w$ and $z$ we have an additional $e^{-\De \chi}$ factor compared with~\eqref{vask}. 
The behavior of $u_k^{(\al)}$ in various Rindler regions can then be obtained from~\eqref{contRindvR}--\eqref{contRindvL} 
\be \label{contBdryR}
 u^{(R)}_k(x) =  \frac{N_k}{\sinh\pi\omega}
	\begin{cases}
		\sinh\pi\om ~(-x^-)^{-\bar q_+} (x^+)^{-q_-} , &x^- < 0, ~x^+ > 0 \; (R) \cr
		-i\sin\pi q_- ~(x^-)^{-\bar q_+} (x^+)^{-q_-} , &x^-> 0,~ x^+ > 0  \; (F) \cr
		0 , &x^-> 0,~ x^+ < 0 \; (L) \cr
		i\sin\pi {\bar q_+} ~(-x^-)^{-\bar q_+} (-x^+)^{-q_-} , &x^-< 0,~ x^+ < 0 \; (P) \cr
	\end{cases}
\ee
and
\be \label{contBdryL}
	u^{(L)}_k(x) =  \frac{N_k}{\sinh\pi\omega}
	\begin{cases}
		0 , &x^- < 0, ~x^+ > 0 \; (R) \cr
		-i\sin q_+ ~(x^-)^{-q_+} (x^+)^{-\bar q_-} , &x^-> 0,~ x^+ > 0 \; (F) \cr
		\sinh\pi\omega ~(x^-)^{- q_+} (-x^+)^{- \bar q_-} , &x^-> 0,~ x^+ < 0 \; (L) \cr
		i \sin\pi {\bar q_-} ~(-x^-)^{- q_+} (-x^+)^{- \bar q_-} , &x^-< 0,~ x^+ < 0 \; (P) \cr
	\end{cases}
	.
\ee

\section{Properties of hypergeometric functions} \label{app:hyper}

Here we collect for convenience various properties of hypergeometric functions used in the main text.

\subsection{Asymptotic behavior of hypergeometric function}

We first discuss the asymptotic behavior of the hypergeometric function $F(a,b;c;z)$ when one or more of its parameters $a,b,c$ 
are taken to be large. 

Below $\lam$ should be understood as a complex parameter with $|\lam| \to \infty$. 

\subsubsection{Case I}


From~\cite{KD2014}, for $y/(y-1) < \ha$ and $\lambda$ not on the negative imaginary axis. 
\be \label{hypAsmpForSplitRind}
	F\le(a, b -i\lambda ; c - i\lambda ; y\ri) = (1-y)^{-a} F\le(a, c-b ; c - i\lambda ; {y \ov y-1}\ri) = (1-y)^{-a} (1 + O(\lambda^{-1})) \ .
\ee
For $\ha < y/(y-1) < 1$ the leading term is the same as~\eqref{hypAsmpForSplitRind}, although 
there are additional terms that are exponentially suppressed at large $|\lam|$.

\subsubsection{Case II} 

When $\lambda$ is on the imaginary axis or in the right half plane and for any real $y \in (1, \infty)$, 
we have~\cite{KD2014}
\be \label{ehb0}
F\le(a+\lam, a-\lam; c ; {1 -y \ov 2}\ri) = {\Ga (c) \Ga (\lam + 1 + a-c) \ov \Ga (a+ \lam)} \le(a_0 \le({\zeta \ov 2}\ri)^{1-c} I_{c-1} (\zeta \lam)
+ O(\Phi_1)  \ri)
\ee
where ($I_{c}$ below is modified Bessel function)
\bega 
\zeta = \log (y + \sqrt{y^2 -1}) , \quad a_0 = 2^{a+\ha-c} (y+1)^{{c \ov 2} - {1 \ov 4} -a} (y-1)^{{1 \ov 4} -{c \ov 2}} \zeta^{c-\ha}  \\
\Phi_1 = |\zeta^{1-c} \lam^{-1} I_{c-1} (\zeta \lam)| +  |\zeta^{-c} \lam^{-1} I_{c} (\zeta \lam)| \ .
\end{gather} 
For $\lam$ in the upper-half complex plane, $- i \lam$ is in the right half plane, while for $\lam$ in the lower-half complex plane, $i \lam$ 

Now consider with  $w \in (0,1)$, 
\bega   \label{hh1}
F (a - i \lam, b-i \lam; c; w^2) 
  =  (1-w^2)^{-b+ i \lam} F\le(b_1 + i \lam_1,b_1-i \lam_1; c; {1 -y \ov 2}\ri) \\
  \label{hh2}
  =  (1-w^2)^{-a+i \lam} F\le(b_2-i \lam_1, b_2+i \lam_1; c; {1 -y \ov 2} \ri)  \\
  b_1 = {c-a + b \ov 2}, \quad \lam_1 = \lam - i {c-a-b \ov 2} , \quad b_2 = {c- b +a \ov 2},  \quad y= {1+ w^2 \ov 1- w^2} \ .
 \end{gather} 
 From~\eqref{ehb0}, we should use~\eqref{hh1} for $\lam$ in the lower-half complex plane and~\eqref{hh2}   for $\lam$ in the upper-half complex plane. We thus  find for ${\rm Im} \lam > 0$ 
\bega
F (a - i \lam, b-i \lam; c; w^2) = (1-w^2)^{-a+i \lam} {\Ga (c) \Ga (a -i \lam + 1 -c) \ov \Ga (a- i \lam)} A_0 
\le({\zeta \ov 2}\ri)^{1-c}  {e^{ - i \zeta \lam_1} \ov \sqrt{- 2 \pi i \zeta \lam_1}} , \\
\zeta = \log (y + \sqrt{y^2 -1})
= \log {1 +w \ov 1-w} , \quad A_0 = 2^{b_2+\ha-c} (y+1)^{{c \ov 2} - {1 \ov 4} -b_2} (y-1)^{{1 \ov 4} -{c \ov 2}} \zeta^{c-\ha}  ,
 \end{gather}  
and for ${\rm Im} \lam < 0$ 
\bega
F (a - i \lam, b-i \lam; c; w^2) = (1-w^2)^{-b+i \lam} {\Ga (c) \Ga (-a +i \lam + 1 ) \ov \Ga (c-a+i \lam)} \tilde A_0 
\le({\zeta \ov 2}\ri)^{1-c}  {e^{ i \zeta \lam_1} \ov \sqrt{2 \pi i \zeta \lam_1}} , \\
\tilde A_0 = 2^{b_1+\ha-c} (y+1)^{{c \ov 2} - {1 \ov 4} -b_1} (y-1)^{{1 \ov 4} -{c \ov 2}} \zeta^{c-\ha}  \ .
 \end{gather}  
In each case we have kept only the leading term and  used~\eqref{ehb0} and the asymptotic expansion of the Bessel function at large argument. Applying the above equations to $F\left(\bar q_-, \bar q_+ ; \Delta ; w^2 \right)$ we then find~\eqref{hyas}.

\subsection{A derivation} \label{app:der1}

Here we give a derivation of~\eqref{enn}. First consider $c = |s|e^{\eta} < 1$ for which we may close the contour in~\eqref{jjj1} in the upper half plane enclosing poles at $\om' = i(n+\ep)$, which gives 
\bega 
J_1  =  (2c)^{-i \om} \sum_{n=0}^\infty {(-1)^n (2c)^n \ov n!} \Ga\le(\bar q_+ + {n \ov 2} \ri) \Ga\le(\bar q_- + {n \ov 2} \ri)   \\
=(2c)^{-i \om} \le[ \Ga (\bar q_+)\Ga (\bar q_-) F \le(\bar q_+, \bar q_-; \ha;  c^2 \ri) \ri. \cr
\le. - 2c \Ga \le(\bar q_+ + \ha \ri)\Ga \le(\bar q_-+ \ha \ri) F \le(\bar q_+ +\ha, \bar q_-+\ha; {3 \ov 2};  c^2 \ri) \ri] \ .
\label{bj0}
\end{gather}
For $c > 1$, we may close the integral of~\eqref{jjj1} in the lower half plane enclosing poles at $\om' = -2i(\bar q_{\pm} + n)$ to find 
\bega
J_1 = (2c)^{-i \om} \sum_{n=0}^\infty {2(-1)^n (2c)^{-2 n} \ov n!} \le[(2c)^{- 2 \bar q_+}  \Ga (iq -n) \Ga (2 (n+\bar q_+)) + (2c)^{- 2 \bar q_-}  \Ga (-iq -n) \Ga (2 (n+\bar q_-)) 
\ri] \cr
= 2 (2c)^{-i \om}\le[(2c)^{- 2 \bar q_+} \Ga (iq ) \Ga (2 \bar q_+) F \le(\bar q_+, \bar q_+ + \ha ; 1 - iq; c^{-2} \ri) \ri. \cr
\le. + (2c)^{- 2 \bar q_-} \Ga (-iq ) \Ga (2 \bar q_-) F \le(\bar q_-, \bar q_- + \ha ; 1 + iq; c^{-2} \ri)\ri] \ .
\label{bj1}
\end{gather}
Finally for $0 < s < s_0,$ we may close the integral for~\eqref{jjj2} in the upper half plane enclosing poles at $\om' = i(n+\ep)$ to find
\bega 
J_2  =  (2c)^{-i \om} \sum_{n=0}^\infty {\sinh\pi(\om+in)(-1)^n (2c)^n \ov n!} \Ga\le(\bar q_+ + {n \ov 2} \ri) \Ga\le(\bar q_- + {n \ov 2} \ri)   \\
=(2c)^{-i \om} \sinh\pi\om \le[ \Ga (\bar q_+)\Ga (\bar q_-) F \le(\bar q_+, \bar q_-; \ha;  c^2 \ri) \ri. \cr
\le. + 2c \Ga \le(\bar q_+ + \ha \ri)\Ga \le(\bar q_-+ \ha \ri) F \le(\bar q_+ +\ha, \bar q_-+\ha; {3 \ov 2};  c^2 \ri) \ri] \ .
\label{bj2}
\end{gather}
Notice that the quantities in square brackets in~\eqref{bj0} and~\eqref{bj2} are identical as functions of $s,$ since for~\eqref{bj0} $s < 0$ so $c = -se^{\eta}$ while for~\eqref{bj2} we have $s > 0,$ so $c = se^{\eta}.$

Applying the following identities to~\eqref{bj0} and \eqref{bj2}
\bega 
F \le(a,b;\ha; c^2 \ri) + 2 c  {\Ga (a+\ha) \Ga (b + \ha) \ov  \Ga (a) \Ga (b)} F \le(a+\ha,b+\ha ;{3 \ov 2}; c^2 \ri) \cr
= {\Ga (a+\ha) \Ga (b + \ha) \ov  \sqrt{\pi} \Ga (a+b+\ha)} F \le(2a,2b; a+b +\ha; {1+c \ov 2} \ri) \\
F \le(a,b;\ha; c^2 \ri) - 2 c   {\Ga (a+\ha) \Ga (b + \ha) \ov  \Ga (a) \Ga (b)}  F \le(a+\ha,b+\ha ;{3 \ov 2}; c^2 \ri) \cr
= {\Ga (a+\ha) \Ga (b + \ha) \ov  \sqrt{\pi} \Ga (a+b+\ha)} F \le(2a,2b; a+b +\ha; {1-c \ov 2} \ri) \ .
\end{gather}
we find for $|s| < s_0$ 
\be \label{ennAppSmallS}
	J_1 = {J_2 \ov \sinh\pi\om} = (2c)^{-i\om} {\Ga(\bar q_+) \Ga(\bar q_-) \Ga (\bar q_+ +\ha) \Ga (\bar q_- + \ha) \ov  \sqrt{\pi} \Ga (\De -i\om +\ha)} F \le(2\bar q_+,2\bar q_-; \De -i\om +\ha; {1+ se^{\eta} \ov 2} \ri) \ ,
\ee
where $J_1$ is for $s<0$ and $J_2$ for $s > 0$. 

We can now show that~\eqref{bj1} yields the exact same result~\eqref{ennAppSmallS}, so~\eqref{ennAppSmallS} also applies for $s < -s_0.$ First notice that the second term in square brackets in~\eqref{bj1} can be obtained from the first by taking $q \to -q$. Since we have $0< c^{-2} < 1$ when~\eqref{bj1} applies, the hypergeometric function in the first term in square brackets in~\eqref{ennAppSmallS} can be re-written as follows using the standard $z \to 1 - {1\ov z}$ identity of the hypergeometric function
\be \label{hypIdenJ1Big}
\begin{aligned}
	F \le(\bar q_+, \bar q_+ + \ha ; 1 - iq; c^{-2} \ri) &= 
	{\pi \Ga(1 - iq) \ov \sin\pi\le(\ha - \De + i\om\ri)} \le[ {c^{2\bar q_+} F \le(\bar q_+, \bar q_-  ; \De - i\om + \ha; 1 -c^{2} \ri) \ov \Ga(1-\bar q_-)\Ga\le(\ha-\bar q_-\ri)\Ga\le(\De -i\om + \ha\ri)} \ri. \cr
	&\le. - \le(1 - c^{-2}\ri)^{\ha -\De + i\om} {c^{2-2\bar q_-} F \le(1-\bar q_+, 1-\bar q_- ; i\om - \De + {3\ov 2}; 1 -c^{2} \ri) \ov \Ga(\bar q_+)\Ga\le(\bar q_+ + \ha\ri)\Ga\le(i\om - \De + {3 \ov 2}\ri)} \ri] \ .
\end{aligned}
\ee
Using product formulas for the gamma function, the Legendre duplication formula and~\eqref{hypIdenJ1Big} the first term in square brackets in~\eqref{bj1} is
\be \label{firstTermbj1}
\begin{aligned}
	&(2c)^{- 2 \bar q_+} \Ga (iq ) \Ga (2 \bar q_+) F \le(\bar q_+, \bar q_+ + \ha ; 1 - iq; c^{-2} \ri) \cr
	&= {\pi^2 \Ga(\bar q_+)\Ga\le(\bar q_+ + \ha\ri) \ov 2\sqrt{\pi} \sin i\pi q \sin\pi \le(\ha - \De +i\om\ri)} \le[ { F \le(\bar q_+, \bar q_-  ; \De - i\om + \ha; 1 -c^{2} \ri) \ov \Ga(1-\bar q_-)\Ga\le(\ha-\bar q_-\ri)\Ga\le(\De -i\om + \ha\ri)} \ri. \cr
	&\le. - \le(1 - c^{-2}\ri)^{\ha -\De + i\om} {c^{2(1 - \De + i\om)} F \le(1-\bar q_+, 1-\bar q_- ; i\om - \De + {3\ov 2}; 1 -c^{2} \ri) \ov \Ga(\bar q_+)\Ga\le(\bar q_+ + \ha\ri)\Ga\le(i\om - \De + {3 \ov 2}\ri)} \ri] \ .
\end{aligned}
\ee
Notice that, when we include the multiplication by the overall factor, the second term in~\eqref{firstTermbj1} is an odd function of $q.$ Since $J_1$ is obtained by adding~\eqref{firstTermbj1} to an identical expression with $q \to -q$ as in~\eqref{bj1}, this second term in~\eqref{firstTermbj1} will cancel out of the full expression for $J_1.$ The result from~\eqref{bj1} is then
\be \label{almostFinalJ1}
\begin{aligned}
	J_1 &= (2c)^{-i\om} {\pi^2 \ov \sqrt{\pi} \sin i\pi q \sin\pi \le(\ha - \De +i\om\ri) \Ga\le(\De -i\om + \ha\ri)} \cr
	&\cdot \le[ {\Ga(\bar q_+)\Ga\le(\bar q_+ + \ha\ri) \ov \Ga(1-\bar q_-)\Ga\le(\ha-\bar q_-\ri)} - {\Ga(\bar q_-)\Ga\le(\bar q_- + \ha\ri) \ov \Ga(1-\bar q_+)\Ga\le(\ha-\bar q_+\ri)}\ri] F \le(\bar q_+, \bar q_-  ; \De - i\om + \ha; 1 -c^{2} \ri) \ .
\end{aligned}
\ee
Using product formulas for gamma functions, the term in square brackets in~\eqref{almostFinalJ1} can be shown to be $\pi^{-2} \Ga(\bar q_+)\Ga\le(\bar q_+ + \ha\ri) \Ga(\bar q_-)\Ga\le(\bar q_- + \ha\ri)  \sin i\pi q \sin\pi \le(\ha - \De +i\om\ri),$ while, since $c > 0$ by definition, we have the quadratic identity for the hypergeometric function
\be \label{quadHypId}
	F \le(\bar q_+, \bar q_-  ; \De - i\om + \ha; 1 -c^{2} \ri) = F \le(2\bar q_+, 2\bar q_-  ; \De - i\om + \ha; {1 - c \ov 2} \ri) \ .
\ee
Thus with these two observations, and recalling that $J_1$ only applies for $s < 0 \Rightarrow c = -se^{\eta},$ the result~\eqref{ennAppSmallS} immediately follows, confirming that~\eqref{ennAppSmallS} applies for all $s < s_0$.

\subsection{Some summation formulas for hypergeometric functions}

In this subsection we quote some useful formulas regarding hypergeometric functions. The fourth formula from section 6.7.1 of~\cite{Prudnikov} is
\be \label{prud4th671}
	\sum_{n=0}^{\infty} {(a)_n (b')_n \ov n! (c')_k} t^n F(a+n , b ; c ; x) = F_2(a,b,b'; c,c'; x, t),
\ee
valid for $|t| + |x| < 1$. Another useful formula is (35) from~\cite{Schlosser} which gives
\be \label{schlossForm}
	F_2(a,b,b'; b,c'; x, t) = (1-x)^{-a} F\le(a,b' ; c'; {t \ov 1-x}\ri) \ .
\ee
The eighth formula from section 6.7.1 of~\cite{Prudnikov},
\be \label{prud8th671}
	\sum_{n=0}^{\infty} {(a)_n (c-b)_n \ov n! (c)_k} t^n F(a+n , b ; c+n ; x) = (1-t)^{-a} F\le(a,b ; c; {x-t \ov 1-t}\ri) \ ,
\ee
valid for $|x|,|t| < 1$, is also useful.

\section{Wavefunctions on the BTZ geometry in the large mass limit}\label{app:WKB}
In this appendix we collect the quantities describing the evolution of a large mass bulk scalar field in the BTZ geometry. We begin by considering finite transverse momentum ($q \neq 0$) and then specialize to zero transverse momentum where the expressions greatly simplify.

\subsection{General $q$} 

We first collect the quantities describing the large mass limit of the bulk mode function in \eqref{kj1}--\eqref{kj2}, which, for convenience we copy below
\bega \label{kj1App}
v_k^{(R)} (X)  = 
\begin{cases}
	\sum_{\gamma = \pm} A^{(\gamma)}_{\rk} (X) e^{ i \nu Z^{(\gamma)}_\rk (X)} \le(1 + O(\nu^{-1}) \ri) \ , &|u| > u_w \equiv {\sqrt{(1-w^2)(1+p^2 w^2)} \ov w} \cr
	A^{(f)}_{\rk} (X) e^{- \nu Z^{(f)}_\rk (X)} \le(1 + O(\nu^{-1}) \ri) \ , &|u| < u_w
\end{cases} \\
\Lam_{k k'}^{R \al} (s)   = \de_{q q'} B_{\rk \rk'}^{R\al} (s) e^{i \nu W^{R\al}_{\rk \rk'} (s)} \le(1 + O(\nu^{-1}) \ri) \ ,
\label{kj2App}
\end{gather} 
where we have taken $X = (\eta,w,\chi) \in R$ above. For $X = (\eta,w,\chi) \in L,$ there is a similar expression
\bega \label{kj1AppL}
v_k^{(L)} (X)  = 
\begin{cases}
	\sum_{\gamma = \pm} A^{(\gamma)}_{\rk} (X) e^{ i \nu \le(2u\eta - 2p\chi + Z^{(\gamma)}_\rk (X)\ri)} \le(1 + O(\nu^{-1}) \ri) \ , &|u| > u_w \equiv {\sqrt{(1-w^2)(1+p^2 w^2)} \ov w} \cr
	A^{(f)}_{\rk} (X) e^{- \nu \le(2ip\chi - 2iu\eta + Z^{(f)}_\rk (X)\ri)} \le(1 + O(\nu^{-1}) \ri) \ , &|u| < u_w
\end{cases} 
\ .
\end{gather} 

At finite $q$, there are four classical turning points in the complex $w$ plane. They are $w = \pm a_q, \pm i b_q$, with 
\be \label{tpFQ}
	a_q = \sqrt{\frac{p^2-u^2-1 + \sqrt{p^2-u^2-1)^2 + 4p^2}}{2p^2}},~~ b_q = \sqrt{-\frac{p^2-u^2-1 - \sqrt{(p^2-u^2-1)^2 + 4p^2}}{2p^2}} \ . 
\ee
The WKB phase is then given by 
\be \label{WKBPhaseFQ}
\begin{aligned}
	&Z^{(\pm)}_\rk (X) = -u\eta + p\chi \pm \le[{i \ov 2} \log\le({a_q\sqrt{w^2 + b_q^2} - i b_q\sqrt{w^2 - a_q^2} \ov a_q\sqrt{w^2 + b_q^2} + i b_q\sqrt{w^2 - a_q^2}}\ri) + |p| \log\le( \sqrt{{w^2 - a_q^2 \ov a_q^2 + b_q^2}} + \sqrt{{w^2 + b_q^2 \ov a_q^2 + b_q^2}}  \ri) \ri. \\
	&\qquad \qquad \qquad \qquad \qquad \qquad \le. - {|u| \ov 2} \log\le( {\sqrt{w^2 + b_q^2}\sqrt{1-a_q^2} + \sqrt{1 + b_q^2}\sqrt{w^2 - a_q^2} \ov \sqrt{w^2 + b_q^2}\sqrt{1-a_q^2} - \sqrt{1 + b_q^2}\sqrt{w^2 - a_q^2}} \ri) \ri] \\
	&Z^{(f)}_\rk (X) = iu\eta -i p\chi + {1 \ov 2} \log\le({a_q\sqrt{w^2 + b_q^2} + b_q\sqrt{a_q^2 - w^2} \ov a_q\sqrt{w^2 + b_q^2} - b_q\sqrt{a_q^2 - w^2}}\ri) -i |p| \log\le( \sqrt{{w^2 + b_q^2 \ov a_q^2 + b_q^2}} + i \sqrt{{a_q^2 - w^2 \ov a_q^2 + b_q^2}}   \ri) \\
	&\qquad \qquad \qquad \qquad \qquad \qquad - {i|u| \ov 2} \log\le( {\sqrt{w^2 + b_q^2}\sqrt{1-a_q^2} -i \sqrt{1 + b_q^2}\sqrt{a_q^2 - w^2} \ov \sqrt{w^2 + b_q^2}\sqrt{1-a_q^2} +i \sqrt{1 + b_q^2}\sqrt{a_q^2 - w^2}} \ri) \ .
\end{aligned}
\ee
The $O(\nu^{-\ha})$ prefactor is given by 
\be \label{WKBPrefactorFQ}
\begin{aligned}
	&A^{(\pm)}_\rk (X) = {w^{\ha} \ov \sqrt{2\nu}} \le( {1 \ov p^2 \le(w^2 - a_q^2\ri)\le(w^2 + b_q^2\ri)}\ri)^{{1 \ov 4}} e^{\pm {i\pi \ov 4}} \\
	&A^{(f)}_\rk (X) = {w^{\ha} \ov \sqrt{2\nu}} \le( {1 \ov p^2 \le(a_q^2 - w^2\ri)\le(w^2 + b_q^2\ri)}\ri)^{{1 \ov 4}} \\
\end{aligned}
\ .
\ee

The large $\nu$ limit of $\Lambda^{R\alpha}_{k k'}(s)$ is described by 
\be \label{LambdaPhasesFQ}
\begin{aligned}
	W^{RR}_{\rk \rk'}(s) &= -{i\pi \ov 2}\ep(s)(|u| - |u'|) +(u' - u)\log|s| - (u' - u)\log(i(u - u')) \\
	&+ {u' \ov 4} \log\le((1+(u'+p')^2)(1+(u'-p')^2)\ri) - {u \ov 4} \log\le((1+(u+p')^2)(1+(u-p')^2)\ri) \\
	&+ {p' \ov 4} \log\le({1+(u'+p')^2 \ov 1+(u'-p')^2}\ri) - {p' \ov 4} \log\le({1+(u+p')^2 \ov 1+(u-p')^2}\ri) \\
	&- {i \ov 4} \log\le({1+i(u'+p') \ov 1-i(u'+p')} \cdot {1+i(u'-p') \ov 1-i(u'-p')} \ri) + {i \ov 4} \log\le({1+i(u+p') \ov 1-i(u+p')} \cdot {1+i(u-p') \ov 1-i(u-p')} \ri) \\
	W^{RL}_{\rk \rk'}(s) &= -{i\pi \ov 2}(2|u+u'| - |u| - |u'|) -(u' + u)\log|s| + (u' + u)\log(i(u + u')) \\
	&- {u' \ov 4} \log\le((1+(u'+p')^2)(1+(u'-p')^2)\ri) - {u \ov 4} \log\le((1+(u+p')^2)(1+(u-p')^2)\ri) \\
	&- {p' \ov 4} \log\le({1+(u'+p')^2 \ov 1+(u'-p')^2}\ri) - {p' \ov 4} \log\le({1+(u+p')^2 \ov 1+(u-p')^2}\ri) \\
	&+ {i \ov 4} \log\le({1+i(u'+p') \ov 1-i(u'+p')} \cdot {1+i(u'-p') \ov 1-i(u'-p')} \ri) + {i \ov 4} \log\le({1+i(u+p') \ov 1-i(u+p')} \cdot {1+i(u-p') \ov 1-i(u-p')} \ri) \\
\end{aligned}
\ee
and
\be \label{LambdaPrefactorsFQ}
\begin{aligned}
	B^{RR}_{\rk \rk'}(s) &= \le(\theta(-s) + \theta(s)\ep(u)\ep(u')\ri) \sqrt{{2\pi \ov i\nu (u - u')}}  \\
	B^{RL}_{\rk \rk'}(s) &= \theta(s) \ep(u')\ep(u+u')\sqrt{{2\pi \ov i\nu (u + u')}} \ , \\
\end{aligned}
\ee
as in \eqref{kj2}.

\subsection{Expressions in the large mass limit at $q=0$}
In this subsection we now specialize to the case $q=0$ (equivalently $p = 0$) where the expressions greatly simplify. Taking the $p \to 0$ limit of \eqref{tpFQ} we find
\be \label{tpZQ}
	a_q \to a_0 = {1 \ov \sqrt{1+u^2}}, \quad b_q \to \infty \ ,
\ee
so there are actually only two (real) turning points in the $p \to 0$ limit.\footnote{All results in this subsection can be obtained from the finite $p$ results by carefully taking $p \to 0$ or instead working with the $p = 0$ effective potential from the outset.}
The bulk wavefunction at a point $X$ in the right exterior is described in the large $\nu$ limit by a WKB phase
\be \label{WKBPhase}
\begin{aligned}
	&Z^{(\pm)}_\rk (X) = -u\eta  \pm \le[{i \ov 2} \log\le({a_0 - i \sqrt{w^2 - a_0^2} \ov a_0 + i \sqrt{w^2 - a_0^2}}\ri) - {|u| \ov 2} \log\le( {\sqrt{1-a^2_0} + \sqrt{w^2 - a_0^2} \ov \sqrt{1-a^2_0} - \sqrt{w^2 - a_0^2}} \ri) \ri] \\
	&Z^{(f)}_\rk (X) = iu\eta + {1 \ov 2} \log\le({a_0 + \sqrt{a_0^2 - w^2} \ov a_0 - \sqrt{a_0^2 - w^2}}\ri) + {i|u| \ov 2} \log\le( {\sqrt{1-a^2_0} + i\sqrt{a_0^2 - w^2} \ov \sqrt{1-a^2_0} - i\sqrt{a_0^2 - w^2}} \ri)
\end{aligned}
\ee
and the $O(\nu^{-\ha})$ prefactor
\be \label{WKBPrefactor}
\begin{aligned}
	&A^{(\pm)}_\rk (X) = {w^{\ha} \ov \sqrt{2\nu}} \le( {w^2 a_0^2 \ov w^2 - a_0^2}\ri)^{{1 \ov 4}} e^{\pm {i\pi \ov 4}} = {w^{\ha} \ov \sqrt{2\nu}} \le( {1 \ov u^2 - u_w^2}\ri)^{{1 \ov 4}} e^{\pm {i\pi \ov 4}} \\
	&A^{(f)}_\rk (X) = {w^{\ha} \ov \sqrt{2\nu}} \le( {w^2 a_0^2 \ov a_0^2 - w^2}\ri)^{{1 \ov 4}} = {w^{\ha} \ov \sqrt{2\nu}} \le( {1 \ov u_w^2 - u^2}\ri)^{{1 \ov 4}} \\
\end{aligned}
\ ,
\ee
as in \eqref{kj1App}.

Using the analytic continuation techniques discussed in appendix~\ref{app:cont}, one can `extend' the BTZ wave functions to the $F$ region of the BTZ black hole. The result is (with $X = (\eta,w,\chi) \in F$)
\be \label{largeNuWFApp}
\begin{aligned}
	v^{(R)}_k(X) &= A^{(\sF)}_{\rk} (X) e^{ i \nu  Z^{(\sF)}_\rk (X) } \le(1 + O(\nu^{-1}) \ri) \\
	v^{(L)}_k(X) &= A^{(\sF)}_{\rk} (X) e^{ i \nu  \le(2u\eta +  Z^{(\sF)}_\rk (X) \ri)} \le(1 + O(\nu^{-1}) \ri) \\
\end{aligned}
 \ ,
\ee
where this `extended' wavefunction in the large mass limit is described by a single WKB phase:
\be \label{WKBPhaseFutInt}
\begin{aligned}
	&Z^{(\sF)}_\rk (X) = -u\eta  - \ep(u) \le[{i \ov 2} \log\le({a_0 - i \sqrt{w^2 - a_0^2} \ov a_0 + i \sqrt{w^2 - a_0^2}}\ri) - {|u| \ov 2} \log\le( {\sqrt{w^2-a^2_0} + \sqrt{1 - a_0^2} \ov \sqrt{w^2-a^2_0} - \sqrt{1 - a_0^2}} \ri) \ri] \ , \\
\end{aligned}
\ee
and corresponding $O(\nu^{-\ha})$ prefactor,
\be \label{WKBPrefactorFutInt}
\begin{aligned}
	&A^{(\sF)}_\rk (X) = {w^{\ha} \ov \sqrt{2\nu}} \le( {w^2 a_0^2 \ov w^2 - a_0^2}\ri)^{{1 \ov 4}} e^{- {i\pi \ov 4} \ep(u)} = {w^{\ha} \ov \sqrt{2\nu}} \le( {1 \ov u^2 - u_w^2}\ri)^{{1 \ov 4}} e^{- {i\pi \ov 4} \ep(u)} \ .
\end{aligned}
\ee
The expressions~\eqref{largeNuWFApp}--\eqref{WKBPrefactorFutInt} apply for all real values of the frequency $u$ since all frequencies become classically allowed inside the horizon.

In \eqref{kj2}, the large $\nu$ limit of $\Lambda^{R\alpha}_{k k'}(s)$ (at $q=0$) is described by 
\be \label{LambdaPhases}
\begin{aligned}
	W^{RR}_{\rk \rk'}(s) &= -{i\pi \ov 2}\ep(s)(|u| - |u'|) +(u' - u)\log|s| - (u' - u)\log(i(u - u')) \\
	&+ {u' \ov 2}\log(1+u'{}^2) - {u \ov 2}\log(1+u^2) -{i \ov 2} \log\le({1+iu' \ov 1-iu'}\ri) +{i \ov 2} \log\le({1+iu \ov 1-iu}\ri) \\
	W^{RL}_{\rk \rk'}(s) &= \theta(s) \le[ -{i\pi \ov 2}(2|u+u'| - |u| - |u'|) -(u' + u)\log|s| + (u' + u)\log(i(u + u')) \ri.\\
	&\le. - {u' \ov 2}\log(1+u'{}^2) - {u \ov 2}\log(1+u^2) + {i \ov 2} \log\le({1+iu' \ov 1-iu'}\ri) + {i \ov 2} \log\le({1+iu \ov 1-iu}\ri)\ri] \ , \\
\end{aligned}
\ee
and the $O(\nu^{-\ha})$ prefactors given in \eqref{LambdaPrefactorsFQ}.

With the expressions above and the help of~\eqref{wkbHelpExp1}--\eqref{wkbHelpExp4} one may compute the exponential factor appearing in the regime $-u_w < u < u_w$ of the integral \eqref{eax} to be
\be \label{GForbidR}
\begin{aligned}
	G^{(f)}_{R} &=  iu \log|c| + {i|u| \ov 2} \log(1-w^2) + i(u' - u)\log(i(u-u')) - {\pi \ov 2}\ep(s)|u| - \log(1-iu) \\
	&-i|u| \log\le(|u| - iw \sqrt{u_w^2 - u^2}\ri) + \log\le(1 + w\sqrt{u_w^2 - u^2}\ri) + i \theta(u)u\log(1+u^2) \\
	&-iu'\log|s| - \ha \log\le({1+iu' \ov 1-iu'}\ri) - {iu' \ov 2}\log(1+u'{}^2) + {\pi \ov 2}\ep(s)|u'| - \log w \ ,
\end{aligned}
\ee
and the analogous expression for the saddle point evaluation of $v^{(RL)}_k(X;s)$ is
\be \label{GForbidL}
\begin{aligned}
	G^{(f)}_{L} &=  iu \log|c| + {i|u| \ov 2} \log(1-w^2) - i(u' + u)\log(i(u+u')) - {\pi \ov 2}(2|u+u'| - |u| -|u'|) - \log(1-iu) \\
	&-i|u| \log\le(|u| - iw \sqrt{u_w^2 - u^2}\ri) + \log\le(1 + w\sqrt{u_w^2 - u^2}\ri) + i \theta(u)u\log(1+u^2) \\
	&+iu'\log|s| + \ha \log\le({1+iu' \ov 1-iu'}\ri) + {iu' \ov 2}\log(1+u'{}^2) - \log w \ ,
\end{aligned}
\ee
recalling $c \equiv s e^{\eta}.$ There are never any genuine saddle points of \eqref{GForbidR} or \eqref{GForbidL}. 

To obtain the expressions~\eqref{GR} and \eqref{GL} from~\eqref{WKBPhase} and~\eqref{LambdaPhases} one must use the identities 
\bega \label{wkbHelpExp1}
a_0 \pm i \sqrt{w^2 - a_0^2} = {1 \ov \sqrt{1+u^2}} (1\pm i w \sqrt{u^2-u_w^2} ) \\
 \sqrt{1-a_0^2} \pm  \sqrt{w^2-a_0^2} = {1 \ov \sqrt{1+u^2}} (|u| \pm  w \sqrt{u^2-u_w^2} ) \\
1\pm i w \sqrt{u^2-u_w^2} = {w^2(1+u^2) \ov 1\mp i w \sqrt{u^2-u_w^2}}\\
|u| \pm w \sqrt{u^2-u_w^2} = {(1-w^2)(1+u^2) \ov |u| \mp w \sqrt{u^2-u_w^2}} \ .
\label{wkbHelpExp4} 
\end{gather}

\section{Table of notation}\label{app:notation}
In this appendix we collect the notation used throughout the paper.

\begin{center}
\begin{tabular}{|c|c|}
\hline
{\bf Symbol}			&	{\bf Meaning} \\
\hline
\hline
$d$					&	Boundary spacetime dimension (bulk spacetime = $d+1$ dimensional)\\
\hline
$H_{R/L}$			&	Right/Left boundary Hamiltonian \\
\hline
$\rt$				&	Dimensionful Schwarzschild/boundary time \\
\hline
$G$					&	Generator of in-falling time evolution \\
\hline
$s$					&	In-falling time parameter \\
\hline
$x = (\rt, \vx)$		&	Boundary point, time and spatial coordinates \\
\hline
$\Sigma$				&	Boundary spatial manifold \\
\hline
$k = (\om, q)$		&	Boundary momenta, frequency and spatial momenta \\
\hline
$X = (r, x)$			&	Bulk point, radial and boundary coordinates \\
\hline
$f~(=f(r))$			&	``Emblackening factor'' in black hole metric \\
\hline
$r_0$				&	Location of the event horizon $(f(r_0) = 0)$ \\
\hline
$w$					&	AdS-Rindler/BTZ radial coordinate \\
\hline
$T$					&	Temperature w.r.t. $\rt$ \\
\hline
$\beta$				&	Inverse temperature $(1/T)$ w.r.t. $\rt$ \\
\hline
$\eta = {2\pi \ov \beta} \rt$	&	Dimensionless time w.r.t. which inverse temperature is $2\pi$ \\
\hline
$U,~V$				&	Null Kruskal coordinates for the black hole \\
\hline
\end{tabular}
\end{center}

\nopagebreak

\begin{center}
\begin{tabular}{|c|c|}
\hline
{\bf Symbol}			&	{\bf Meaning} \\
\hline
\hline
$R/L/F/P$			&	Right/Left/Future/Past regions of the eternal black hole \\
{}					&	(Section~\ref{sec:bdgen}: Left/Right/Future/Past regions of Minkowski plane) \\
\hline
$\sH^{({\rm Fock})}_{\rm BH}$	& Bulk Hilbert space of small excitations on the eternal black hole geometry \\
\hline
$\phi(X)$			&	Free bulk scalar field of mass $m$ \\
\hline
$\phi_{R/L/F/P}$		&	Restriction of bulk field to $R/L/F/P$ bulk sub-region \\
\hline
$v^{(R/L)}_{k}(X)$	&	Bulk mode function for field in $R/L$ sub-region \\
\hline
$a^{(R/L)}_{k}$		&	Bulk/boundary oscillators associated to $R/L$ sub-region \\
\hline
$\de_{qq'}$			&	Momentum delta function (Kronecker delta for discrete, $2\pi \de(q-q')$ for continuous) \\
\hline
$J$					&	Bulk $\sC \sP \sT$ operator \\
\hline
$\al,~ \beta$		&	Indices taking on values $L,~R$ \\
\hline
$\ep(\om)$			&	Sign function, $\ep(\om) = 1$ if $\om > 0$, $\ep(\om) = -1$ if $\om < 0$ \\
\hline
$w^{(\al)}_k$		&	Analytic continuation of exterior mode function \\
{}					&	Continuation in lower/upper half $U,~V$ planes for positive/negative $\om$ \\
\hline
$c^{(\al)}_k$		&	Oscillator associated to analytically continued mode function \\
\hline
$\ket{HH}$			&	Hartle-Hawking state for bulk quantum fields \\
\hline
$\widetilde{\sY}_{R/L}$	&	Bulk subalgebra in right/left exterior region \\
\hline
$\ket{0}_{R/L}$		&	Right Schwarzschild vacuum state \\
\hline
$\sB$				&	Operator algebra defined in the finite $N$ CFT \\
\hline
$\sO(x)$				&	Single-trace CFT operator \\
\hline
$\sH$				&	CFT Hilbert space \\
\hline
$\sA_{R/L, TFD}$		&	Algebra of single-trace operators in right/left CFT defined around the TFD state \\
\hline
$\ktfd$				&	Thermal field double state of two CFTs at temperature $T = 1/\beta$ \\
\hline
$E_n$				&	$n^{th}$ energy in CFT spectrum (eigenvalue of $H_{R/L}$) \\
\hline
$\ket{n}$			&	$n^{th}$ energy eigenstate of CFT \\
\hline
$\theta$				&	$\sC \sP \sT$ operation on the CFT \\
\hline
$\ket{\tilde{n}} = \theta \ket{n}$	&	$\sC \sP \sT$ conjugation of $n^{th}$ CFT energy eigenstate \\
\hline
$\rho_{\beta}$		&	Thermal density operator on CFT at inverse temperature $\beta$ \\
\hline
\end{tabular}
\end{center}

\begin{center}
\begin{tabular}{|c|c|}
\hline
{\bf Symbol}			&	{\bf Meaning} \\
\hline
\hline
$\sJ$				&	Gelfand ideal in GNS construction \\
\hline
$\sH^{(\rm GNS)}_{TFD}$	&	GNS Hilbert space built from single-trace operators and $\ktfd$ \\
\hline
$\ket{\Omega_0}$		&	GNS vacuum state \\
\hline
$\pi(a)$				&	GNS representation of single-trace operator $a$ on $\sH^{(\rm GNS)}_{TFD}$ \\
\hline
$\sY_{R/L}$			&	GNS representation of $\sA_{R/L, TFD}$ on $\sH^{(\rm GNS)}_{TFD}$ \\
\hline
$u^{(\al)}_k$		&	Local mode function for boundary generalized free field \\
\hline
$\mathbf{\De}$		&	Modular operator for a finite $N$ CFT algebra \\
\hline
$t~(= \rt / \beta)$	&	Modular time \\
\hline
$\De_0$				&	Modular operator for a single-trace subalgebra \\
\hline
$\De$				&	Scaling dimensional of single-trace operator dual to $\phi$ \\
\hline
$\nu = \De - {d \ov 2}$	&	Scaling parameter for large mass limit in bulk \\
\hline
$q_{\pm} = {{\De + i(\om \pm q)} \ov 2}$	&	Combination of momenta relevant for calculations \\
\hline
$\sA_0$				&	Single-trace algebra about the CFT vacuum \\
\hline
$\sH^{({\rm GNS})}_0$&	GNS Hilbert space of small excitations above the CFT vacuum \\
\hline
$\sY$				&	Representation of $\sA_0$ on $\sH^{({\rm GNS})}_0$ \\
\hline
$\widetilde{\sY}$	&	Algebras of bulk field about pure AdS \\
\hline
$\sH^{({\rm Fock})}_0$&	Hilbert space of small excitations above pure AdS \\
\hline
$\sR/\sL/\sF/\sP$	&	Right/Left/Future/Past AdS-Rindler regions of the Poincare patch of AdS \\
\hline
$S_R$				&	Entanglement entropy of boundary spatial subregion $R$ in the CFT \\
\hline
$\hat{R}$			&	Boundary causal completion of boundary subregion $R$ \\
\hline
$\sY_{\hat{R}}$		&	Restriction of single-trace algebra to boundary subregion $\hat{R}$ \\
\hline
$\hat{\sY}_R \equiv \le(\sY_{\hat{R}}\ri)''$	&	Weak closure of single-trace algebra in boundary subregion $\hat{R}$ \\
\hline
$\ga_R$				&	Ryu-Takayanagi surface for boundary spatial subregion $R$ \\
\hline
$E_R$				&	Homology hypersurface associated to $\ga_R$ and boundary spatial subregion $R$ \\
\hline
$\sS_{E_R}$			&	Entanglement entropy of bulk subregion $E_R$ in bulk EFT \\
\hline
\end{tabular}
\end{center}

\begin{center}
\begin{tabular}{|c|c|}
\hline
{\bf Symbol}			&	{\bf Meaning} \\
\hline
\hline
$\sM$				&	Generic von Neumann algebra used in half-sided modular inclusions \\
\hline
$J_{\sM},~\De_{\sM},~K_{\sM}$	&	Modular conjugation, operator, and Hamiltonian associated to $\sM$ \\
\hline
$\sN$				&	Subalgebra of $\sM$ obeying half-sided modular inclusion property \\
\hline
$U(s) \equiv e^{i G s}$	&	Unitary operator implementing half-sided modular translation \\
\hline
$\sigma_s(a) $	&	Adjoint action of $U(s)^{\da}$ on operator $a$ \\
\hline
$\Lambda^{\al \beta}_{kk'}(s)$	&	Decomposition of $\sigma_s\le(a^{(\al)}_k\ri)$ in terms of $a^{(\beta)}_{k'}$ \\
\hline
$\Sigma^{\al \beta}_{kk'}(s)$	&	Decomposition of $\sigma_s\le(c^{(\al)}_k\ri)$ in terms of $c^{(\beta)}_{k'}$ \\
\hline
$A^{\al \beta}_{kk'}(s)~\le(B^{\al \beta}_{kk'}(s)\ri)$	&	Positive (Negative) frequency part of $\Sigma^{\al \beta}_{kk'}(s)$ \\
\hline
$e^{i \ga_k}$		&	Undetermined phase in the action of $U(s)$ on GFFs \\
\hline
$(\eta, \chi)$		&	Dimensionless Rindler/BTZ boundary coordinates \\
\hline
$z$					&	Poincare AdS radial coordinate \\
\hline
$\zeta$				&	BTZ black hole tortoise coordinate \\
\hline
$f_k(w)$				&	AdS-Rindler/BTZ radial mode function \\
\hline
$K(X, y)$			&	Smearing function to describe $\phi(X)$ in terms of $\pi\le(\sO(y)\ri)$ \\
\hline
$\phi^{(R/L)}_q(\eta, w)$	&	Bulk field of fixed angular momentum $q$ in right/left exterior of BTZ black hole \\
\hline
$K_q(\eta, w; \eta')$	&	Smearing function for $\phi^{(R)}_q(\eta, w)$ in terms of $\pi\le(\sO_q(\eta')\ri)$ \\
\hline
$\Phi(X;s) \equiv \sigma_s\le(\phi(X)\ri)$	&	Evolution of a bulk field at $X$ under adjoint action of $U(s)^{\da}$ \\
\hline
$\sX_{\eta_0}$		&	Boundary GFF algebra supported at $\eta \leq \eta_0$ \\
\hline
$\widetilde{\sX}_{\eta_0}$	&	Bulk field algebra supported for $U \leq -e^{-\eta_0},~ V \geq 0$ \\
\hline
$e^{i \de_k}$		&	Phase shift of bulk mode function at the horizon \\
\hline
$s_0$				&	Value of in-falling time at which the evolved operator crosses the horizon \\
\hline
$\rk = (u,p) \equiv (\om, q)/\nu$	&	Re-scaled frequency and momentum for large $\nu$ limit \\
\hline
\end{tabular}
\end{center}

\end{document}